# Multi-field asymptotic homogenization of thermo-piezoelectric materials with periodic microstructure

Francesca Fantoni[1,*], Andrea Bacigalupo[1,*], Marco Paggi[1]

[1] IMT School for Advanced Studies Lucca, Piazza S.Francesco 19, 55100 Lucca, Italy

January 12, 2017

**Abstract**

This study proposes a multi-field asymptotic homogenization for the analysis of thermo-piezoelectric materials with periodic microstructures. The effect of the microstructural heterogeneity is taken into account by means of periodic perturbation functions, which derive from the solution of non homogeneous recursive cell problems defined over the unit periodic cell. A strong coupling is present between the micro displacement field and the micro electric potential field, since the mechanical and the electric problems are fully coupled in the asymptotically expanded microscale field equations. The micro displacement, the electric potential, and the relative temperature fields have been related to the macroscopic quantities and to their gradients in the derived down-scaling relations. Average field equations of infinite order have been obtained and the closed form of the overall constitutive tensors has been determined for the equivalent first-order homogenized continuum. A formal solution of such equations has been derived by means of an asymptotic expansion of the macro fields. The accuracy of the proposed formulation is assessed in relation to illustrative examples of a bi-material periodic microstructure subjected to harmonic body forces, free charge densities, and heat sources, whose periodicity is much greater than the characteristic microstructural size. The good agreement obtained between the solution of the homogenized model and the finite element solution of the original heterogeneous material problem confirms the validity of the proposed formulation.

## 1 Introduction

The use of composite materials has become increasingly attractive to fulfill industry needs due to their mechanical and physico-chemical properties. Namely composites can be tailored to meet specific design requirements. They increase strength, durability, corrosion resistance, damage tolerance, and reduce weight. They find applications in numerous engineering sectors such as for example aerospace, energy, automotive, marine, electrical, chemical, and biomedical.

In such a context, electromechanical devices (transducers, resonators, acoustic wave sensors among the others) are noteworthy. They involve the use of piezoelectric materials, exhibiting a linear coupling between the electric and the mechanical fields. Due to their crystalline structure, which becomes electrically polarized if mechanically stressed (direct piezoelectric effect), they also experience mechanical deformations in the presence of an electric field (inverse piezoelectric effect) (Yang, 2004). Piezoelectric materials include also pyroelectric materials, distinguished by thermo-electrical interactions. Pyroelectricity characterizes the dependence of spontaneous polarization of crystalline materials to temperature. It is the property whereby a change of temperature generates the presence of a charge on the surface of the pyroelectric material itself and the direction of the pyroelectric current depends on the nature of thermal gradients (Moulson and Herbert, 2003; Batra and Aggarwal, 2013). A correct characterization of piezoelectric, pyroelectric, and, more generally, temperature-dependent properties of such material systems is of great importance to enable applications in sensing, transduction, actuation, energy harvesting, thermal imaging, gas analysis, radiometry, and in the biomedical field, allowing to reach the optimal device design in any operating conditions. Production of such devices is expanding in small formats (nanometer scale), compatibly with microfabrication processes for micro-electronic applications (MEMS-based devices), since thin films offer distinct advantages for thermal to electrical energy conversion over bulk samples (Kimata, 2013). Such materials are microstructured, being characterized by a periodicity or a quasi-periodicity.

---

[*]Corresponding authors: Tel:+39 0583 4326613,
E-mail addresses: francesca.fantoni@imtlucca.it; andrea.bacigalupo@imtlucca.it

Multiscale techniques constitute a powerful method to study thermo-piezoelectric materials whose macroscopic behaviour is influenced by multi-field phenomena occurring at the microscale. In order to obtain an accurate and synthetic description of the macroscopic thermo-piezoelectric properties without performing a computationally expensive modeling of the whole heterogeneous medium, homogenization techniques can be efficiently employed. For composites with a periodic microstructure, in particular, homogenization represents a suitable tool for modeling the effects of the microscopic phases on the overall properties. Such techniques developed for determining the overall static and dynamic properties of periodic composite materials can be generally distinguished in asympthotic methods (Allaire, 1992; Andrianov et al., 2008; Auriault and Bonnet, 1985; Bacigalupo, 2014; Bakhvalov and Panasenko, 1984; Bensoussan et al., 1978; Gambin and Kröner, 1989; Sanchez-Palencia, 1974; Tran et al., 2012), variational-asymptotic techniques (Bacigalupo and Gambarotta, 2014a,b; Bacigalupo et al., 2014; Peerlings and Fleck, 2004; Smyshlyaev and Cherednichenko, 2000), and computational approaches based on a polynomial approximation of the macroscopic fields (Addessi et al., 2013; Bacca et al., 2013a,b,c; Bacigalupo and De Bellis, 2015; Bacigalupo and Gambarotta, 2010; Bigoni and Drugan, 2007; Forest and Sab, 1998; Forest and Trinh, 2011; Miehe et al., 1999; Kouznetsova et al., 2002, 2004). In all these formulations, the equivalent homogenized macroscopic medium is described trough a first-order continuum or, alternatively, by means of a nonlocal one.

An extension of the above mentioned techniques to a multi-field case has concerned thermomechanics (Aboudi et al., 2001; Kanouté et al., 2009; Zhang et al., 2007), piezoelectricity (Deraemaeker and Nasser, 2010), and thermo-diffusive phenomena (Bacigalupo et al., 2014, 2016a,b). In (Salvadori et al., 2014) the effects of the microstructure on macroscopic elastic and thermodiffusive properties and on the coupling between them has been investigated in relation to lithium-ion batteries and solid oxide fuel cells.

In this framework, the present work describes an asymptotic homogenization technique for static analysis of thermo-piezoelectric materials with a periodic microstructure, making a progress from the approaches proposed in (Bacigalupo, 2014; Bacigalupo and Gambarotta, 2014b; Bakhvalov and Panasenko, 1984; Smyshlyaev and Cherednichenko, 2000).

In this study, the asymptotic expansion of the micro-displacement, electric potential, and relative temperature fields allows to obtain from the expression of the micro-scale field equations a series of recursive differential problems defined over the periodic unit cell in terms of the microscopic variables. Solvability conditions for such non homogeneous recursive cell problems lead to the derivation of down-scaling relations, which correlate the microscopic fields to the macroscopic ones and to their gradients by means of perturbation functions. Those functions depend exclusively on the geometrical and physico-mechanical properties of the material under consideration and account for the influence of microstructural inhomogeneities on the displacement, electric potential, and relative temperature, and on their coupling. The exact expressions of the overall constitutive tensors have been determined for the class of periodic thermo-piezoelectric materials considered herein, expressed in terms of perturbation functions and constitutive tensors at the microscale. Average field equations of infinite order have been derived following the method proposed in (Bacigalupo, 2014; Smyshlyaev and Cherednichenko, 2000), substituting the down-scaling relations into the micro-field equations. Their formal solution has been obtained by performing an asymptotic expansion of the macro fields in terms of the characteristic microstructural size. Finally, truncating such expansions to the zeroth order, field equations for the equivalent Cauchy thermo-piezoelectric medium are derived.

Section 2 of the paper is devoted to describe the governing equations at the microscale. Cell problems and their solution are accurately presented in Section 3, which shows the existence of a strong coupling between the mechanical and the electrical fields and the related perturbation functions. Section 4 presents the down-scaling and the up-scaling relations, which allow to express the macroscopic displacement, electric potential, and temperature fields in terms of the corresponding microscopic ones. Average field equations of infinite order are described in Section 5, together with the closed form of the overall thermo-piezoelectric constitutive parameters and the field equations of the equivalent first-order continuum.

The homogenization technique proposed in the paper is finally applied to the case of a bi-material periodic microstructure subjected to periodic body forces, charge densities, and heat sources. The solution of the first-order homogenized problem is then compared with the numerical results obtained by a finite element analysis of the heterogeneous problem, in order to assess the validity of the proposed homogenization procedure and its accuracy. In this regard, a thermo-piezoelectric element has been formulated and implemented in the finite element program FEAP (Zienkiewicz and Taylor, 1977) to numerically solve the coupled thermo-electromechanical problem. Finally, conclusions are proposed in Section 7.

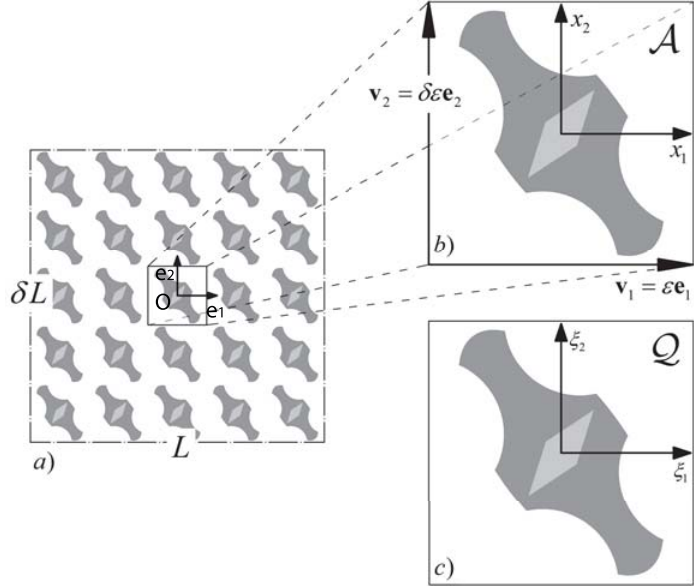

Figure 1: *(a) Periodic microstructure of the heterogeneous medium with structural characteristic size L; (b)Periodic cell $\mathcal{A}$ with microstructural characteristic size $\varepsilon$ and periodicity vectors $\mathbf{v}_1$ and $\mathbf{v}_2$; (c) Periodic unit cell $\mathcal{Q}$.*

## 2 Field equations of the periodic heterogeneous material and kinematic multi-scale description

Consider a heterogeneous composite material characterized by a periodic microstructure under the assumption of small strains. The continuum is described as a linear thermo-piezoelectric Cauchy medium (Mindlin, 1974) subject to stresses induced by body forces, free charge densities, and temperature changes. Restricting the notation to the two-dimensional case for the sake of simplicity and without losing generality, the position vector $\mathbf{x} = x_1\,\mathbf{e}_1 + x_2\,\mathbf{e}_2$ characterizes each material point in the orthogonal reference system $\{O, \mathbf{e}_1, \mathbf{e}_2\}$ with origin at point $O$, as shown in Figure 1(a).

The continuum is described by the micro displacement field $\mathbf{u}(\mathbf{x}) = u_i\,\boldsymbol{e}_i$, the micro electrical potential field $\phi(\mathbf{x})$, and the micro relative temperature field $\theta(\mathbf{x}) = T(\mathbf{x}) - T_0$, where $T_0$ is a reference stress-free temperature. The entire periodic medium can be obtained by spanning a periodic cell $\mathcal{A} = [0, \varepsilon] \times [0, \delta\varepsilon]$ by the two orthogonal periodicity vectors defined as $\mathbf{v}_1 = d_1\,\mathbf{e}_1 = \varepsilon\,\mathbf{e}_1$, and $\mathbf{v}_2 = d_2\,\mathbf{e}_2 = \delta\varepsilon\,\mathbf{e}_2$, being $\varepsilon$ the characteristic size of the cell $\mathcal{A}$ (Figure 1(b)). Because of the $\mathcal{A}$-periodicity of the material, the micro-elasticity tensor $\mathbb{C}^{(m,\varepsilon)}(\mathbf{x}) = C_{ijkl}^{(m,\varepsilon)}\boldsymbol{e}_i \otimes \boldsymbol{e}_j \otimes \boldsymbol{e}_k \otimes \boldsymbol{e}_l$, together with the micro-dielectric permittivity tensor (at constant strain) $\boldsymbol{\beta}^{(m,\varepsilon)}(\mathbf{x}) = \beta_{ij}^{(m,\varepsilon)}\boldsymbol{e}_i \otimes \boldsymbol{e}_j$, and the micro-heat conduction tensor $\boldsymbol{K}^{(m,\varepsilon)}(\mathbf{x}) = K_{ij}^{(m,\varepsilon)}\boldsymbol{e}_i \otimes \boldsymbol{e}_j$ obey the following properties:

$$\mathbb{C}^{(m,\varepsilon)}(\mathbf{x} + \mathbf{v}_i) = \mathbb{C}^{(m,\varepsilon)}(\mathbf{x}), \ \ i=1,2, \ \ \forall \mathbf{x} \in \mathcal{A}, \tag{1a}$$

$$\boldsymbol{\beta}^{(m,\varepsilon)}(\mathbf{x} + \mathbf{v}_i) = \boldsymbol{\beta}^{(m,\varepsilon)}(\mathbf{x}), \ \ i=1,2, \ \ \forall \mathbf{x} \in \mathcal{A}, \tag{1b}$$

$$\boldsymbol{K}^{(m,\varepsilon)}(\mathbf{x} + \mathbf{v}_i) = \boldsymbol{K}^{(m,\varepsilon)}(\mathbf{x}), \ \ i=1,2, \ \ \forall \mathbf{x} \in \mathcal{A}, \tag{1c}$$

where the superscript $m$ refers to the *microscale*. Analogous relations hold for the micro-coupling tensors, namely the piezoelectric stress/charge tensor $\boldsymbol{e}^{(m,\varepsilon)}(\mathbf{x}) = e_{ijk}^{(m,\varepsilon)}\boldsymbol{e}_i \otimes \boldsymbol{e}_j \otimes \boldsymbol{e}_k$, the thermal dilatation tensor $\boldsymbol{\alpha}^{(m,\varepsilon)}(\mathbf{x}) = \alpha_{ij}^{(m,\varepsilon)} \boldsymbol{e}_i \otimes \boldsymbol{e}_j$, and the pyroelectric vector $\boldsymbol{\gamma}^{(m,\varepsilon)}(\mathbf{x}) = \gamma_i^{(m,\varepsilon)} \boldsymbol{e}_i$:

$$\boldsymbol{e}^{(m,\varepsilon)}(\mathbf{x} + \mathbf{v}_i) = \boldsymbol{e}^{(m,\varepsilon)}(\mathbf{x}), \quad i = 1, 2, \quad \forall \mathbf{x} \in \mathcal{A}, \tag{2a}$$

$$\boldsymbol{\alpha}^{(m,\varepsilon)}(\mathbf{x} + \mathbf{v}_i) = \boldsymbol{\alpha}^{(m,\varepsilon)}(\mathbf{x}), \quad i = 1, 2, \quad \forall \mathbf{x} \in \mathcal{A}, \tag{2b}$$

$$\boldsymbol{\gamma}^{(m,\varepsilon)}(\mathbf{x} + \mathbf{v}_i) = \boldsymbol{\gamma}^{(m,\varepsilon)}(\mathbf{x}), \quad i = 1, 2, \quad \forall \mathbf{x} \in \mathcal{A}. \tag{2c}$$

Rescaling the size of the periodic cell $\mathcal{A}$ by the characteristic length $\varepsilon$, it is possible to reproduce the periodic microstructure by a non dimensional unit cell $\mathcal{Q} = [0,1] \times [0,\delta]$ shown in Figure 1-(c). In this regard, two variables, the macroscopic (slow) one $\mathbf{x} \in \mathcal{A}$, and the microscopic (fast) one $\boldsymbol{\xi} = \mathbf{x}/\varepsilon \in \mathcal{Q}$ allow the separation of the scales distinguishing from the macro and the micro scales, see for example (Allaire, 1992; Bacigalupo, 2014; Bakhvalov and Panasenko, 1984; Gambin and Kröner, 1989; Peerlings and Fleck, 2004; Smyshlyaev and Cherednichenko, 2000; Tran et al., 2012). Therefore, the constitutive tensors (1a)-(2c) are made only dependent on the microscopic variable $\boldsymbol{\xi}$. They result $\mathcal{Q}$-periodic and defined over $\mathcal{Q}$ as

$$\begin{aligned} \mathbb{C}^{(m,\varepsilon)}(\mathbf{x}) = \mathbb{C}^m(\boldsymbol{\xi} = \mathbf{x}/\varepsilon), \quad \boldsymbol{\beta}^{(m,\varepsilon)}(\mathbf{x}) = \boldsymbol{\beta}^m(\boldsymbol{\xi} = \mathbf{x}/\varepsilon), \quad \boldsymbol{K}^{(m,\varepsilon)}(\mathbf{x}) = \boldsymbol{K}^m(\boldsymbol{\xi} = \mathbf{x}/\varepsilon), \\ \boldsymbol{e}^{(m,\varepsilon)}(\mathbf{x}) = \boldsymbol{e}^m(\boldsymbol{\xi} = \mathbf{x}/\varepsilon), \quad \boldsymbol{\alpha}^{(m,\varepsilon)}(\mathbf{x}) = \boldsymbol{\alpha}^m(\boldsymbol{\xi} = \mathbf{x}/\varepsilon), \quad \boldsymbol{\gamma}^{(m,\varepsilon)}(\mathbf{x}) = \boldsymbol{\gamma}^m(\boldsymbol{\xi} = \mathbf{x}/\varepsilon). \end{aligned} \tag{3}$$

Constitutive relations determining the micro-stress $\boldsymbol{\sigma}(\mathbf{x})$, the micro-electric displacement $\mathbf{D}(\mathbf{x})$, and the micro-heat flux $\mathbf{q}(\mathbf{x})$ are given by the following formulae (Mindlin, 1974):

$$\boldsymbol{\sigma}(\mathbf{x}) = \mathbb{C}^m\left(\frac{\mathbf{x}}{\varepsilon}\right) \boldsymbol{\varepsilon}(\mathbf{x}) + \boldsymbol{e}^m\left(\frac{\mathbf{x}}{\varepsilon}\right) \nabla \phi(\mathbf{x}) - \boldsymbol{\alpha}^m\left(\frac{\mathbf{x}}{\varepsilon}\right) \theta(\mathbf{x}), \tag{4a}$$

$$\mathbf{D}(\mathbf{x}) = \tilde{\boldsymbol{e}}^m\left(\frac{\mathbf{x}}{\varepsilon}\right) \boldsymbol{\varepsilon}(\mathbf{x}) - \boldsymbol{\beta}^m\left(\frac{\mathbf{x}}{\varepsilon}\right) \nabla \phi(\mathbf{x}) + \boldsymbol{\gamma}^m\left(\frac{\mathbf{x}}{\varepsilon}\right) \theta(\mathbf{x}), \tag{4b}$$

$$\mathbf{q}(\mathbf{x}) = -\boldsymbol{K}^m\left(\frac{\mathbf{x}}{\varepsilon}\right) \nabla \theta(\mathbf{x}), \tag{4c}$$

where $\boldsymbol{\varepsilon} = \mathrm{sym}\nabla\mathbf{u}$ represents the micro-strain tensor and $\tilde{\boldsymbol{e}}^m = \tilde{e}^m_{ijk} \boldsymbol{e}_i \otimes \boldsymbol{e}_j \otimes \boldsymbol{e}_k$, with $\tilde{e}^m_{ijk} = e^m_{jki}$. Under the assumption of a quasi-static process, the time derivative of $\mathbf{u}(\mathbf{x})$ and $\theta(\mathbf{x})$ has been neglected in the heat flux equation (4c) (Mindlin, 1974; Nowacki, 1986). The stress field $\boldsymbol{\sigma}(\mathbf{x})$, the electric displacement field $\mathbf{D}(\mathbf{x})$, and the heat flux $\mathbf{q}(\mathbf{x})$ satisfy, respectively, the following local balance equations

$$\nabla \cdot \boldsymbol{\sigma}(\mathbf{x}) + \mathbf{b}(\mathbf{x}) = \mathbf{0}, \quad \nabla \cdot \mathbf{D}(\mathbf{x}) - \rho_e(\mathbf{x}) = 0, \quad \nabla \cdot \mathbf{q}(\mathbf{x}) + r(\mathbf{x}) = 0, \tag{5}$$

where body forces $\mathbf{b}(\mathbf{x})$, free charge densities $\rho_e(\mathbf{x})$, and heat sources $r(\mathbf{x})$ are made dependent only on the slow variable $\mathbf{x}$. It is assumed that volume forces are $\mathcal{L}$-periodic with $\mathcal{L} = [0, L] \times [0, \delta L]$, and have vanishing mean values on $\mathcal{L}$. The structural length $L$ has to be much greater than the microstructural length $\varepsilon$ ($L >> \varepsilon$) for the validity of the scales separability condition, so that $\mathcal{L}$ can be considered as a true representative portion of the whole body. The following partial differential equations result from the substitution of constitutive relations (4a)-(4c) into the local balance equations (5)

$$\nabla \cdot \left(\mathbb{C}^m\left(\frac{\mathbf{x}}{\varepsilon}\right) \nabla \mathbf{u}(\mathbf{x})\right) + \nabla \cdot \left(\boldsymbol{e}^m\left(\frac{\mathbf{x}}{\varepsilon}\right) \nabla \phi(\mathbf{x})\right) - \nabla \cdot \left(\boldsymbol{\alpha}^m\left(\frac{\mathbf{x}}{\varepsilon}\right) \theta(\mathbf{x})\right) + \mathbf{b}(\mathbf{x}) = \mathbf{0}, \tag{6a}$$

$$\nabla \cdot \left(\tilde{\boldsymbol{e}}^m\left(\frac{\mathbf{x}}{\varepsilon}\right) \nabla \mathbf{u}(\mathbf{x})\right) - \nabla \cdot \left(\boldsymbol{\beta}^m\left(\frac{\mathbf{x}}{\varepsilon}\right) \nabla \phi(\mathbf{x})\right) + \nabla \cdot \left(\boldsymbol{\gamma}^m\left(\frac{\mathbf{x}}{\varepsilon}\right) \theta(\mathbf{x})\right) - \rho_e(\mathbf{x}) = 0, \tag{6b}$$

$$\nabla \cdot \left(\boldsymbol{K}^m\left(\frac{\mathbf{x}}{\varepsilon}\right) \nabla \theta(\mathbf{x})\right) + r(\mathbf{x}) = 0. \tag{6c}$$

Denoting with $[[f]] = f^i(\Sigma) - f^j(\Sigma)$ the jump of the values of function $f$ at the interface $\Sigma$ between two different phases $i$ and $j$ in the periodic cell $\mathcal{A}$, the following continuity conditions hold for a perfectly bounded interface

$$[[\mathbf{u}(\mathbf{x})]]|_{\boldsymbol{x}\in\Sigma} = \mathbf{0}, \quad \left[\left[\left(\mathbb{C}^m\left(\frac{\mathbf{x}}{\varepsilon}\right) \nabla \mathbf{u}(\mathbf{x}) + \boldsymbol{e}^m\left(\frac{\mathbf{x}}{\varepsilon}\right) \nabla \phi(\mathbf{x}) - \boldsymbol{\alpha}^m\left(\frac{\mathbf{x}}{\varepsilon}\right) \theta(\mathbf{x})\right) \cdot \mathbf{n}\right]\right]\bigg|_{\boldsymbol{x}\in\Sigma} = \mathbf{0}, \tag{7a}$$

$$[[\phi(\mathbf{x})]]|_{\boldsymbol{x}\in\Sigma} = \mathbf{0}, \quad \left[\left[\left(\tilde{\boldsymbol{e}}^m\left(\frac{\mathbf{x}}{\varepsilon}\right) \nabla \mathbf{u}(\mathbf{x}) - \boldsymbol{\beta}^m\left(\frac{\mathbf{x}}{\varepsilon}\right) \nabla \phi(\mathbf{x}) + \boldsymbol{\gamma}^m\left(\frac{\mathbf{x}}{\varepsilon}\right) \theta(\mathbf{x})\right) \cdot \mathbf{n}\right]\right]\bigg|_{\boldsymbol{x}\in\Sigma} = \mathbf{0}, \tag{7b}$$

$$[[\theta(\mathbf{x})]]|_{\boldsymbol{x}\in\Sigma} = \mathbf{0}, \quad \left[\left[\boldsymbol{K}^m\left(\frac{\mathbf{x}}{\varepsilon}\right) \nabla \theta(\mathbf{x}) \cdot \mathbf{n}\right]\right]\bigg|_{\boldsymbol{x}\in\Sigma} = \mathbf{0}, \tag{7c}$$

where $\mathbf{n}$ denotes the outward normal to the interface $\Sigma$. Because of the $\mathcal{Q}$-periodicity of microscopic constitutive tensors in equations (6a)-(6c) and interface conditions (7a)-(7c), and taking into account the $\mathcal{L}$-periodicity of volume forces, the micro-fields depend on both the slow variable $\mathbf{x}$ and the fast one $\boldsymbol{\xi}$ and can be expressed in the following form

$$\mathbf{u} = \mathbf{u}\left(\mathbf{x}; \boldsymbol{\xi} = \frac{\mathbf{x}}{\varepsilon}\right), \quad \phi = \phi\left(\mathbf{x}; \boldsymbol{\xi} = \frac{\mathbf{x}}{\varepsilon}\right), \quad \theta = \theta\left(\mathbf{x}; \boldsymbol{\xi} = \frac{\mathbf{x}}{\varepsilon}\right).$$

Deriving the solution of the set of PDEs (6a)-(6c) could be particularly complex both analytically and numerically because of the $\mathcal{Q}$-periodicity of their coefficients. The use of homogenization techniques to replace the continuum with its microstructure by an equivalent homogeneous one can be very effective. In this regard, the formulation of an equivalent first-order thermo-piezoelectric continuum will be derived in what follows and the exact expressions of the overall constitutive tensors will be determined. This allows to replace the solution of equations (6a)-(6c) of the heterogeneous medium with the solution of an equivalent homogeneous material whose solutions are close to those of (6a)-(6c), but whose coefficients are not affected by rapid oscillations due to the underlying microstructure. By means of the asymptotic homogenization technique, the computational cost of solving the microscopic field equations (6a)-(6c) can therefore be overcome and the global behaviour of the composite structure can be accurately described. In the equivalent homogenized continuum, the macroscopic fields in each material point $\mathbf{x}$ are indicated as $\mathbf{U}(\mathbf{x}) = U_i\, \boldsymbol{e}_i$ for the displacement, $\Phi(\mathbf{x})$ for the electric potential, and $\Theta(\mathbf{x})$ for the relative temperature.

## 3 Asymptotic expansions of microscopic field equations

The three micro-fields governing the problem can be represented trough an asymptotic expansion in powers of the microstructural length scale $\varepsilon$ that mantains the micro and the macro scales separated by keeping the slow variable $\mathbf{x}$ distinguished from the fast one $\boldsymbol{\xi} = \mathbf{x}/\varepsilon$. Following the approach presented in (Bakhvalov and Panasenko, 1984; Bensoussan et al., 1978; Sanchez-Palencia, 1974), the asymptotic expansions take the general form

$$u_h\left(\mathbf{x}, \frac{\mathbf{x}}{\varepsilon}\right) = \sum_{l=0}^{+\infty} \varepsilon^l u_h^{(l)}\left(\mathbf{x}, \frac{\mathbf{x}}{\varepsilon}\right) = u_h^{(0)}\left(\mathbf{x}, \frac{\mathbf{x}}{\varepsilon}\right) + \varepsilon u_h^{(1)}\left(\mathbf{x}, \frac{\mathbf{x}}{\varepsilon}\right) + \varepsilon^2 u_h^{(2)}\left(\mathbf{x}, \frac{\mathbf{x}}{\varepsilon}\right) + O(\varepsilon^3), \tag{8a}$$

$$\phi\left(\mathbf{x}, \frac{\mathbf{x}}{\varepsilon}\right) = \sum_{l=0}^{+\infty} \varepsilon^l \phi^{(l)}\left(\mathbf{x}, \frac{\mathbf{x}}{\varepsilon}\right) = \phi^{(0)}\left(\mathbf{x}, \frac{\mathbf{x}}{\varepsilon}\right) + \varepsilon \phi^{(1)}\left(\mathbf{x}, \frac{\mathbf{x}}{\varepsilon}\right) + \varepsilon^2 \phi^{(2)}\left(\mathbf{x}, \frac{\mathbf{x}}{\varepsilon}\right) + O(\varepsilon^3), \tag{8b}$$

$$\theta\left(\mathbf{x}, \frac{\mathbf{x}}{\varepsilon}\right) = \sum_{l=0}^{+\infty} \varepsilon^l \theta^{(l)}\left(\mathbf{x}, \frac{\mathbf{x}}{\varepsilon}\right) = \theta^{(0)}\left(\mathbf{x}, \frac{\mathbf{x}}{\varepsilon}\right) + \varepsilon \theta^{(1)}\left(\mathbf{x}, \frac{\mathbf{x}}{\varepsilon}\right) + \varepsilon^2 \theta^{(2)}\left(\mathbf{x}, \frac{\mathbf{x}}{\varepsilon}\right) + O(\varepsilon^3). \tag{8c}$$

Exploiting the property $\frac{D}{Dx_j}f(\mathbf{x}; \boldsymbol{\xi} = \frac{\mathbf{x}}{\varepsilon}) = \left(\frac{\partial f}{\partial x_j} + \frac{1}{\varepsilon}\frac{\partial f}{\partial \xi_j}\right)\Big|_{\boldsymbol{\xi}=\frac{\mathbf{x}}{\varepsilon}} = \left(\frac{\partial f}{\partial x_j} + \frac{1}{\varepsilon}f_{,j}\right)\Big|_{\boldsymbol{\xi}=\frac{\mathbf{x}}{\varepsilon}}$, expansions (8a)-(8c) can be substituted into the microscopic field equations (6a)-(6c). In particular, equation (6a) takes the form

$$\left\{\varepsilon^{-2}\left[\left(C_{ijkl}^m\, u_{k,l}^{(0)}\right)_{,j} + \left(e_{ijk}^m\, \phi_{,k}^{(0)}\right)_{,j}\right] + \right.$$

$$+ \ \varepsilon^{-1}\left\{\left[C_{ijkl}^m\left(\frac{\partial u_k^{(0)}}{\partial x_l} + u_{k,l}^{(1)}\right)\right]_{,j} + \frac{\partial}{\partial x_j}\left(C_{ijkl}^m\, u_{k,l}^{(0)}\right) + \left[e_{ijk}^m\left(\frac{\partial \phi^{(0)}}{\partial x_k} + \phi_{,k}^{(1)}\right)\right]_{,j} + \frac{\partial}{\partial x_j}\left(e_{ijk}^m\, \phi_{,k}^{(0)}\right) + \right.$$

$$- \ \left(\alpha_{ij}^m\, \theta^{(0)}\right)_{,j}\right\} +$$

$$+ \ \left\{\left[C_{ijkl}^m\left(\frac{\partial u_k^{(1)}}{\partial x_l} + u_{k,l}^{(2)}\right)\right]_{,j} + \frac{\partial}{\partial x_j}\left[C_{ijkl}^m\left(\frac{\partial u_k^{(0)}}{\partial x_l} + u_{k,l}^{(1)}\right)\right] + \left[e_{ijk}^m\left(\frac{\partial \phi^{(1)}}{\partial x_k} + \phi_{,k}^{(2)}\right)\right]_{,j} + \right.$$

$$+ \ \frac{\partial}{\partial x_j}\left[e_{ijk}^m\left(\frac{\partial \phi^{(0)}}{\partial x_k} + \phi_{,k}^{(1)}\right)\right] - \left(\alpha_{ij}^m\, \theta^{(1)}\right)_{,j} - \frac{\partial}{\partial x_j}\left(\alpha_{ij}^m \theta^{(0)}\right)\right\} +$$

$$
\begin{aligned}
+ \quad & \varepsilon \left\{ \left[ C^m_{ijkl} \left( \frac{\partial u^{(2)}_k}{\partial x_l} + u^{(3)}_{k,l} \right) \right]_{,j} + \frac{\partial}{\partial x_j}\left[ C^m_{ijkl}\left(\frac{\partial u^{(1)}_k}{\partial x_l}+u^{(2)}_{k,l}\right)\right] + \left[e^m_{ijk}\left(\frac{\partial \phi^{(2)}}{\partial x_k}+\phi^{(3)}_{,k}\right)\right]_{,j} + \right. \\
+ \quad & \left. \frac{\partial}{\partial x_j}\left[e^m_{ijk}\left(\frac{\partial \phi^{(1)}}{\partial x_k}+\phi^{(2)}_{,k}\right)\right] - \left(\alpha^m_{ij}\theta^{(2)}\right)_{,j} - \frac{\partial}{\partial x_j}\left(\alpha^m_{ij}\theta^{(1)}\right) \right\} + O(\varepsilon^2) \right\}\bigg|_{\boldsymbol{\xi}=\frac{\mathbf{x}}{\varepsilon}} + b_i(\mathbf{x})=0. \quad (9)
\end{aligned}
$$

Similarly, equation (6b) can be written as

$$
\begin{aligned}
& \left\{ \varepsilon^{-2}\left[\left(e^m_{kli}u^{(0)}_{k,l}\right)_{,i} - \left(\beta^m_{il}\phi^{(0)}_{,l}\right)_{,i}\right] + \right. \\
+ \quad & \varepsilon^{-1}\left\{\left[e^m_{kli}\left(\frac{\partial u^{(0)}_k}{\partial x_l}+u^{(1)}_{k,l}\right)\right]_{,i} + \frac{\partial}{\partial x_i}\left(e^m_{kli}u^{(0)}_{k,l}\right) - \left[\beta^m_{il}\left(\frac{\partial \phi^{(0)}}{\partial x_l}+\phi^{(1)}_{,l}\right)\right]_{,i} - \frac{\partial}{\partial x_i}\left(\beta^m_{il}\phi^{(0)}_{,l}\right) + \left(\gamma^m_i\theta^{(0)}\right)_{,i}\right\} \\
+ \quad & \left\{\left[e^m_{kli}\left(\frac{\partial u^{(1)}_k}{\partial x_l}+u^{(2)}_{k,l}\right)\right]_{,i} + \frac{\partial}{\partial x_i}\left[e^m_{kli}\left(\frac{\partial u^{(0)}_k}{\partial x_l}+u^{(1)}_{k,l}\right)\right] - \left[\beta^m_{il}\left(\frac{\partial \phi^{(1)}}{\partial x_l}+\phi^{(2)}_{,l}\right)\right]_{,i} + \right. \\
- \quad & \left. \frac{\partial}{\partial x_i}\left[\beta^m_{il}\left(\frac{\partial \phi^{(0)}}{\partial x_l}+\phi^{(1)}_{,l}\right)\right] + \left(\gamma^m_i\theta^{(1)}\right)_{,i} + \frac{\partial}{\partial x_i}\left(\gamma^m_i\theta^{(0)}\right)\right\} + \\
+ \quad & \varepsilon\left\{\left[e^m_{kli}\left(\frac{\partial u^{(2)}_k}{\partial x_l}+u^{(3)}_{k,l}\right)\right]_{,i} + \frac{\partial}{\partial x_i}\left[e^m_{kli}\left(\frac{\partial u^{(1)}_k}{\partial x_l}+u^{(2)}_{k,l}\right)\right] - \left[\beta^m_{il}\left(\frac{\partial \phi^{(2)}}{\partial x_l}+\phi^{(3)}_{,l}\right)\right]_{,i} + \right. \\
- \quad & \left. \frac{\partial}{\partial x_i}\left[\beta^m_{il}\left(\frac{\partial \phi^{(1)}}{\partial x_l}+\phi^{(2)}_{,l}\right)\right] + \left(\gamma^m_i\theta^{(2)}\right)_{,i} + \frac{\partial}{\partial x_i}\left(\gamma^m_i\theta^{(1)}\right)\right\} + O(\varepsilon^2)\right\}\bigg|_{\boldsymbol{\xi}=\frac{\mathbf{x}}{\varepsilon}} - \rho_e(\mathbf{x})=0, \quad (10)
\end{aligned}
$$

and finally, eq. (6c) results

$$
\begin{aligned}
& \left\{\varepsilon^{-2}\left(K^m_{ij}\theta^{(0)}_{,j}\right)_{,i} + \varepsilon^{-1}\left\{\left[K^m_{ij}\left(\frac{\partial \theta^{(0)}}{\partial x_j}+\theta^{(1)}_{,j}\right)\right]_{,i} + \frac{\partial}{\partial x_i}\left(K^m_{ij}\theta^{(0)}_{,j}\right)\right\} + \right. \\
+ \quad & \left[K^m_{ij}\left(\frac{\partial \theta^{(1)}}{\partial x_j}+\theta^{(2)}_{,j}\right)\right]_{,i} + \frac{\partial}{\partial x_i}\left[K^m_{ij}\left(\frac{\partial \theta^{(0)}}{\partial x_j}+\theta^{(1)}_{,j}\right)\right] + \\
+ \quad & \varepsilon\left\{\left[K^m_{ij}\left(\frac{\partial \theta^{(2)}}{\partial x_j}+\theta^{(3)}_{,j}\right)\right]_{,i} + \frac{\partial}{\partial x_i}\left[K^m_{ij}\left(\frac{\partial \theta^{(1)}}{\partial x_j}+\theta^{(2)}_{,j}\right)\right]\right\} + \\
+ \quad & O(\varepsilon^2)\}\big|_{\boldsymbol{\xi}=\frac{\mathbf{x}}{\varepsilon}} + r(\mathbf{x})=0. \quad (11)
\end{aligned}
$$

Interface conditions (7a)-(7c) can be expressed in terms of the fast variable $\boldsymbol{\xi}$, since the micro fields $\boldsymbol{u}(\mathbf{x};\boldsymbol{\xi})$, $\phi(\mathbf{x};\boldsymbol{\xi})$, and $\theta(\mathbf{x};\boldsymbol{\xi})$ are assumed to be $\mathcal{Q}$-periodic regular functions of the variable $\mathbf{x}$ (Bakhvalov and Panasenko, 1984). Denoting with $\Sigma_1$ the interface between two different phases in the unit cell $\mathcal{Q}$, and taking into account asymptotic expansions (8a)-(8c), interface conditions (7a) are rephrased as

$$
\begin{aligned}
& \left[\!\left[u^{(0)}_h\right]\!\right]\big|_{\boldsymbol{\xi}\in\Sigma_1} + \varepsilon\left[\!\left[u^{(1)}_h\right]\!\right]\big|_{\boldsymbol{\xi}\in\Sigma_1} + \varepsilon^2\left[\!\left[u^{(2)}_h\right]\!\right]\big|_{\boldsymbol{\xi}\in\Sigma_1} + ... = 0, \\
& \frac{1}{\varepsilon}\left[\!\left[\left(C^m_{ijkl}u^{(0)}_{k,l}+e^m_{ijk}\phi^{(0)}_{,k}\right)n_j\right]\!\right]\big|_{\boldsymbol{\xi}\in\Sigma_1} + \left[\!\left[\left\{C^m_{ijkl}\left(\frac{\partial u^{(0)}_k}{\partial x_l}+u^{(1)}_{k,l}\right)+e^m_{ijk}\left(\frac{\partial \phi^{(0)}}{\partial x_k}+\phi^{(1)}_{,k}\right)\right.\right.\right. \\
& \left.\left.\left. -\alpha^m_{ij}\theta^{(0)}\right\}n_j\right]\!\right]\big|_{\boldsymbol{\xi}\in\Sigma_1} + \varepsilon\left[\!\left[\left\{C^m_{ijkl}\left(\frac{\partial u^{(1)}_k}{\partial x_l}+u^{(2)}_{k,l}\right)+e^m_{ijk}\left(\frac{\partial \phi^{(1)}}{\partial x_k}+\phi^{(2)}_{,k}\right)-\alpha^m_{ij}\theta^{(1)}\right\}n_j\right]\!\right]\big|_{\boldsymbol{\xi}\in\Sigma_1} + \\
& \varepsilon^2\left[\!\left[\left\{C^m_{ijkl}\left(\frac{\partial u^{(2)}_k}{\partial x_l}+u^{(3)}_{k,l}\right)+e^m_{ijk}\left(\frac{\partial \phi^{(2)}}{\partial x_k}+\phi^{(3)}_{,k}\right)-\alpha^m_{ij}\theta^{(2)}\right\}n_j\right]\!\right]\big|_{\boldsymbol{\xi}\in\Sigma_1} + ... = 0. \quad (12)
\end{aligned}
$$

Analogously, interface conditions (7b) read

$$
\left[\!\left[\phi^{(0)}\right]\!\right]\big|_{\boldsymbol{\xi}\in\Sigma_1} + \varepsilon\left[\!\left[\phi^{(1)}\right]\!\right]\big|_{\boldsymbol{\xi}\in\Sigma_1} + \varepsilon^2\left[\!\left[\phi^{(2)}\right]\!\right]\big|_{\boldsymbol{\xi}\in\Sigma_1} + ... = 0,
$$

$$\frac{1}{\varepsilon}\left[\left[\left(e^m_{kli}\, u^{(0)}_{k,l} - \beta^m_{il}\, \phi^{(0)}_{,l}\right) n_i\right]\right]\Big|_{\boldsymbol{\xi}\in\Sigma_1} + \left[\left[\left\{e^m_{kli}\left(\frac{\partial u^{(0)}_k}{\partial x_l} + u^{(1)}_{k,l}\right) - \beta^m_{il}\left(\frac{\partial \phi^{(0)}}{\partial x_l} + \phi^{(1)}_{,l}\right)\right.\right.\right.$$

$$\left.\left.\left. +\gamma^m_i\, \theta^{(0)}\right\} n_i\right]\right]\Big|_{\boldsymbol{\xi}\in\Sigma_1} + \varepsilon\left[\left[\left\{e^m_{kli}\left(\frac{\partial u^{(1)}_k}{\partial x_l} + u^{(2)}_{k,l}\right) - \beta^m_{il}\left(\frac{\partial \phi^{(1)}}{\partial x_l} + \phi^{(2)}_{,l}\right) + \gamma^m_i\, \theta^{(1)}\right\} n_i\right]\right]\Big|_{\boldsymbol{\xi}\in\Sigma_1} +$$

$$\varepsilon^2\left[\left[\left\{e^m_{kli}\left(\frac{\partial u^{(2)}_k}{\partial x_l} + u^{(3)}_{k,l}\right) - \beta^m_{il}\left(\frac{\partial \phi^{(2)}}{\partial x_l} + \phi^{(3)}_{,l}\right) + \gamma^m_i\, \theta^{(2)}\right\} n_i\right]\right]\Big|_{\boldsymbol{\xi}\in\Sigma_1} + ... = 0, \qquad (13)$$

and the interface conditions (7c) become

$$\left[\left[\theta^{(0)}\right]\right]\Big|_{\boldsymbol{\xi}\in\Sigma_1} + \varepsilon\left[\left[\theta^{(1)}\right]\right]\Big|_{\boldsymbol{\xi}\in\Sigma_1} + \varepsilon^2\left[\left[\theta^{(2)}\right]\right]\Big|_{\boldsymbol{\xi}\in\Sigma_1} + ... = 0,$$

$$\frac{1}{\varepsilon}\left[\left[\left(K^m_{ij}\, \theta^{(0)}_{,j}\right) n_i\right]\right]\Big|_{\boldsymbol{\xi}\in\Sigma_1} + \left[\left[\left\{K^m_{ij}\left(\frac{\partial \theta^{(0)}}{\partial x_j} + \theta^{(1)}_{,j}\right)\right\} n_i\right]\right]\Big|_{\boldsymbol{\xi}\in\Sigma_1} +$$

$$+\varepsilon\left[\left[\left\{K^m_{ij}\left(\frac{\partial \theta^{(1)}}{\partial x_j} + \theta^{(2)}_{,j}\right)\right\} n_i\right]\right]\Big|_{\boldsymbol{\xi}\in\Sigma_1} + \varepsilon^2\left[\left[\left\{K^m_{ij}\left(\frac{\partial \theta^{(2)}}{\partial x_j} + \theta^{(3)}_{,j}\right)\right\} n_i\right]\right]\Big|_{\boldsymbol{\xi}\in\Sigma_1} + ... = 0. \qquad (14)$$

From equations (9)-(11) one can notice that a *strong* coupling exists between the micro-displacement $\mathbf{u}(\mathbf{x};\boldsymbol{\xi})$ and the micro-electric potential $\phi(\mathbf{x};\boldsymbol{\xi})$ because the mechanical and the electric problems remain coupled at any order $\varepsilon$ in the asymptotically expanded microscale field equations. Hence, the proposed homogenization model can be regarded as a generalization of the multi-field asymptotic technique proposed in (Bacigalupo et al., 2016a) for thermo-diffusive materials with *weak* coupling between the micro/macro fields and the single-field standard asymptotic technique for static problems, see for example (Bacigalupo, 2014; Bakhvalov and Panasenko, 1984). Differential problems deriving from equations (9)-(11) are now made explicit at the different orders $\varepsilon$ for both the thermal and the piezoelectric fields. They bring to the formulation of cell problems described in Section 3.1.

**Heat diffusion problem**
Starting from the heat diffusion problem described by the field equation (11), at the order $\varepsilon^{-2}$ one has the following differential problem

$$\left(K^m_{ij}\, \theta^{(0)}_{,j}\right)_{,i} = h^{(0)}(\mathbf{x}), \qquad (15)$$

with interface conditions

$$\left[\left[\theta^{(0)}\right]\right]\Big|_{\boldsymbol{\xi}\in\Sigma_1} = 0, \quad \left[\left[\left(K^m_{ij}\theta^{(0)}_{,j}\right) n_i\right]\right]\Big|_{\boldsymbol{\xi}\in\Sigma_1} = 0. \qquad (16)$$

Solvability condition of (15) in the class of $\mathcal{Q}$-periodic functions, together with interface conditions (16), implies that $h^0(\mathbf{x}) = 0$, see (Bakhvalov and Panasenko, 1984), and the solution $\theta^{(0)}$ corresponds to the macroscopic temperature, namely:

$$\theta^{(0)}(\mathbf{x};\boldsymbol{\xi}) = \Theta(\mathbf{x}), \qquad (17)$$

which depends only on the slow variable $\mathbf{x}$.

Taking into account the expression (17) of the solution at the expansion order $\varepsilon^{-2}$, at the order $\varepsilon^{-1}$ equation (11) yields

$$\left(K^m_{ij}\, \theta^{(1)}_{,j}\right)_{,i} + K^m_{ij,i}\frac{\partial \Theta}{\partial x_j} = h^{(1)}(\mathbf{x}), \qquad (18)$$

with interface conditions

$$\left[\left[\theta^{(1)}\right]\right]\Big|_{\boldsymbol{\xi}\in\Sigma_1} = 0, \quad \left[\left[K^m_{ij}\left(\frac{\partial \Theta}{\partial x_j} + \theta^{(1)}_{,j}\right) n_i\right]\right]\Big|_{\boldsymbol{\xi}\in\Sigma_1} = 0. \qquad (19)$$

Due to the $\mathcal{Q}$-periodicity of components $K^m_{ij}$, one has $h^1(\mathbf{x}) = \left\langle K^m_{ij,i}\right\rangle = 0$ for the solvability of the differential problem (18), where $\langle(\cdot)\rangle = \frac{1}{\delta}\int_{\mathcal{Q}}(\cdot)d\boldsymbol{\xi}$. In consideration of equation (17), the solution $\theta^{(1)}$ at the order $\varepsilon^{-1}$ takes the form

$$\theta^{(1)}(\mathbf{x};\boldsymbol{\xi}) = M^{(1)}_{q_1}(\boldsymbol{\xi})\frac{\partial \Theta(\mathbf{x})}{\partial x_{q_1}}, \qquad (20)$$

where the perturbation function $M_{q_1}^{(1)}$ expresses the influence of the fast variable $\boldsymbol{\xi} = \frac{\mathbf{x}}{\varepsilon}$.
Perturbation functions are $\mathcal{Q}$-periodic and it is assumed that they have vanishing mean values over the unit cell $\mathcal{Q}$. Therefore, $M_{q_1}^{(1)}(\boldsymbol{\xi})$ satisfies the following normalization condition

$$\left\langle M_{q_1}^{(1)} \right\rangle = \frac{1}{\delta} \int_{\mathcal{Q}} M_{q_1}^{(1)}(\boldsymbol{\xi}) \, d\boldsymbol{\xi} = 0, \tag{21}$$

which is a general property imposed for all the perturbation functions.

Bearing in mind the solutions (17) and (20) of the differential problems at orders $\varepsilon^{-2}$ and $\varepsilon^{-1}$, respectively, equation (11) yields at the order $\varepsilon^0$

$$\left[ K_{ij}^m \left( M_{q_1}^{(1)} \frac{\partial^2 \Theta}{\partial x_{q_1} \partial x_j} + \theta_{,j}^{(2)} \right) \right]_{,i} + \frac{\partial}{\partial x_i} \left[ K_{ij}^m \left( \frac{\partial \Theta}{\partial x_j} + M_{q_1,j}^{(1)} \frac{\partial \Theta}{\partial x_{q_1}} \right) \right] = h^{(2)}(\mathbf{x}), \tag{22}$$

with interface conditions

$$\left[\left[\theta^{(2)}\right]\right]\Big|_{\boldsymbol{\xi} \in \Sigma_1} = 0, \quad \left[\left[K_{ij}^m \left( M_{q_1}^{(1)} \frac{\partial^2 \Theta}{\partial x_{q_1} \partial x_j} + \theta_{,j}^{(2)} \right) n_i\right]\right]\Big|_{\boldsymbol{\xi} \in \Sigma_1} = 0. \tag{23}$$

Solvability condition of the differential problem (22) entails that

$$h^{(2)}(\mathbf{x}) = \left\langle \left( K_{iq_2}^m M_{q_1}^{(1)} \right)_{,i} \right\rangle \frac{\partial^2 \Theta(\mathbf{x})}{\partial x_{q_1} \partial x_{q_2}} + \left\langle K_{q_1 q_2}^m + K_{q_2 j}^m M_{q_1,j} \right\rangle \frac{\partial^2 \Theta(\mathbf{x})}{\partial x_{q_1} \partial x_{q_2}} = \left\langle K_{q_1 q_2}^m + K_{q_2 j}^m M_{q_1,j} \right\rangle \frac{\partial^2 \Theta(\mathbf{x})}{\partial x_{q_1} \partial x_{q_2}},$$

since $\left\langle \left( K_{iq_2}^m M_{q_1}^{(1)} \right)_{,i} \right\rangle = 0$ for the divergence theorem and the $\mathcal{Q}$-periodicity of $K_{iq_2}^m$ and $M_{q_1}^{(1)}$. Consequently, the solution $\theta^{(2)}$ of the heat diffusion differential problem at the order $\varepsilon^{-2}$ takes the following form

$$\theta^{(2)}(\mathbf{x}; \boldsymbol{\xi}) = M_{q_1 q_2}^{(2)}(\boldsymbol{\xi}) \frac{\partial^2 \Theta(\mathbf{x})}{\partial x_{q_1} \partial x_{q_2}}, \tag{24}$$

thus introducing the perturbation function $M_{q_1 q_2}^{(2)}$.

**Piezoelectric problem**

For what regards the piezoelectric problem governed by field equations (9)-(10), at the order $\varepsilon^{-2}$ one has

$$\begin{aligned}
\left( C_{ijkl}^m u_{k,l}^{(0)} \right)_{,j} + \left( e_{ijk}^m \phi_{,k}^{(0)} \right)_{,j} &= f_i^{(0)}(\mathbf{x}), \\
\left( e_{kli}^m u_{k,l}^{(0)} \right)_{,i} - \left( \beta_{il}^m \phi_{,l}^{(0)} \right)_{,i} &= g^{(0)}(\mathbf{x}),
\end{aligned} \tag{25}$$

with interface conditions

$$\left[\left[u_k^{(0)}\right]\right]\Big|_{\boldsymbol{\xi} \in \Sigma_1} = 0, \quad \left[\left[\phi^{(0)}\right]\right]\Big|_{\boldsymbol{\xi} \in \Sigma_1} = 0,$$
$$\left[\left[\left( C_{ijkl}^m u_{k,l}^{(0)} + e_{ijk}^m \phi_{,k}^{(0)} \right) n_j\right]\right]\Big|_{\boldsymbol{\xi} \in \Sigma_1} = 0, \quad \left[\left[\left( e_{kli}^m u_{k,l}^{(0)} - \beta_{il}^m \phi_{,l}^{(0)} \right) n_i\right]\right]\Big|_{\boldsymbol{\xi} \in \Sigma_1} = 0. \tag{26}$$

Analogously to heat conduction, the solvability condition for a $\mathcal{Q}$-periodic function implies that $f_i^{(0)}(\mathbf{x}) = 0$ and $g^{(0)}(\mathbf{x}) = 0$, and therefore the solution of problem (25) does not depend on the fast variable $\boldsymbol{\xi}$, taking the form

$$u_k^{(0)}(\mathbf{x}; \boldsymbol{\xi}) = U_k(\mathbf{x}), \quad \phi^{(0)}(\mathbf{x}; \boldsymbol{\xi}) = \Phi(\mathbf{x}). \tag{27}$$

In force of solutions (27) of the piezoelectric differential problem at the order $\varepsilon^{-2}$, from equations (9) and (10), at the order $\varepsilon^{-1}$ one has

$$\begin{aligned}
\left( C_{ijkl}^m u_{k,l}^{(1)} \right)_{,j} + C_{ijkl,j}^m \frac{\partial U_k}{\partial x_l} + \left( e_{ijk}^m \phi_{,k}^{(1)} \right)_{,j} + e_{ijk,j}^m \frac{\partial \Phi}{\partial x_k} - \alpha_{ij,j}^m \Theta &= f_i^{(1)}(\mathbf{x}), \\
\left( e_{kli}^m u_{k,l}^{(1)} \right)_{,i} + e_{kli,i}^m \frac{\partial U_k}{\partial x_l} - \left( \beta_{il}^m \phi_{,l}^{(1)} \right)_{,i} - \beta_{il,i}^m \frac{\partial \Phi}{\partial x_l} + \gamma_{i,i}^m \Theta &= g^{(1)}(\mathbf{x}),
\end{aligned} \tag{28}$$

with interface conditions

$$\left[\left[u_k^{(1)}\right]\right]\bigg|_{\boldsymbol{\xi}\in\Sigma_1} = 0, \quad \left[\left[\phi^{(1)}\right]\right]\bigg|_{\boldsymbol{\xi}\in\Sigma_1} = 0,$$

$$\left[\left[\left(C_{ijkl}^m\left(\frac{\partial U_k}{\partial x_l} + u_{k,l}^{(1)}\right) + e_{ijk}^m\left(\frac{\partial \Phi}{\partial x_k} + \phi_{,k}^{(1)}\right) - \alpha_{ij}^m\,\Theta\right)n_j\right]\right]\bigg|_{\boldsymbol{\xi}\in\Sigma_1} = 0,$$

$$\left[\left[\left(e_{kli}^m\left(\frac{\partial U_k}{\partial x_l} + u_{k,l}^{(1)}\right) - \beta_{il}^m\left(\frac{\partial \Phi}{\partial x_l} + \phi_{,l}^{(1)}\right) + \gamma_i^m\,\Theta\right)n_i\right]\right]\bigg|_{\boldsymbol{\xi}\in\Sigma_1} = 0. \quad (29)$$

In this case, the solvability condition implies that functions $f_i^{(1)}(\mathbf{x})$ and $g^{(1)}(\mathbf{x})$ take the form

$$f_i^{(1)}(\mathbf{x}) = \langle C_{ijkl,j}^m \rangle \frac{\partial U_k(\mathbf{x})}{\partial x_l} + \langle e_{ijk,j}^m \rangle \frac{\partial \Phi(\mathbf{x})}{\partial x_k} - \langle \alpha_{ij,j}^m \rangle \Theta(\mathbf{x}) = 0,$$

$$g^{(1)}(\mathbf{x}) = \langle e_{kli,i}^m \rangle \frac{\partial U_k(\mathbf{x})}{\partial x_l} - \langle \beta_{il,i}^m \rangle \frac{\partial \Phi(\mathbf{x})}{\partial x_l} + \langle \gamma_{i,i}^m \rangle \Theta(\mathbf{x}) = 0,$$

and, consequently, the solutions of the differential problem (28) read

$$u_k^{(1)}(\mathbf{x};\boldsymbol{\xi}) = N_{kpq_1}^{(1)}(\boldsymbol{\xi})\frac{\partial U_p(\mathbf{x})}{\partial x_{q_1}} + \tilde{N}_{kq_1}^{(1)}(\boldsymbol{\xi})\frac{\partial \Phi(\mathbf{x})}{\partial x_{q_1}} + \hat{N}_k^{(1)}(\boldsymbol{\xi})\,\Theta(\mathbf{x}),$$

$$\phi^{(1)}(\mathbf{x};\boldsymbol{\xi}) = W_{q_1}^{(1)}(\boldsymbol{\xi})\frac{\partial \Phi(\mathbf{x})}{\partial x_{q_1}} + \tilde{W}_{pq_1}^{(1)}(\boldsymbol{\xi})\frac{\partial U_p(\mathbf{x})}{\partial x_{q_1}} + \hat{W}_k^{(1)}(\boldsymbol{\xi})\,\Theta(\mathbf{x}). \quad (30)$$

Finally, taking into account solutions (27) and (30) of the piezoelectric differential problems at $\varepsilon^{-2}$ and $\varepsilon^{-1}$, differential problems (9)-(10) at the order $\varepsilon^0$ are expressed as

$$+ \left(C_{ijkl}^m u_{k,l}^{(2)}\right)_{,j} + \left[\left(C_{ijkq_2}^m N_{kpq_1}^{(1)}\right)_{,j} + C_{iq_1pq_2}^m + C_{iq_2kl}^m N_{kpq_1,l}^{(1)} + \left(e_{ijq_2}^m \tilde{W}_{pq_1}^{(1)}\right)_{,j} + e_{iq_2k}^m \tilde{W}_{pq_1,k}^{(1)}\right]\frac{\partial^2 U_p}{\partial x_{q_1}\partial x_{q_2}}$$

$$+ \left(e_{ijk}^m \phi_{,k}^{(2)}\right)_{,j} + \left[\left(C_{ijkq_2}^m \tilde{N}_{kq_1}^{(1)}\right)_{,j} + C_{iq_2kl}^m \tilde{N}_{kq_1,l}^{(1)} + \left(e_{ijq_2}^m W_{q_1}^{(1)}\right)_{,j} + e_{iq_1q_2}^m + e_{iq_2k}^m W_{q_1,k}^{(1)}\right]\frac{\partial^2 \Phi}{\partial x_{q_1}\partial x_{q_2}}$$

$$+ \left[\left(C_{ijkq_1}^m \hat{N}_k^{(1)}\right)_{,j} + C_{iq_1kl}^m \hat{N}_{k,l}^{(1)} + \left(e_{ijq_1}^m \hat{W}^{(1)}\right)_{,j} + e_{iq_1k}^m \hat{W}_{,k}^{(1)} - \left(\alpha_{ij}^m M_{q_1}^{(1)}\right)_{,j} - \alpha_{iq_1}^m\right]\frac{\partial \Theta}{\partial x_{q_1}} = f_i^{(2)}(\mathbf{x}),$$

$$+ \left(e_{kli}^m u_{k,l}^{(2)}\right)_{,i} + \left[\left(e_{kq_2i}^m N_{kpq_1}^{(1)}\right)_{,i} + e_{klq_2}^m N_{kpq_1,l}^{(1)} + e_{pq_2q_1} - \left(\beta_{iq_2}^m \tilde{W}_{pq_1}^{(1)}\right)_{,i} - \beta_{q_2l}^m \tilde{W}_{pq_1,l}^{(1)}\right]\frac{\partial^2 U_p}{\partial x_{q_1}\partial x_{q_2}}$$

$$- \left(\beta_{il}^m \phi_{,l}^{(2)}\right)_{,i} + \left[\left(e_{kq_2i}^m \tilde{N}_{kq_1}^{(1)}\right)_{,i} + e_{klq_2}^m \tilde{N}_{kq_1,l}^{(1)} - \left(\beta_{iq_2}^m W_{q_1}^{(1)}\right)_{,i} - \beta_{q_1q_2}^m - \beta_{q_2l}^m W_{q_1,l}^{(1)}\right]\frac{\partial^2 \Phi}{\partial x_{q_1}\partial x_{q_2}}$$

$$+ \left[\left(e_{kq_1i}^m \hat{N}_k^{(1)}\right)_{,i} + e_{klq_1}^m \hat{N}_{k,l}^{(1)} - \left(\beta_{iq_1}^m \hat{W}^{(1)}\right)_{,i} - \beta_{q_1l}^m \hat{W}_{,l}^{(1)} + \left(\gamma_i^m M_{q_1}^{(1)}\right)_{,i} - \gamma_{q_1}^m\right]\frac{\partial \Theta}{\partial x_{q_1}} = g^{(2)}(\mathbf{x}), \quad (31)$$

with interface conditions

$$\left[\left[u_k^{(2)}\right]\right]\bigg|_{\boldsymbol{\xi}\in\Sigma_1} = 0, \quad \left[\left[\phi^{(2)}\right]\right]\bigg|_{\boldsymbol{\xi}\in\Sigma_1} = 0,$$

$$\left[\left[\left\{\left(C_{ijkl}^m u_{k,l}^{(2)}\right) + \left(e_{ijk}^m \phi_{,k}^{(2)}\right) + \left(C_{ijkl}^m N_{kpq_1}^{(1)} + e_{ijl}^m \tilde{W}_{pq_1}^{(1)}\right)\frac{\partial^2 U_p}{\partial x_{q_1}\partial x_l} + \left(C_{ijkl}^m \tilde{N}_{kq_1}^{(1)} + e_{ijl}^m W_{q_1}^{(1)}\right)\frac{\partial^2 \Phi}{\partial x_{q_1}\partial x_l}\right.\right.$$

$$+ \left.\left.\left(C_{ijkl}^m \hat{N}_k^{(1)} + e_{ijl}^m \hat{W}^{(1)} - \alpha_{ij} M_l^{(1)}\right)\frac{\partial \Theta}{\partial x_l}\right\}n_j\right]\right]\bigg|_{\boldsymbol{\xi}\in\Sigma_1} = 0,$$

$$\left[\left[\left\{\left(e_{kli}^m u_{k,l}^{(2)}\right) - \left(\beta_{il}^m \phi_{,l}^{(2)}\right) + \left(e_{kli}^m N_{kpq_1}^{(1)} - \beta_{il}^m \tilde{W}_{pq_1}^{(1)}\right)\frac{\partial^2 U_p}{\partial x_{q_1}\partial x_l} + \left(e_{kli}^m \tilde{N}_{kq_1}^{(1)} - \beta_{il}^m W_{q_1}^{(1)}\right)\frac{\partial^2 \Phi}{\partial x_{q_1}\partial x_l}\right.\right.$$

$$+ \left.\left.\left(e_{kli}^m \hat{N}_k^{(1)} - \beta_{il}^m \hat{W}^{(1)} + \gamma_i M_l^{(1)}\right)\frac{\partial \Theta}{\partial x_l}\right\}n_i\right]\right]\bigg|_{\boldsymbol{\xi}\in\Sigma_1} = 0. \quad (32)$$

Here, the following relations hold because of the solvability condition for problem (31)

$$f_i^{(2)}(\mathbf{x}) = \left\langle C_{iq_1pq_2}^m + C_{iq_1pq_2}^m N_{kpq_1,l}^{(1)} + e_{iq_2k}^m \tilde{W}_{pq_1,k}^{(1)}\right\rangle \frac{\partial^2 U_p(\mathbf{x})}{\partial x_{q_1}\partial x_{q_2}} +$$

$$+ \left\langle C^m_{iq_2kl}\,\tilde{N}^{(1)}_{kq_1,l} + e^m_{iq_1q_2} + e^m_{iq_2k}\,W^{(1)}_{q_1,k} \right\rangle \frac{\partial^2 \Phi(\mathbf{x})}{\partial x_{q_1}\partial x_{q_2}} +$$

$$+ \left\langle C^m_{iq_1kl}\,\hat{N}^{(1)}_{k,l} + e^m_{iq_1k}\,\hat{W},k^{(1)} - \alpha^m_{iq_1} \right\rangle \frac{\partial \Theta(\mathbf{x})}{\partial x_{q_1}},$$

$$g^{(2)}(\mathbf{x}) = \left\langle e^m_{klq_2}\,N^{(1)}_{kpq_1,l} + e^m_{pq_2q_1} - \beta^m_{q_2l}\,\tilde{W}^{(1)}_{pq_1,l} \right\rangle \frac{\partial^2 U_p(\mathbf{x})}{\partial x_{q_1}\partial x_{q_2}} +$$

$$+ \left\langle e^m_{klq_2}\,\tilde{N}^{(1)}_{kq_1,l} - \beta^m_{q_1q_2} - \beta^m_{q_2l}\,W^{(1)}_{q_1,l} \right\rangle \frac{\partial^2 \Phi(\mathbf{x})}{\partial x_{q_1}\partial x_{q_2}} +$$

$$+ \left\langle e^m_{klq_1}\,\hat{N}^{(1)}_{k,l} - \beta^m_{q_1l}\,\hat{W},l^{(1)} + \gamma^m_{q_1} \right\rangle \frac{\partial \Theta(\mathbf{x})}{\partial x_{q_1}},$$

the solutions $u_k^{(2)}$ and $\phi^{(2)}$ take the following form

$$u_k^{(2)}(\mathbf{x};\boldsymbol{\xi}) = N^{(2)}_{kpq_1q_2}(\boldsymbol{\xi})\frac{\partial^2 U_p}{\partial x_{q_1}\partial x_{q_2}} + \tilde{N}^{(2)}_{kq_1q_2}(\boldsymbol{\xi})\frac{\partial^2 \Phi}{\partial x_{q_1}\partial x_{q_2}} + \hat{N}^{(2)}_{kq_1}(\boldsymbol{\xi})\frac{\partial \Theta}{\partial x_{q_1}},$$

$$\phi^{(2)}(\mathbf{x};\boldsymbol{\xi}) = W^{(2)}_{q_1q_2}(\boldsymbol{\xi})\frac{\partial^2 \Phi}{\partial x_{q_1}\partial x_{q_2}} + \tilde{W}^{(2)}_{pq_1q_2}(\boldsymbol{\xi})\frac{\partial U_p}{\partial x_{q_1}\partial x_{q_2}} + \hat{W}^{(2)}_{q_1}(\boldsymbol{\xi})\frac{\partial \Theta}{\partial x_{q_1}}. \tag{33}$$

The general expression of higher order solutions $u_k^{(m)}(\mathbf{x};\boldsymbol{\xi})$, $\phi^{(m)}(\mathbf{x};\boldsymbol{\xi})$ and $\theta^{(m)}(\mathbf{x};\boldsymbol{\xi})$, with $m \in \mathbb{Z}$ and $m \geq 1$, is reported in Appendix A.

### 3.1 Cell problems and perturbation functions

Solutions of recursive differential problems (9)-(11) allow to write non-homogeneous cell problems at the different orders of $\varepsilon$ in terms of the perturbation functions. Perturbation functions exclusively depend on geometrical and physico-mechanical properties of the material reflecting the effects of the material inhomogeneities on displacements, electric potential and temperature. In the following, the form of such cell problems is discussed at the different orders of $\varepsilon$, for heat diffusion and piezoelectric problems.

**Heat diffusion problem**
From equation (18) and in consideration of the solution (20), the cell problem at the order $\varepsilon^{-1}$ takes the form

$$\left(K^m_{ij}\,M^{(1)}_{q_1,j}\right)_{,i} + K^m_{iq_1,i} = 0, \tag{34}$$

with interface conditions expressed in terms of perturbation function $M^{(1)}_{q_1}$ as

$$\left.\left[\left[M^{(1)}_{q_1}\right]\right]\right|_{\boldsymbol{\xi}\in\Sigma_1} = 0,$$
$$\left.\left[\left[K^m_{ij}\left(M^{(1)}_{q_1,j} + \delta_{jq_1}\right)n_i\right]\right]\right|_{\boldsymbol{\xi}\in\Sigma_1} = 0. \tag{35}$$

Once the perturbation function $M^{(1)}_{q_1}$ is determined, from equation (22) and taking into account the solution (24), one obtains the cell problem at the order $\varepsilon^0$, where symmetrization with respect to indices $q_1$ and $q_2$ is introduced. This procedure leads to

$$\left(K^m_{ij}\,M^{(2)}_{q_1q_2,j}\right)_{,i} + \frac{1}{2}\left[\left(K^m_{iq_2}\,M^{(1)}_{q_1}\right)_{,i} + K^m_{q_1q_2} + K^m_{q_2j}\,M^{(1)}_{q_1,j} + \left(K^m_{iq_2}\,M^{(1)}_{q_2}\right)_{,i} + K^m_{q_2q_1} + K^m_{q_1j}\,M^{(1)}_{q_2,j}\right] =$$
$$\frac{1}{2}\left\langle K^m_{q_1q_2} + K^m_{q_2j}\,M^{(1)}_{q_1,j} + K^m_{q_2q_1} + K^m_{q_1j}\,M^{(1)}_{q_2,j}\right\rangle, \tag{36}$$

with interface conditions in terms of perturbation functions taking the following form

$$\left.\left[\left[M^{(2)}_{q_1q_2}\right]\right]\right|_{\boldsymbol{\xi}\in\Sigma_1} = 0,$$

$$\left[\!\!\left[K^m_{ij}\left[M^{(2)}_{q_1q_2,j}+\frac{1}{2}\left(\delta_{jq_2}M^{(1)}_{q_1}+\delta_{jq_1}M^{(1)}_{q_2}\right)\right]n_i\right]\!\!\right]\bigg|_{\boldsymbol{\xi}\in\Sigma_1}=0. \tag{37}$$

The solution of the cell problem (36)-(37) allows to derive perturbation function $M^{(2)}_{q_1q_2}$.

**Piezoelectric problem**

For what concerns the piezoelectric problem, from equation (28) and taking into account the form (30) of the solutions $u^{(1)}_k$ and $\phi^{(1)}$, the strong coupling among the relative perturbation functions leads to three cell problems at the order $\varepsilon^{-1}$ whose solutions lead to the perturbation functions $N^{(1)}_{kpq_1}, \tilde{N}^{(1)}_{kq_1}, \hat{N}^{(1)}_k, W^{(1)}_{q_1}, \tilde{W}^{(1)}_{pq_1}$, and $\hat{W}^{(1)}$. In particular, perturbation functions $N^{(1)}_{kpq_1}$ and $\tilde{W}^{(1)}_{pq_1}$ are determined from the following cell problem

$$\begin{cases} \left(C^m_{ijkl}N^{(1)}_{kpq_1,l}\right)_{,j}+\left(e^m_{ijk}\tilde{W}^{(1)}_{pq_1,k}\right)_{,j}+C^m_{ijpq_1,j}=0 \\ \left(e^m_{kli}N^{(1)}_{kpq_1,l}\right)_{,i}-\left(\beta^m_{il}\tilde{W}^{(1)}_{pq_1,l}\right)_{,i}+e^m_{pq_1i,i}=0 \end{cases}, \tag{38}$$

with interface conditions

$$\left[\!\!\left[N^{(1)}_{kpq_1}\right]\!\!\right]\bigg|_{\boldsymbol{\xi}\in\Sigma_1}=0,$$
$$\left[\!\!\left[\tilde{W}^{(1)}_{pq_1}\right]\!\!\right]\bigg|_{\boldsymbol{\xi}\in\Sigma_1}=0,$$
$$\left[\!\!\left[\left\{C^m_{ijkl}\left(\delta_{kp}\delta_{lq_1}+N^{(1)}_{kpq_1,l}\right)+e^m_{ijk}\tilde{W}^{(1)}_{pq_1,k}\right\}n_j\right]\!\!\right]\bigg|_{\boldsymbol{\xi}\in\Sigma_1}=0,$$
$$\left[\!\!\left[\left\{e^m_{kli}\left(\delta_{kp}\delta_{lq_1}+N^{(1)}_{kpq_1,l}\right)-\beta^m_{il}\tilde{W}^{(1)}_{pq_1,l}\right\}n_i\right]\!\!\right]\bigg|_{\boldsymbol{\xi}\in\Sigma_1}=0. \tag{39}$$

Perturbation functions $\tilde{N}^{(1)}_{kq_1}$ and $W^{(1)}_{q_1}$ are the solutions of the following cell problem

$$\begin{cases} \left(C^m_{ijkl}\tilde{N}^{(1)}_{kq_1,l}\right)_{,j}+\left(e^m_{ijk}W^{(1)}_{q_1,k}\right)_{,j}+e^m_{ijq_1,j}=0 \\ \left(e^m_{kli}\tilde{N}^{(1)}_{kq_1,l}\right)_{,i}-\left(\beta^m_{il}W^{(1)}_{q_1,l}\right)_{,i}-\beta^m_{iq_1,i}=0 \end{cases}, \tag{40}$$

whose interface conditions are expressed as

$$\left[\!\!\left[\tilde{N}^{(1)}_{kq_1}\right]\!\!\right]\bigg|_{\boldsymbol{\xi}\in\Sigma_1}=0,$$
$$\left[\!\!\left[W^{(1)}_{q_1}\right]\!\!\right]\bigg|_{\boldsymbol{\xi}\in\Sigma_1}=0,$$
$$\left[\!\!\left[\left\{C^m_{ijkl}\tilde{N}^{(1)}_{kq_1,l}+e^m_{ijk}\left(\delta_{kq_1}+W^{(1)}_{q_1,k}\right)\right\}n_j\right]\!\!\right]\bigg|_{\boldsymbol{\xi}\in\Sigma_1}=0,$$
$$\left[\!\!\left[\left\{e^m_{kli}\tilde{N}^{(1)}_{kq_1,l}-\beta^m_{il}\left(\delta_{lq_1}+W^{(1)}_{q_1,l}\right)\right\}n_i\right]\!\!\right]\bigg|_{\boldsymbol{\xi}\in\Sigma_1}=0. \tag{41}$$

Finally, perturbation functions $\hat{N}^{(1)}_k$ and $\hat{W}^{(1)}$ are provided by the following cell problem at the order $\varepsilon^{-1}$

$$\begin{cases} \left(C^m_{ijkl}\hat{N}^{(1)}_{k,l}\right)_{,j}+\left(e^m_{ijk}\hat{W}^{(1)}_{,k}\right)_{,j}-\alpha^m_{ij,j}=0 \\ \left(e^m_{kli}\hat{N}^{(1)}_{k,l}\right)_{,i}-\left(\beta^m_{il}\hat{W}^{(1)}_{,l}\right)_{,i}+\gamma^m_{i,i}=0 \end{cases}, \tag{42}$$

with interface conditions

$$\left[\!\!\left[\hat{N}^{(1)}_k\right]\!\!\right]\bigg|_{\boldsymbol{\xi}\in\Sigma_1}=0,$$
$$\left[\!\!\left[\hat{W}^{(1)}\right]\!\!\right]\bigg|_{\boldsymbol{\xi}\in\Sigma_1}=0,$$
$$\left[\!\!\left[\left\{C^m_{ijkl}\hat{N}^{(1)}_{k,l}+e^m_{ijk}\hat{W}^{(1)}_{,k}-\alpha^m_{ij}\right\}n_j\right]\!\!\right]\bigg|_{\boldsymbol{\xi}\in\Sigma_1}=0,$$

$$\left[\left[\left\{e_{kli}^m \hat{N}_{k,l}^{(1)} - \beta_{il}^m \hat{W}_{,l}^{(1)} + \gamma_i^m\right\} n_i\right]\right]\bigg|_{\boldsymbol{\xi}\in\Sigma_1} = 0. \tag{43}$$

Analogously to what done at the order $\varepsilon^{-1}$, from the differential problem (31) and recalling the solutions (33), one derives the following three cell problems at the order $\varepsilon^0$. Specifically, the following cell problem, symmetrized with respect to indices $q_1$ and $q_2$, allows to derive perturbation functions $N_{kpq_1q_2}^{(2)}$ and $\tilde{W}_{pq_1q_2}^{(2)}$

$$\begin{cases}
\left(C_{ijkl}^m N_{kpq_1q_2,l}^{(2)}\right)_{,j} + \left(e_{ijk}^m \tilde{W}_{pq_1q_2,k}^{(2)}\right)_{,j} + \frac{1}{2}\bigg[\left(C_{ijkq_2}^m N_{kpq_1}^{(1)}\right)_{,j} + C_{iq_1pq_2}^m + C_{iq_2kl}^m N_{kpq_1,l}^{(1)} + \\
+ \left(e_{ijq_2}^m \tilde{W}_{pq_1}^{(1)}\right)_{,j} + e_{iq_2k}^m \tilde{W}_{pq_1,k}^{(1)} + \left(C_{ijkq_1}^m N_{kpq_2}^{(1)}\right)_{,j} + C_{iq_2pq_1}^m + C_{iq_1kl}^m N_{kpq_2,l}^{(1)} + \left(e_{ijq_1}^m \tilde{W}_{pq_2}^{(1)}\right)_{,j} + \\
+ e_{iq_1k}^m \tilde{W}_{pq_2,k}^{(1)}\bigg] = \frac{1}{2}\left\langle C_{iq_1pq_2}^m + C_{iq_2kl}^m N_{kpq_1,l}^{(1)} + e_{iq_2k}^m \tilde{W}_{pq_1,k}^{(1)} + C_{iq_2pq_1}^m + C_{iq_1kl}^m N_{kpq_2,l}^{(1)} + \\
+ e_{iq_1k}^m \tilde{W}_{pq_2,k}^{(1)}\right\rangle \\[4pt]
\left(e_{kli}^m N_{kpq_1q_2,l}^{(2)}\right)_{,i} - \left(\beta_{il}^m \tilde{W}_{pq_1q_2,l}^{(2)}\right)_{,i} + \frac{1}{2}\bigg[\left(e_{kq_2i}^m N_{kpq_1}^{(1)}\right)_{,i} + e_{klq_2}^m N_{kpq_1,l}^{(1)} + e_{pq_2q_1}^m + \\
- \left(\beta_{iq_2}^m \tilde{W}_{pq_1}^{(1)}\right)_{,i} - \beta_{q_2l}^m \tilde{W}_{pq_1,l}^{(1)} + \left(e_{kq_1i}^m N_{kpq_2}^{(1)}\right)_{,i} + e_{klq_1}^m N_{kpq_2,l}^{(1)} + e_{pq_1q_2}^m - \left(\beta_{iq_1}^m \tilde{W}_{pq_2}^{(1)}\right)_{,i} + \\
- \beta_{q_1l}^m \tilde{W}_{pq_2,l}^{(1)}\bigg] = \frac{1}{2}\left\langle e_{klq_2}^m N_{kpq_1,l}^{(1)} + e_{pq_2q_1}^m - \beta_{q_2l}^m \tilde{W}_{pq_1,l}^{(1)} + e_{klq_1}^m N_{kpq_2,l}^{(1)} + e_{pq_1q_2}^m + \\
- \beta_{q_1l}^m \tilde{W}_{pq_2,l}^{(1)}\right\rangle
\end{cases} \tag{44}$$

with interface conditions

$$\left[\left[N_{kpq_1q_2}^{(2)}\right]\right]\bigg|_{\boldsymbol{\xi}\in\Sigma_1} = 0,$$

$$\left[\left[\tilde{W}_{pq_1q_2}^{(2)}\right]\right]\bigg|_{\boldsymbol{\xi}\in\Sigma_1} = 0,$$

$$\left[\left[\left\{C_{ijkl}^m N_{kpq_1q_2,l}^{(2)} + e_{ijk}^m \tilde{W}_{pq_1q_2,k}^{(2)} + \frac{1}{2}\left(C_{ijkq_2}^m N_{kpq_1}^{(1)} + C_{ijkq_1}^m N_{kpq_2}^{(1)} + e_{ijq_2}^m \tilde{W}_{pq_1}^{(1)} + e_{ijq_1}^m \tilde{W}_{pq_2}^{(1)}\right)\right\} n_j\right]\right]\bigg|_{\boldsymbol{\xi}\in\Sigma_1} = 0,$$

$$\left[\left[\left\{e_{kli}^m N_{kpq_1q_2,l}^{(2)} - \beta_{il}^m \tilde{W}_{pq_1q_2,l}^{(2)} + \frac{1}{2}\left(e_{kq_2i}^m N_{kpq_1}^{(1)} - \beta_{iq_2}^m \tilde{W}_{pq_1}^{(1)} + e_{kq_1i}^m N_{kpq_2}^{(1)} - \beta_{iq_1}^m \tilde{W}_{pq_2}^{(1)}\right)\right\} n_i\right]\right]\bigg|_{\boldsymbol{\xi}\in\Sigma_1} = 0. \tag{45}$$

Perturbation functions $\tilde{N}_{kq_1q_2}^{(2)}$ and $W_{q_1q_2}^{(2)}$ are the solutions of the following cell problem, whose expression is reported, once again, in the symmetrized form with respect to $q_1$ and $q_2$ indices

$$\begin{cases}
\left(C_{ijkl}^m \tilde{N}_{kq_1q_2,l}^{(2)}\right)_{,j} + \left(e_{ijk}^m W_{q_1q_2,k}^{(2)}\right)_{,j} + \frac{1}{2}\bigg[\left(C_{ijkq_2}^m \tilde{N}_{kq_1}^{(1)}\right)_{,j} + C_{iq_2kl}^m \tilde{N}_{kq_1,l}^{(1)} + \left(e_{ijq_2}^m W_{q_1}^{(1)}\right)_{,j} + \\
+ e_{iq_1q_2}^m + e_{iq_2k}^m W_{q_1,k}^{(1)} + \left(C_{ijkq_1}^m \tilde{N}_{kq_2}^{(1)}\right)_{,j} + C_{iq_1kl}^m \tilde{N}_{kq_2,l}^{(1)} + \left(e_{ijq_1}^m W_{q_2}^{(1)}\right)_{,j} + e_{iq_2q_1}^m + e_{iq_1k}^m W_{q_2,k}^{(1)}\bigg] = \\
= \frac{1}{2}\left\langle C_{iq_2kl}^m \tilde{N}_{kq_1,l}^{(1)} + e_{iq_1q_2}^m + e_{iq_2k}^m W_{q_1,k}^{(1)} + C_{iq_1kl}^m \tilde{N}_{kq_2,l}^{(1)} + e_{iq_2q_1}^m + e_{iq_1k}^m W_{q_2,k}^{(1)}\right\rangle \\[4pt]
\left(e_{kli}^m \tilde{N}_{kq_1q_2,l}^{(2)}\right)_{,i} - \left(\beta_{il}^m W_{q_1q_2,l}^{(2)}\right)_{,i} + \frac{1}{2}\bigg[\left(e_{kq_2i}^m \tilde{N}_{kq_1}^{(1)}\right)_{,i} + e_{klq_2}^m \tilde{N}_{kq_1,l}^{(1)} - \left(\beta_{iq_2}^m W_{q_1}^{(1)}\right)_{,i} - \beta_{q_1q_2}^m + \\
- \beta_{q_2l}^m W_{q_1,l}^{(1)} + \left(e_{kq_1i}^m \tilde{N}_{kq_2}^{(1)}\right)_{,i} + e_{klq_1}^m \tilde{N}_{kq_2,l}^{(1)} - \left(\beta_{iq_1}^m W_{q_2}^{(1)}\right)_{,i} - \beta_{q_2q_1}^m - \beta_{q_1l}^m W_{q_2,l}^{(1)}\bigg] = \\
= \frac{1}{2}\left\langle e_{klq_2}^m \tilde{N}_{kq_1,l}^{(1)} - \beta_{q_1q_2}^m - \beta_{q_2l}^m W_{q_1,l}^{(1)} + e_{klq_1}^m \tilde{N}_{kq_2,l}^{(1)} - \beta_{q_2q_1}^m - \beta_{q_1l}^m W_{q_2,l}^{(1)}\right\rangle
\end{cases} \tag{46}$$

with interface conditions expressed as

$$\left[\left[\tilde{N}_{kq_1q_2}^{(2)}\right]\right]\bigg|_{\boldsymbol{\xi}\in\Sigma_1} = 0,$$

$$\left[\left[W_{q_1q_2}^{(2)}\right]\right]\bigg|_{\boldsymbol{\xi}\in\Sigma_1} = 0,$$

$$\left[\left[\left\{C^m_{ijkl}\,\tilde{N}^{(2)}_{kq_1q_2,l}+e^m_{ijk}\,W^{(2)}_{q_1q_2,k}+\frac{1}{2}\left(C^m_{ijkq_2}\,\tilde{N}^{(1)}_{kq_1}+e^m_{ijq_2}\,W^{(1)}_{q_1}+C^m_{ijkq_1}\,\tilde{N}^{(1)}_{kq_2}+e^m_{ijq_1}\,W^{(1)}_{q_2}\right)\right\}n_j\right]\right]\bigg|_{\boldsymbol{\xi}\in\Sigma_1}=0,$$

$$\left[\left[\left\{e^m_{kli}\,\tilde{N}^{(2)}_{kq_1q_2,l}-\beta^m_{il}\,W^{(2)}_{q_1q_2,l}+\frac{1}{2}\left(e^m_{kq_2i}\,\tilde{N}^{(1)}_{kq_1}-\beta^m_{iq_2}\,W^{(1)}_{q_1}+e^m_{kq_1i}\,\tilde{N}^{(1)}_{kq_2}-\beta^m_{iq_1}\,W^{(1)}_{q_2}\right)\right\}n_i\right]\right]\bigg|_{\boldsymbol{\xi}\in\Sigma_1}=0. \quad (47)$$

Finally, perturbation functions $\hat{N}^{(2)}_{kq_1}$ and $\hat{W}^{(2)}_{q_1}$ are provided by the solution of the cell problem

$$\begin{cases}\left(C^m_{ijkl}\,\hat{N}^{(2)}_{kq_1,l}\right)_{,j}+\left(e^m_{ijk}\,\hat{W}^{(2)}_{q_1,k}\right)_{,j}+\left(C^m_{ijkq_1}\,\hat{N}^{(1)}_k\right)_{,j}+C^m_{iq_1kl}\,\hat{N}^{(1)}_{k,l}+\left(e^m_{ijq_1}\,\hat{W}^{(1)}\right)_{,j}+\\+e^m_{iq_1k}\,\hat{W}^{(1)}_{,k}-\left(\alpha^m_{ij}\,M^{(1)}_{q_1}\right)_{,j}-\alpha^m_{iq_1}=\left\langle C^m_{iq_1kl}\,\hat{N}^{(1)}_{k,l}+e_{iq_1k}\,\hat{W}^{(1)}_{,k}-\alpha^{(m)}_{iq_1}\right\rangle\\\left(e^m_{kli}\,\hat{N}^{(2)}_{kq_1,l}\right)_{,i}-\left(\beta^m_{il}\,\hat{W}^{(2)}_{q_1,l}\right)_{,i}+\left(e^m_{kq_1i}\,\hat{N}^{(1)}_k\right)_{,i}+e^m_{klq_1}\,\hat{N}^{(1)}_{k,l}-\left(\beta^m_{iq_1}\,\hat{W}^{(1)}\right)_{,i}+\\-\beta^m_{q_1l}\,\hat{W}^{(1)}_{,l}+\left(\gamma^m_i\,M^{(1)}_{q_1}\right)_{,i}+\gamma^m_{q_1}=\left\langle e^m_{klq_1}\,\hat{N}^{(1)}_{k,l}-\beta^m_{q_1l}\,\hat{W}^{(1)}_{,l}+\gamma^m_{q_1}\right\rangle\end{cases},$$
(48)

with interface conditions

$$\left[\left[\hat{N}^{(2)}_{kq_1}\right]\right]\bigg|_{\boldsymbol{\xi}\in\Sigma_1}=0,$$
$$\left[\left[\hat{W}^{(2)}_{q_1}\right]\right]\bigg|_{\boldsymbol{\xi}\in\Sigma_1}=0,$$
$$\left[\left[\left\{C^m_{ijkl}\,\hat{N}^{(2)}_{kq_1,l}+e^m_{ijk}\,\hat{W}^{(2)}_{q_1,k}+C^m_{ijkq_1}\,\hat{N}^{(1)}_k+e^m_{ijq_1}\,\hat{W}^{(1)}-\alpha^m_{ij}\,M^{(1)}_{q_1}\right\}n_j\right]\right]\bigg|_{\boldsymbol{\xi}\in\Sigma_1}=0,$$
$$\left[\left[\left\{e^m_{kli}\,\hat{N}^{(2)}_{kq_1,l}-\beta^m_{il}\,\hat{W}^{(2)}_{q_1,l}+e^m_{kq_1i}\,\hat{N}^{(1)}_k-\beta^m_{iq_1}\,\hat{W}^{(1)}+\gamma^m_i\,M^{(1)}_{q_1}\right\}n_i\right]\right]\bigg|_{\boldsymbol{\xi}\in\Sigma_1}=0. \quad (49)$$

The form of higher order (at $\varepsilon^m$ with $m\in\mathbb{Z}$ and $m\geq 1$) heat diffusion and piezoelectric cell problems can be found in Appendix A.1.

## 4  Down-scaling and up-scaling relations

From the solution of the above cell problems at different $\varepsilon$ described in Section 3.1, it is possible to express the microscopic fields $\mathbf{u}(\mathbf{x};\boldsymbol{\xi})$, $\phi(\mathbf{x};\boldsymbol{\xi})$ and $\theta(\mathbf{x};\boldsymbol{\xi})$ as asymptotic expansions of powers of the microscopic length $\varepsilon$ in terms of the macroscopic fields $\mathbf{U}(\mathbf{x})$, $\Phi(\mathbf{x})$, and $\Theta(\mathbf{x})$ and their gradients and in terms of the $\mathcal{Q}$-periodic perturbation functions. In particular, from expansions (8a) and (8b), taking into account the form of solutions (27), (30), (33), and (81) of cell problems at the different orders of $\varepsilon$, one derives the down-scaling relations of the micro-displacement field and the micro-electric potential field

$$u_k(\mathbf{x};\boldsymbol{\xi})=\left[U_k(\mathbf{x})+\sum_{l=1}^{+\infty}\varepsilon^l\sum_{|q|=l}\left(N^{(l)}_{kpq}(\boldsymbol{\xi})\frac{\partial^l U_p(\mathbf{x})}{\partial x_q}+\tilde{N}^{(l)}_{kq}(\boldsymbol{\xi})\frac{\partial^l\Phi(\mathbf{x})}{\partial x_q}\right)+\sum_{l=1}^{+\infty}\varepsilon^l\sum_{|q|=l-1}\hat{N}^{(l)}_{kq}(\boldsymbol{\xi})\frac{\partial^{l-1}\Theta(\mathbf{x})}{\partial x_q}\right]\bigg|_{\boldsymbol{\xi}=\frac{\mathbf{x}}{\varepsilon}},$$
(50a)

$$\phi(\mathbf{x};\boldsymbol{\xi})=\left[\phi(\mathbf{x})+\sum_{l=1}^{+\infty}\varepsilon^l\sum_{|q|=l}\left(W^{(l)}_q(\boldsymbol{\xi})\frac{\partial^l\Phi(\mathbf{x})}{\partial x_q}+\tilde{W}^{(l)}_{pq}(\boldsymbol{\xi})\frac{\partial^l U_p(\mathbf{x})}{\partial x_q}\right)+\sum_{l=1}^{+\infty}\varepsilon^l\sum_{|q|=l-1}\hat{W}^{(l)}_q(\boldsymbol{\xi})\frac{\partial^{l-1}\Theta(\mathbf{x})}{\partial x_q}\right]\bigg|_{\boldsymbol{\xi}=\frac{\mathbf{x}}{\varepsilon}}.$$
(50b)

Analogously, from expansion (8c) and the solutions (17), (20), (24), and (78), the down scaling relations of the micro relative temperature field are expressed as

$$\theta(\mathbf{x};\boldsymbol{\xi})=\left[\Theta(\mathbf{x})+\sum_{l=1}^{+\infty}\varepsilon^l\sum_{|q|=l}\left(M^{(l)}_q(\boldsymbol{\xi})\frac{\partial^l\Theta(\mathbf{x})}{\partial x_q}\right)\right]\bigg|_{\boldsymbol{\xi}=\frac{\mathbf{x}}{\varepsilon}}. \quad (51)$$

In equations (50a)-(51), $q = q_1, q_2, ...q_l$ is a multi-index, being $|q|$ the length of $q$, and $\frac{\partial^l(\cdot)}{\partial x_q} = \frac{\partial^l(\cdot)}{\partial x_{q_1}...\partial x_{q_l}}$. The $\mathcal{Q}$-periodic perturbation functions $N_{kpq}^{(l)}, \tilde{N}_{pq}^{(l)}, \hat{N}_{kq}^{(l)}, W_q^{(l)}, \tilde{W}_{pq}^{(l)}, \hat{W}_q^{(l)}, M_q^{(l)}$ reflect the influence of microstructural inhomogeneities of the material trough their dependency on the fast variable $\boldsymbol{\xi} = \mathbf{x}/\varepsilon$, while the macro-fields $\mathbf{U}(\mathbf{x})$, $\Phi(\mathbf{x})$, and $\Theta(\mathbf{x})$ are $\mathcal{L}$-periodic functions and therefore depend only on the slow variable $\mathbf{x}$.

Up-scaling relations, allow to define the macroscopic fields in terms of the relative microscopic ones. In particular, the macroscopic fields can be defined as the mean values of microscopic quantities (50a)-(51) over the unit cell $\mathcal{Q}$:

$$U_k(\mathbf{x}) \doteq \left\langle u_k(\mathbf{x}, \frac{\mathbf{x}}{\varepsilon} + \boldsymbol{\varsigma}) \right\rangle,$$
$$\Phi(\mathbf{x}) \doteq \left\langle \phi(\mathbf{x}, \frac{\mathbf{x}}{\varepsilon} + \boldsymbol{\varsigma}) \right\rangle,$$
$$\Theta(\mathbf{x}) \doteq \left\langle \theta(\mathbf{x}, \frac{\mathbf{x}}{\varepsilon} + \boldsymbol{\varsigma}) \right\rangle, \tag{52}$$

where it has been used a new variable $\boldsymbol{\varsigma} \in \mathcal{Q}$, such that the vector $\varepsilon\boldsymbol{\varsigma} \in \mathcal{A}$ defines the translation of the medium with respect to the $\mathcal{L}$-periodic volume forces $\mathbf{b}(\mathbf{x}), \rho_e(\mathbf{x})$, and $r(\mathbf{x})$ (Bacigalupo, 2014; Smyshlyaev and Cherednichenko, 2000). It can be proved that a $\mathcal{Q}$-periodic function $g(\boldsymbol{\xi} + \boldsymbol{\varsigma})$ satisfies the invariance property

$$\langle g(\boldsymbol{\xi} + \boldsymbol{\varsigma}) \rangle = \frac{1}{\delta} \int_{\mathcal{Q}} g(\boldsymbol{\xi} + \boldsymbol{\varsigma}) \, d\boldsymbol{\varsigma} = \frac{1}{\delta} \int_{\mathcal{Q}} g(\boldsymbol{\xi} + \boldsymbol{\varsigma}) \, d\boldsymbol{\xi}. \tag{53}$$

Equation (53), together with the normalization condition of type (21), valid for all the perturbation functions, leads to the up-scaling relations (52).

## 5 From average field equations of infinite order to homogenized thermo-piezoelectric first-order continuum

Substituting the down scaling relations (50a)-(51) into the micro-field equations (6a)-(6c) and reordering the different powers of $\varepsilon$, one obtains the following average field equations of infinite order

$$n_{ipq_1q_2}^{(2)} \frac{\partial^2 U_p(\mathbf{x})}{\partial x_{q_1} \partial x_{q_2}} + \sum_{n=0}^{+\infty} \varepsilon^{n+1} \sum_{|q|=n+3} n_{ipq}^{(n+3)} \frac{\partial^{n+3} U_p(\mathbf{x})}{\partial x_q} +$$
$$+ \tilde{n}_{iq_1q_2}^{(2)} \frac{\partial^2 \Phi(\mathbf{x})}{\partial x_{q_1} \partial x_{q_2}} + \sum_{n=0}^{+\infty} \varepsilon^{n+1} \sum_{|q|=n+3} \tilde{n}_{iq}^{(n+3)} \frac{\partial^{n+3} \Phi(\mathbf{x})}{\partial x_q} +$$
$$- \hat{n}_{iq_1}^{(2)} \frac{\partial \Theta(\mathbf{x})}{\partial x_{q_1}} - \sum_{n=0}^{+\infty} \varepsilon^{n+1} \sum_{|q|=n+2} \hat{n}_{iq}^{(n+3)} \frac{\partial^{n+2} \Theta(\mathbf{x})}{\partial x_q} + b_i(\mathbf{x}) = 0, \tag{54a}$$

$$\tilde{w}_{pq_1q_2}^{(2)} \frac{\partial^2 U_p(\mathbf{x})}{\partial x_{q_1} \partial x_{q_2}} + \sum_{n=0}^{+\infty} \varepsilon^{n+1} \sum_{|q|=n+3} \tilde{w}_{pq}^{(n+3)} \frac{\partial^{n+3} U_p(\mathbf{x})}{\partial x_q} +$$
$$- w_{q_1q_2}^{(2)} \frac{\partial^2 \Phi(\mathbf{x})}{\partial x_{q_1} \partial x_{q_2}} - \sum_{n=0}^{+\infty} \varepsilon^{n+1} \sum_{|q|=n+3} w_q^{(n+3)} \frac{\partial^{n+3} \Phi(\mathbf{x})}{\partial x_q} +$$
$$+ \hat{w}_{q_1}^{(2)} \frac{\partial \Theta(\mathbf{x})}{\partial x_{q_1}} + \sum_{n=0}^{+\infty} \varepsilon^{n+1} \sum_{|q|=n+2} \hat{w}_q^{(n+3)} \frac{\partial^{n+2} \Theta(\mathbf{x})}{\partial x_q} - \rho_e(\mathbf{x}) = 0, \tag{54b}$$

$$m_{q_1q_2}^{(2)} \frac{\partial^2 \Theta(\mathbf{x})}{\partial x_{q_1} \partial x_{q_2}} + \sum_{n=0}^{+\infty} \varepsilon^{n+1} \sum_{|q|=n+3} m_q^{(n+3)} \frac{\partial^{n+3} \Theta(\mathbf{x})}{\partial x_q} + r(\mathbf{x}) = 0. \tag{54c}$$

In equations (54a)-(54c) the coefficients of the gradients of the macroscopic fields are the known terms of the corresponding cell problem. Therefore, at the order $\varepsilon^0$ one has

$$n^{(2)}_{ipq_1q_2} = \frac{1}{2}\left\langle C^m_{iq_1pq_2} + C^m_{iq_2kl}\,N^{(1)}_{kpq_1,l} + e^m_{iq_2k}\,\tilde{W}^{(1)}_{pq_1,k} + C^m_{iq_2pq_1} + C^m_{iq_1kl}\,N^{(1)}_{kpq_2,l} + e^m_{iq_1k}\,\tilde{W}^{(1)}_{pq_2,k}\right\rangle, \tag{55a}$$

$$w^{(2)}_{q_1q_2} = \frac{1}{2}\left\langle \beta^m_{q_1q_2} + \beta^m_{q_2l}\,W^{(1)}_{q_1,l} - e^m_{klq_2}\,\tilde{N}^{(1)}_{kq_1,l} + \beta^m_{q_2q_1} + \beta^m_{q_1l}\,W^{(1)}_{q_2,l} - e^m_{q_1kl}\,\tilde{N}^{(1)}_{kq_2,l}\right\rangle, \tag{55b}$$

$$m^{(2)}_{q_1q_2} = \frac{1}{2}\left\langle K^m_{q_1q_2} + K^m_{q_2j}\,M^{(1)}_{q_1,j} + K^m_{q_2q_1} + K^m_{q_1j}\,M^{(1)}_{q_2,j}\right\rangle, \tag{55c}$$

$$\tilde{n}^{(2)}_{iq_1q_2} = \frac{1}{2}\left\langle C^m_{iq_2kl}\,\tilde{N}^{(1)}_{kq_1,l} + e^m_{iq_1q_2} + e^m_{iq_2k}\,W^{(1)}_{q_1,k} + C^m_{iq_1kl}\,\tilde{N}^{(1)}_{kq_2,l} + e^m_{iq_2q_1} + e^m_{iq_1k}\,W^{(1)}_{q_2,k}\right\rangle, \tag{55d}$$

$$\tilde{w}^{(2)}_{pq_1q_2} = \frac{1}{2}\left\langle e^m_{klq_2}\,N^{(1)}_{kpq_1,l} + e^m_{pq_2q_1} - \beta^m_{q_2l}\,\tilde{W}^{(1)}_{pq_1,l} + e^m_{klq_1}\,N^{(1)}_{kpq_2,l} + e^m_{q_2pq_1} - \beta^m_{q_1l}\,\tilde{W}^{(1)}_{pq_2,l}\right\rangle, \tag{55e}$$

$$\hat{n}^{(2)}_{iq_1} = \left\langle \alpha^m_{iq_1} - C^m_{iq_1kl}\,\hat{N}^{(1)}_{k,l} - e^m_{iq_1k}\,\hat{W}^{(1)}_{,k}\right\rangle, \tag{55f}$$

$$\hat{w}^{(2)}_{q_1} = \left\langle e^m_{klq_1}\,\hat{N}^{(1)}_{k,l} - \beta^m_{q_1l}\,\hat{W}^{(1)}_{,l} + \gamma^m_{q_1}\right\rangle. \tag{55g}$$

The corresponding quantities at the order $\varepsilon^m$, with $m \in \mathbb{Z}$ and $m \geq 1$, are

$$n^{(m+2)}_{ipq_1\ldots q_{m+2}} = \frac{1}{2^{m+2}}\sum_{\mathcal{P}(q)}\left\langle C^m_{iq_{m+1}kq_{m+2}}\,N^{(m)}_{kpq_1\ldots q_m} + C^m_{iq_{m+2}kl}\,N^{(m+1)}_{kpq_1\ldots q_{m+1},l} + e^m_{iq_{m+2}k}\,\tilde{W}^{(m+1)}_{pq_1\ldots q_{m+1},k}\right\rangle, \tag{56a}$$

$$w^{(m+2)}_{q_1\ldots q_{m+2}} = \frac{1}{2^{m+2}}\sum_{\mathcal{P}(q)}\left\langle \beta^m_{q_{m+1}q_{m+2}}\,W^{(m)}_{q_1\ldots q_m} + \beta^m_{q_{m+2}l}\,W^{(m+1)}_{q_1\ldots q_{m+1},l} - e^m_{klq_{m+2}}\,\tilde{N}^{(m+1)}_{kq_1\ldots q_{m+1},l}\right\rangle, \tag{56b}$$

$$m^{(m+2)}_{q_1\ldots q_{m+2}} = \frac{1}{2^{m+2}}\sum_{\mathcal{P}(q)}\left\langle K^m_{q_{m+1}q_{m+2}}\,M^{(m)}_{q_1\ldots q_m} + K^m_{q_{m+2}j}\,M^{(m+1)}_{q_1\ldots q_{m+1},j}\right\rangle, \tag{56c}$$

$$\tilde{n}^{(m+2)}_{iq_1\ldots q_{m+2}} = \frac{1}{2^{m+2}}\sum_{\mathcal{P}(q)}\left\langle C^m_{iq_{m+2}kl}\,\tilde{N}^{(m+1)}_{kq_1\ldots q_{m+1},l} + e^m_{iq_{m+1}q_{m+2}}\,W^{(m)}_{q_1\ldots q_m} + e^m_{iq_{m+2}k}\,W^{(m)}_{pq_1\ldots q_m,k}\right\rangle, \tag{56d}$$

$$\tilde{w}^{(m+2)}_{pq_1\ldots q_{m+2}} = \frac{1}{2^{m+2}}\sum_{\mathcal{P}(q)}\left\langle e^m_{klq_{m+2}}\,N^{(m+1)}_{kpq_1\ldots q_{m+1},l} + e^m_{kq_{m+2}q_{m+1}}\,N^{(m)}_{kpq_1\ldots q_m} - \beta^m_{q_{m+2}l}\,\tilde{W}^{(m+1)}_{pq_1\ldots q_{m+1},l}\right\rangle, \tag{56e}$$

$$\hat{n}^{(m+2)}_{iq_1\ldots q_{m+1}} = \frac{1}{2^{m+1}}\sum_{\mathcal{P}(q)}\left\langle \alpha^m_{iq_{m+1}}\,M^{(m)}_{q_1\ldots q_m} - C^m_{iq_{m+1}kl}\,\hat{N}^{(m+1)}_{kq_1\ldots q_m,l} - e^m_{iq_{m+1}k}\,\hat{W}^{(m+1)}_{q_1\ldots q_m,k}\right\rangle, \tag{56f}$$

$$\hat{w}^{(m+2)}_{q_1\ldots q_{m+1}} = \frac{1}{2^{m+1}}\sum_{\mathcal{P}(q)}\left\langle e^m_{klq_{m+1}}\,\hat{N}^{(m+1)}_{kq_1\ldots q_m,l} - \beta^m_{q_{m+1}l}\,\hat{W}^{(m+1)}_{q_1\ldots q_m,l} + \gamma^m_{q_{m+1}}\,M^{(m)}_{q_1\ldots q_m}\right\rangle, \tag{56g}$$

where symbol $\mathcal{P}(q)$ denotes all the possible permutations of the multi-index $q$.

The average field equations of infinite order (54a)-(54c) can be formally solved by performing an asymptotic expansion of the macro fields $U_k(\mathbf{x})$, $\Phi(\mathbf{x})$, and $\Theta(\mathbf{x})$, in powers of $\varepsilon$, namely

$$U_k(\mathbf{x}) = \sum_{j=0}^{+\infty} \varepsilon^j\,U^{(j)}_k(\mathbf{x}), \tag{57a}$$

$$\Phi(\mathbf{x}) = \sum_{j=0}^{+\infty} \varepsilon^j\,\Phi^{(j)}(\mathbf{x}), \tag{57b}$$

$$\Theta(\mathbf{x}) = \sum_{j=0}^{+\infty} \varepsilon^j\,\Theta^{(j)}(\mathbf{x}). \tag{57c}$$

From the substitution of equations (57a)-(57c) into (54a)-(54c), three sets of recursive differential problems can be obtained with regard to the terms of the asymptotic expansions of the macro fields $U^{(j)}_k$, $\Phi^{(j)}$, and $\Theta^{(j)}$.

**Heat diffusion problem**

For the heat diffusion problem (54c) one has

$$m_{q_1q_2}^{(2)}\left(\frac{\partial^2\Theta^{(0)}}{\partial x_{q_1}\partial x_{q_2}} + \varepsilon\frac{\partial^2\Theta^{(1)}}{\partial x_{q_1}\partial x_{q_2}} + \varepsilon^2\frac{\partial^2\Theta^{(2)}}{\partial x_{q_1}\partial x_{q_2}} + ...\right) + \varepsilon\, m_{q_1...q_3}^{(3)}\left(\frac{\partial^3\Theta^{(0)}}{\partial x_{q_1}...\partial x_{q_3}} + \varepsilon\frac{\partial^3\Theta^{(1)}}{\partial x_{q_1}...\partial x_{q_3}} + \right.$$
$$\left. + \varepsilon^2\frac{\partial^3\Theta^{(2)}}{\partial x_{q_1}...\partial x_{q_3}} + ...\right) + \varepsilon^2\, m_{q_1...q_4}^{(4)}\left(\frac{\partial^4\Theta^{(0)}}{\partial x_{q_1}...\partial x_{q_4}} + \varepsilon\frac{\partial^4\Theta^{(1)}}{\partial x_{q_1}...\partial x_{q_4}} + \varepsilon^2\frac{\partial^4\Theta^{(2)}}{\partial x_{q_1}...\partial x_{q_4}} + ...\right) + ... + r(\mathbf{x}) = 0,$$
(58)

which gives the following macroscopic recursive problems at the different orders of $\varepsilon$, namely at $\varepsilon^0$

$$m_{q_1q_2}^{(2)}\frac{\partial^2\Theta^{(0)}}{\partial x_{q_1}\partial x_{q_2}} + r(\mathbf{x}) = 0, \qquad (59)$$

at $\varepsilon^1$

$$m_{q_1q_2}^{(2)}\frac{\partial^2\Theta^{(1)}}{\partial x_{q_1}\partial x_{q_2}} + m_{q_1...q_3}^{(3)}\frac{\partial^3\Theta^{(0)}}{\partial x_{q_1}...\partial x_{q_3}} = 0, \qquad (60)$$

and, at $\varepsilon^m$ with $m \in \mathbb{Z}$ and $m \geq 1$

$$m_{q_1q_2}^{(2)}\frac{\partial^2\Theta^{(m)}}{\partial x_{q_1}\partial x_{q_2}} + \sum_{p=3}^{m+2}\sum_{|j|=p} m_j^{(p)}\frac{\partial^p\Theta^{(m-p+2)}}{\partial x_j} = 0, \qquad (61)$$

where $j$ is a multi-index.

**Piezoelectric problem**

Analogously, from the piezoelectric problems (54a) and (54b), one derives two sets of recursive differential problems, expressed as

$$n_{ipq_1q_2}^{(2)}\left(\frac{\partial^2 U_p^{(0)}}{\partial x_{q_1}\partial x_{q_2}} + \varepsilon\frac{\partial^2 U_p^{(1)}}{\partial x_{q_1}\partial x_{q_2}} + \varepsilon^2\frac{\partial^2 U_p^{(2)}}{\partial x_{q_1}\partial x_{q_2}} + ...\right) + \varepsilon\, n_{ipq_1...q_3}^{(3)}\left(\frac{\partial^3 U_p^{(0)}}{\partial x_{q_1}...\partial x_{q_3}} + \varepsilon\frac{\partial^3 U_p^{(1)}}{\partial x_{q_1}...\partial x_{q_3}} + \right.$$
$$\left. +\varepsilon^2\frac{\partial^3 U_p^{(2)}}{\partial x_{q_1}...\partial x_{q_3}} + ...\right) + \varepsilon^2\, n_{ipq_1...q_4}^{(4)}\left(\frac{\partial^4 U_p^{(0)}}{\partial x_{q_1}...\partial x_{q_4}} + \varepsilon\frac{\partial^4 U_p^{(1)}}{\partial x_{q_1}...\partial x_{q_4}} + \varepsilon^2\frac{\partial^4 U_p^{(2)}}{\partial x_{q_1}...\partial x_{q_4}} + ...\right) + ... +$$
$$+\tilde{n}_{iq_1q_2}^{(2)}\left(\frac{\partial^2\Phi^{(0)}}{\partial x_{q_1}\partial x_{q_2}} + \varepsilon\frac{\partial^2\Phi^{(1)}}{\partial x_{q_1}\partial x_{q_2}} + \varepsilon^2\frac{\partial^2\Phi^{(2)}}{\partial x_{q_1}\partial x_{q_2}} + ...\right) + \varepsilon\,\tilde{n}_{iq_1...q_3}^{(3)}\left(\frac{\partial^3\Phi^{(0)}}{\partial x_{q_1}...\partial x_{q_3}} + \varepsilon\frac{\partial^3\Phi^{(1)}}{\partial x_{q_1}...\partial x_{q_3}} + \right.$$
$$\left. +\varepsilon^2\frac{\partial^3\Phi^{(2)}}{\partial x_{q_1}...\partial x_{q_3}} + ...\right) + \varepsilon^2\,\tilde{n}_{iq_1...q_4}^{(4)}\left(\frac{\partial^4\Phi^{(0)}}{\partial x_{q_1}...\partial x_{q_4}} + \varepsilon\frac{\partial^4\Phi^{(1)}}{\partial x_{q_1}...\partial x_{q_4}} + \varepsilon^2\frac{\partial^4\Phi^{(2)}}{\partial x_{q_1}...\partial x_{q_4}} + ...\right) + ... +$$
$$-\hat{n}_{iq_1}^{(2)}\left(\frac{\partial\Theta^{(0)}}{\partial x_{q_1}} + \varepsilon\frac{\partial\Theta^{(1)}}{\partial x_{q_1}} + \varepsilon^2\frac{\partial\Theta^{(2)}}{\partial x_{q_1}} + ...\right) - \varepsilon\,\hat{n}_{iq_1q_2}^{(3)}\left(\frac{\partial^2\Theta^{(0)}}{\partial x_{q_1}\partial x_{q_2}} + \varepsilon\frac{\partial^2\Theta^{(1)}}{\partial x_{q_1}\partial x_{q_2}} + \right.$$
$$\left. +\varepsilon^2\frac{\partial^2\Theta^{(2)}}{\partial x_{q_1}\partial x_{q_2}} + ...\right) - \varepsilon^2\,\hat{n}_{iq_1...q_3}^{(4)}\left(\frac{\partial^3\Theta^{(0)}}{\partial x_{q_1}...\partial x_{q_3}} + \varepsilon\frac{\partial^3\Theta^{(1)}}{\partial x_{q_1}...\partial x_{q_3}} + \varepsilon^2\frac{\partial^3\Theta^{(2)}}{\partial x_{q_1}...\partial x_{q_3}} + ...\right) + ... + b_i(\mathbf{x}) = 0,$$
(62a)

$$\tilde{w}_{pq_1q_2}^{(2)}\left(\frac{\partial^2 U_p^{(0)}}{\partial x_{q_1}\partial x_{q_2}} + \varepsilon\frac{\partial^2 U_p^{(1)}}{\partial x_{q_1}\partial x_{q_2}} + \varepsilon^2\frac{\partial^2 U_p^{(2)}}{\partial x_{q_1}\partial x_{q_2}} + ...\right) + \varepsilon\,\tilde{w}_{pq_1...q_3}^{(3)}\left(\frac{\partial^3 U_p^{(0)}}{\partial x_{q_1}...\partial x_{q_3}} + \varepsilon\frac{\partial^3 U_p^{(1)}}{\partial x_{q_1}...\partial x_{q_3}} + \right.$$
$$\left. +\varepsilon^2\frac{\partial^3 U_p^{(2)}}{\partial x_{q_1}...\partial x_{q_3}} + ...\right) + \varepsilon^2\,\tilde{w}_{pq_1...q_4}^{(4)}\left(\frac{\partial^4 U_p^{(0)}}{\partial x_{q_1}...\partial x_{q_4}} + \varepsilon\frac{\partial^4 U_p^{(1)}}{\partial x_{q_1}...\partial x_{q_4}} + \varepsilon^2\frac{\partial^4 U_p^{(2)}}{\partial x_{q_1}...\partial x_{q_4}} + ...\right) + ... +$$
$$-w_{q_1q_2}^{(2)}\left(\frac{\partial^2\Phi^{(0)}}{\partial x_{q_1}\partial x_{q_2}} + \varepsilon\frac{\partial^2\Phi^{(1)}}{\partial x_{q_1}\partial x_{q_2}} + \varepsilon^2\frac{\partial^2\Phi^{(2)}}{\partial x_{q_1}\partial x_{q_2}} + ...\right) - \varepsilon\, w_{q_1...q_3}^{(3)}\left(\frac{\partial^3\Phi^{(0)}}{\partial x_{q_1}...\partial x_{q_3}} + \varepsilon\frac{\partial^3\Phi^{(1)}}{\partial x_{q_1}...\partial x_{q_3}} + \right.$$
$$\left. +\varepsilon^2\frac{\partial^3\Phi^{(2)}}{\partial x_{q_1}...\partial x_{q_3}} + ...\right) - \varepsilon^2\, w_{q_1...q_4}^{(4)}\left(\frac{\partial^4\Phi^{(0)}}{\partial x_{q_1}...\partial x_{q_4}} + \varepsilon\frac{\partial^4\Phi^{(1)}}{\partial x_{q_1}...\partial x_{q_4}} + \varepsilon^2\frac{\partial^4\Phi^{(2)}}{\partial x_{q_1}...\partial x_{q_4}} + ...\right) + ... +$$
$$+\hat{w}_{q_1}^{(2)}\left(\frac{\partial\Theta^{(0)}}{\partial x_{q_1}} + \varepsilon\frac{\partial\Theta^{(1)}}{\partial x_{q_1}} + \varepsilon^2\frac{\partial\Theta^{(2)}}{\partial x_{q_1}} + ...\right) + \varepsilon\,\hat{w}_{q_1q_2}^{(3)}\left(\frac{\partial^2\Theta^{(0)}}{\partial x_{q_1}\partial x_{q_2}} + \varepsilon\frac{\partial^2\Theta^{(1)}}{\partial x_{q_1}\partial x_{q_2}} + \right.$$

$$+\varepsilon^2 \frac{\partial^2 \Theta^{(2)}}{\partial x_{q_1} \partial x_{q_2}} + ...\Bigg) + \varepsilon^2\, \hat{w}^{(4)}_{q_1...q_3} \left( \frac{\partial^3 \Theta^{(0)}}{\partial x_{q_1}...\partial x_{q_3}} + \varepsilon \frac{\partial^3 \Theta^{(1)}}{\partial x_{q_1}...\partial x_{q_3}} + \varepsilon^2 \frac{\partial^3 \Theta^{(2)}}{\partial x_{q_1}...\partial x_{q_3}} + ... \right) + ... - \rho_e(\mathbf{x}) = 0. \tag{62b}$$

Equations (62a) and (62b), at the order $\varepsilon^0$, bring to the following macro problems

$$n^{(2)}_{ipq_1q_2} \frac{\partial^2 U_p^{(0)}}{\partial x_{q_1}\partial x_{q_2}} + \tilde{n}^{(2)}_{iq_1q_2} \frac{\partial^2 \Phi^{(0)}}{\partial x_{q_1}\partial x_{q_2}} - \hat{n}^{(2)}_{iq_1} \frac{\partial \Theta^{(0)}}{\partial x_{q_1}} + b_i(\mathbf{x}) = 0, \tag{63a}$$

$$\tilde{w}^{(2)}_{pq_1q_2} \frac{\partial^2 U_p^{(0}}{\partial x_{q_1}\partial x_{q_2}} - w^{(2)}_{q_1q_2} \frac{\partial^2 \Phi^{(0)}}{\partial x_{q_1}\partial x_{q_2}} + \hat{w}^{(2)}_{q_1} \frac{\partial \Theta^{(0)}}{\partial x_{q_1}} - \rho_e(\mathbf{x}) = 0. \tag{63b}$$

At the order $\varepsilon^1$, one has

$$n^{(2)}_{ipq_1q_2} \frac{\partial^2 U_p^{(1)}}{\partial x_{q_1}\partial x_{q_2}} + \tilde{n}^{(2)}_{iq_1q_2} \frac{\partial^2 \Phi^{(1)}}{\partial x_{q_1}\partial x_{q_2}} - \hat{n}^{(2)}_{iq_1} \frac{\partial \Theta^{(1)}}{\partial x_{q_1}} +$$
$$+ n^{(3)}_{ipq_1...q_3} \frac{\partial^3 U_p^{(0)}}{\partial x_{q_1}...\partial x_{q_3}} + \tilde{n}^{(3)}_{iq_1...q_3} \frac{\partial^3 \Phi^{(0)}}{\partial x_{q_1}...\partial x_{q_3}} - \hat{n}^{(3)}_{iq_1q_2} \frac{\partial^2 \Theta^{(0)}}{\partial x_{q_1}\partial x_{q_2}} = 0, \tag{64a}$$

$$\tilde{w}^{(2)}_{pq_1q_2} \frac{\partial^2 U_p^{(1)}}{\partial x_{q_1}\partial x_{q_2}} - w^{(2)}_{q_1q_2} \frac{\partial^2 \Phi^{(1)}}{\partial x_{q_1}\partial x_{q_2}} + \hat{w}^{(2)}_{q_1} \frac{\partial \Theta^{(1)}}{\partial x_{q_1}} +$$
$$+ \tilde{w}^{(3)}_{pq_1...q_3} \frac{\partial^3 U_p^{(0)}}{\partial x_{q_1}...\partial x_{q_3}} - w^{(3)}_{q_1...q_3} \frac{\partial^3 \Phi^{(0)}}{\partial x_{q_1}...\partial x_{q_3}} + \hat{w}^{(3)}_{q_1q_2} \frac{\partial^2 \Theta^{(0)}}{\partial x_{q_1}\partial x_{q_2}} = 0, \tag{64b}$$

and finally, at the order $\varepsilon^m$ (with $m \in \mathbb{Z}$ and $m \geq 1$) equations (62a) and (62b) provide

$$n^{(2)}_{ipq_1q_2} \frac{\partial^2 U_p^{(m)}}{\partial x_{q_1}\partial x_{q_2}} + \tilde{n}^{(2)}_{iq_1q_2} \frac{\partial^2 \Phi^{(m)}}{\partial x_{q_1}\partial x_{q_2}} - \hat{n}^{(2)}_{iq_1} \frac{\partial \Theta^{(m)}}{\partial x_{q_1}} + \sum_{r=3}^{m+2} \sum_{|j|=r} \left( n^{(r)}_{ipj} \frac{\partial^r U_p^{(m-r+2)}}{\partial x_j} + \tilde{n}^{(r)}_{ij} \frac{\partial^r \Phi^{(m-r+2)}}{\partial x_j} \right) +$$
$$- \sum_{r=3}^{m+2} \sum_{|j|=r-1} \hat{n}^{(r)}_{ij} \frac{\partial^{r-1} \Theta^{(m-r+2)}}{\partial x_j} = 0, \tag{65a}$$

$$\tilde{w}^{(2)}_{pq_1q_2} \frac{\partial^2 U_p^{(m)}}{\partial x_{q_1}\partial x_{q_2}} - w^{(2)}_{q_1q_2} \frac{\partial^2 \Phi^{(m)}}{\partial x_{q_1}\partial x_{q_2}} + \hat{w}^{(2)}_{q_1} \frac{\partial \Theta^{(m)}}{\partial x_{q_1}} + \sum_{r=3}^{m+2} \sum_{|j|=r} \left( \tilde{w}^{(r)}_{pj} \frac{\partial^r U_p^{(m-r+2)}}{\partial x_j} - w^{(r)}_j \frac{\partial^r \Phi^{(m-r+2)}}{\partial x_j} \right) +$$
$$+ \sum_{r=3}^{m+2} \sum_{|j|=r-1} \hat{w}^{(r)}_j \frac{\partial^{r-1} \Theta^{(m-r+2)}}{\partial x_j} = 0. \tag{65b}$$

The $\mathcal{L}$-periodic solutions of differential problems (59)-(65b) are required to fulfill the following normalization conditions

$$\frac{1}{\delta L^2} \int_{\mathcal{L}} U_p^{(m)}(\mathbf{x})\, d\mathbf{x} = 0, \quad \frac{1}{\delta L^2} \int_{\mathcal{L}} \Phi^{(m)}(\mathbf{x})\, d\mathbf{x} = 0, \quad \frac{1}{\delta L^2} \int_{\mathcal{L}} \Theta^{(m)}(\mathbf{x})\, d\mathbf{x} = 0, \tag{66}$$

for each $m \in \mathbb{Z}$. Truncating expansions (57a)-(57c) of the macro fields at the zeroth order, namely

$$U_k(\mathbf{x}) \approx U_k^{(0)}(\mathbf{x}), \quad \Phi(\mathbf{x}) \approx \Phi^{(0)}(\mathbf{x}), \quad \Theta(\mathbf{x}) \approx \Theta^{(0)}(\mathbf{x}),$$

one obtains the equivalent first-order (Cauchy) homogeneous continuum of the investigated periodic thermo-piezoelectric medium. Thanks to the symmetry and positive definiteness of tensors $\boldsymbol{n}^{(2)} = n^{(2)}_{ipq_1q_2} \mathbf{e}_i \otimes \mathbf{e}_p \otimes \mathbf{e}_{q_1} \otimes \mathbf{e}_{q_2}$, $\boldsymbol{w}^{(2)} = w^{(2)}_{q_1q_2} \mathbf{e}_{q_1} \otimes \mathbf{e}_{q_2}$, $\boldsymbol{m}^{(2)} = m^{(2)}_{q_1q_2} \mathbf{e}_{q_1} \otimes \mathbf{e}_{q_2}$, and the equality between the components of tensors $\tilde{\boldsymbol{n}}^{(2)} = \tilde{n}^{(2)}_{iq_1q_2} \mathbf{e}_i \otimes \mathbf{e}_{q_1} \otimes \mathbf{e}_{q_2}$ and $\tilde{\boldsymbol{w}}^{(2)} = \tilde{w}^{(2)}_{iq_1q_2} \mathbf{e}_i \otimes \mathbf{e}_{q_1} \otimes \mathbf{e}_{q_2}$ (see the detailed demonstration of such properties in Appendix B), the zeroth order problems (63a),(63b) and (59) can be written in terms of the components $C_{iq_1pq_2}$, $\beta_{q_1q_2}$, $K_{q_1q_2}$, $e_{iq_1q_2}$, $\alpha_{iq_1}$, and $\gamma_{q_1}$ of the overall constitutive tensors, namely:

$$C_{iq_1pq_2} \frac{\partial^2 U_p(\mathbf{x})}{\partial x_{q_1}\partial x_{q_2}} + e_{iq_1q_2} \frac{\partial^2 \Phi(\mathbf{x})}{\partial x_{q_1}\partial x_{q_2}} - \alpha_{iq_1} \frac{\partial \Theta(\mathbf{x})}{\partial x_{q_1}} + b_i(\mathbf{x}) = 0, \tag{67a}$$

$$e_{pq_1q_2}\frac{\partial^2 U_p(\mathbf{x})}{\partial x_{q_1}\partial x_{q_2}} - \beta_{q_1q_2}\frac{\partial^2 \Phi(\mathbf{x})}{\partial x_{q_1}\partial x_{q_2}} + \gamma_{q_1}\frac{\partial \Theta(\mathbf{x})}{\partial x_{q_1}} - \rho_e(\mathbf{x}) = 0, \tag{67b}$$

$$K_{q_1q_2}\frac{\partial^2 \Theta(\mathbf{x})}{\partial x_{q_1}\partial x_{q_2}} + r(\mathbf{x}) = 0. \tag{67c}$$

In Appendix B, the relations between the components of tensors $\boldsymbol{n}^{(2)}, \boldsymbol{w}^{(2)}, \boldsymbol{m}^{(2)}, \tilde{\boldsymbol{n}}^{(2)}$, and $\tilde{\boldsymbol{w}}^{(2)}$, as expressed in equations (55a)-(55e), and the components of the corresponding overall constitutive tensors $\mathbb{C}, \boldsymbol{\beta}, \boldsymbol{K}$, and $\boldsymbol{e}$, are derived in detail. Such relations are expressed as

$$n^{(2)}_{ipq_1q_2} = \frac{1}{2}\left(C_{pq_1iq_2} + C_{pq_2iq_1}\right), \quad w^{(2)}_{q_1q_2} = \beta_{q_1q_2}, \quad m^{(2)}_{q_1q_2} = K_{q_1q_2}, \quad \tilde{n}^{(2)}_{iq_1q_2} = \tilde{w}^{(2)}_{iq_1q_2} = e_{iq_1q_2}.$$

## 6 Benchmark test: homogenization of a two-phase thermo-piezoelectric material

In order to assess the capabilities of the presented first-order homogenization technique, the general formulation derived in the previous sections is now tested in the case of a two-dimensional infinite thermo-piezoelectric material subjected to $\mathcal{L}$-periodic body forces $\mathbf{b}(\mathbf{x})$, free charge densities $\rho_e(\mathbf{x})$, and heat sources $r(\mathbf{x})$, for the problem geometry shown in figure 2(a).

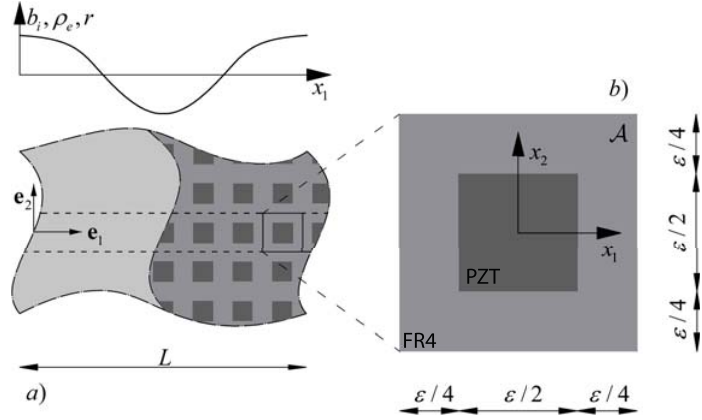

Figure 2: (a) Heterogeneous model and homogenized one subject to $\mathcal{L}$-periodic volume forces $b_i(x_1)$, free charge densities $\rho_e(x_1)$, and heat sources $r(x_1)$; (b)Periodic cell $\mathcal{A}$ with characteristic size $\varepsilon$ made of FR4 and a ceramic PZT inclusion.

The analytical solution of the homogenized model in terms of macro displacement $\mathbf{U}(\mathbf{x})$, electric potential $\Phi(\mathbf{x})$, and temperature $\Theta(\mathbf{x})$ is compared with the results provided by a finite element analysis of the corresponding heterogeneous model.

The periodic cell $\mathcal{A}$ of the considered thermo-piezoelectric material is a 10mm× 10mm cell with a 5mm× 5mm inclusion (see figure 2(b)). The two different phases constituting the periodic cell are assumed to be homogeneous. The geometry of such periodic cell reproduces the one of a pyroelectric cell typically used as an energy harvester and organized in array of elements (Hsiao and Jia-Wai, 2015; Hsiao et al., 2015). The inclusion is made of a material like the Lead Zirconate Titanate (PZT-5H) which has marked piezoelectric and pyroelectric properties. Such a material is characterized by the following constitutive tensors (Guo et al., 2003; Kommepalli et al., 2010; Malmonge et al., 2003; Umemiya et al., 2006; Yang, 2004) that, accordingly to the notation detailed in Appendix C, equation (116), are expressed as

$$\mathbb{C}_{PZT} = \begin{pmatrix} 11.7 & 8.41 & 0 \\ 8.41 & 12.6 & 0 \\ 0 & 0 & 2\cdot 2.3 \end{pmatrix} 10^{10}\,\frac{\text{N}}{\text{m}^2}, \quad \boldsymbol{\beta}_{PZT} = \begin{pmatrix} 1.302 & 0 \\ 0 & 1.505 \end{pmatrix} 10^{-8}\,\frac{\text{C}}{\text{V m}},$$

$$\boldsymbol{K}_{PZT} = \begin{pmatrix} 1.5 & 0 \\ 0 & 1.5 \end{pmatrix}\,\frac{\text{W}}{\text{m K}}, \quad \tilde{\boldsymbol{e}}_{PZT} = \begin{pmatrix} 23.3 & -6.5 & 0 \\ 0 & 0 & \sqrt{2}\cdot 17 \end{pmatrix}\,\frac{\text{C}}{\text{m}^2},$$

$$\boldsymbol{\alpha}_{PZT} = \begin{pmatrix} 1.71 \\ 1.71 \\ 0 \end{pmatrix} 10^6 \, \frac{\text{N}}{\text{m}^2\,\text{K}}, \qquad \boldsymbol{\gamma}_{PZT} = \begin{pmatrix} 5 \\ 5 \end{pmatrix} 10^{-4} \, \frac{\text{C}}{\text{m}^2\,K}.$$

The matrix of the periodic cell is made of a glass-reinforced epoxy laminate sheet (FR-4) having negligible piezoelectric and pyroelectric properties (i.e. $\boldsymbol{e}_{FR4} = \boldsymbol{0}$, $\boldsymbol{\gamma}_{FR4} = \boldsymbol{0}$) described by the following constitutive tensors (Azar and Graebner, 1996; Wang et al., 2001):

$$\mathbb{C}_{FR4} = \begin{pmatrix} 1.75 & 0.53 & 0 \\ 0.53 & 1.75 & 0 \\ 0 & 0 & 2 \cdot 6.15 \end{pmatrix} 10^{10} \, \frac{\text{N}}{\text{m}^2}, \qquad \boldsymbol{\beta}_{FR4} = \begin{pmatrix} 4.16 & 0 \\ 0 & 4.16 \end{pmatrix} 10^{-11} \, \frac{\text{C}}{\text{V m}},$$

$$\boldsymbol{K}_{FR4} = \begin{pmatrix} 0.81 & 0 \\ 0 & 0.81 \end{pmatrix} \frac{\text{W}}{\text{m K}}, \qquad \boldsymbol{\alpha}_{FR4} = \begin{pmatrix} 4.5 & 0 \\ 0 & 4.5 \end{pmatrix} 10^5 \, \frac{\text{N}}{\text{m}^2\,\text{K}}.$$

Perturbation functions $N_{kpq_1}^{(1)}, \tilde{N}_{kq_1}^{(1)}, \hat{N}_k^{(1)}, W_{q_1}^{(1)}, \tilde{W}_{pq_1}^{(1)}, \hat{W}^{(1)}$, and $M_{q_1}^{(1)}$ have been derived trough the numerical resolution obtained by a finite element procedure of the cell problems (34), (38), (40) and (42) at the order $\varepsilon^{-1}$. Some of the obtained perturbation functions over the unit cell $\mathcal{Q}$ are represented in figure 3, which shows that perturbation functions are $\mathcal{Q}$-periodic, smooth along the boundaries of the unit cell, and have vanishing mean values over $\mathcal{Q}$ for the imposed normalization condition of type (21).

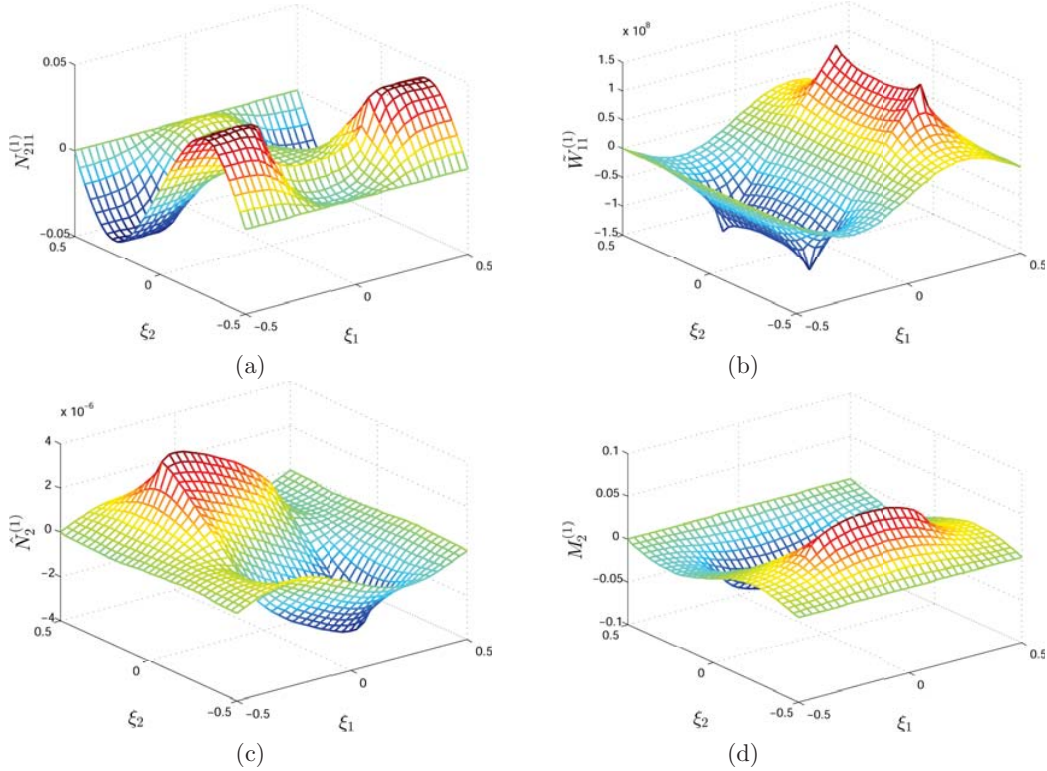

Figure 3: *Perturbation functions obtained trough the finite element solution of cell problems at the order $\varepsilon^{-1}$ over the unit cell $\mathcal{Q}$ made of FR4 with a PZT-5H inclusion whose topology is the one of figure 2-b: (a) $N_{211}^{(1)}$; (b) $\tilde{W}_{11}^{(1)}$; (c) $\hat{N}_2^{(1)}$; (d) $M_2^{(1)}$.*

Once perturbation functions are determined, the overall constitutive tensors of the first-order homogenized medium have been computed by means of closed-forms equations (55a)-(56a) and they result

$$\mathbb{C} = \begin{pmatrix} 2.56 & 0.73 & 0 \\ 0.73 & 2.50 & 0 \\ 0 & 0 & 2 \cdot 0.81 \end{pmatrix} 10^{10} \, \frac{\text{N}}{\text{m}^2}, \qquad \boldsymbol{\beta} = \begin{pmatrix} 7.194 & 0 \\ 0 & 7.198 \end{pmatrix} 10^{-11} \, \frac{\text{C}}{\text{V m}},$$

$$\boldsymbol{K} = \begin{pmatrix} 9.415 & 0 \\ 0 & 9.415 \end{pmatrix} 10^{-1} \frac{\mathrm{W}}{\mathrm{m\,K}}, \quad \tilde{\boldsymbol{e}} = \begin{pmatrix} 0.0146 & -5.8674 & 0 \\ 0 & 0 & \sqrt{2}\cdot 0.0076 \end{pmatrix} \frac{\mathrm{C}}{\mathrm{m}^2},$$

$$\boldsymbol{\alpha} = \begin{pmatrix} 4.49 \\ 5.58 \\ -\sqrt{2}\cdot 0.31 \end{pmatrix} 10^5 \frac{\mathrm{N}}{\mathrm{m}^2\,\mathrm{K}}, \quad \boldsymbol{\gamma} = \begin{pmatrix} 8.88 \\ 5.94 \end{pmatrix} 10^{-7} \frac{\mathrm{C}}{\mathrm{m}^2\,K}. \tag{68}$$

One considers the piezo-electric tensor of the inclusion in the form $\boldsymbol{e}_{INC} = \eta\,\boldsymbol{e}_{PZT}$ as a function of the piezo-electric properties of the ceramic material PZT through a piezo-electric multiplicative factor $\eta$ with $0 \leq \eta \leq 2$, and such that, for $\eta = 0$, the inclusion presents vanishing piezo-electric features. In such conditions, the components of the overall constitutive tensors $\mathbb{C}, \boldsymbol{\beta}, \boldsymbol{\alpha}$, and $\boldsymbol{\gamma}$, expressed respectively by equations (55a), (55b), (55f), and (56a), vary with respect to $\eta$, as represented in figure 4. In particular, figure 4 depicts the dimensionless components of the overall constitutive tensors, defined as

$$\tilde{C}_{iq_1 pq_2} = \frac{C_{iq_1 pq_2}}{C_{1111}|_{\eta=0}}, \quad \tilde{\beta}_{q_1 q_2} = \frac{\beta_{q_1 q_2}}{\beta_{11}|_{\eta=0}}, \quad \tilde{\alpha}_{iq_1} = \frac{\alpha_{iq_1}}{\alpha_{11}|_{\eta=0}}, \quad \tilde{\gamma}_{q_1} = \frac{\gamma_{q_1}}{\gamma_1|_{\eta=0}}. \tag{69}$$

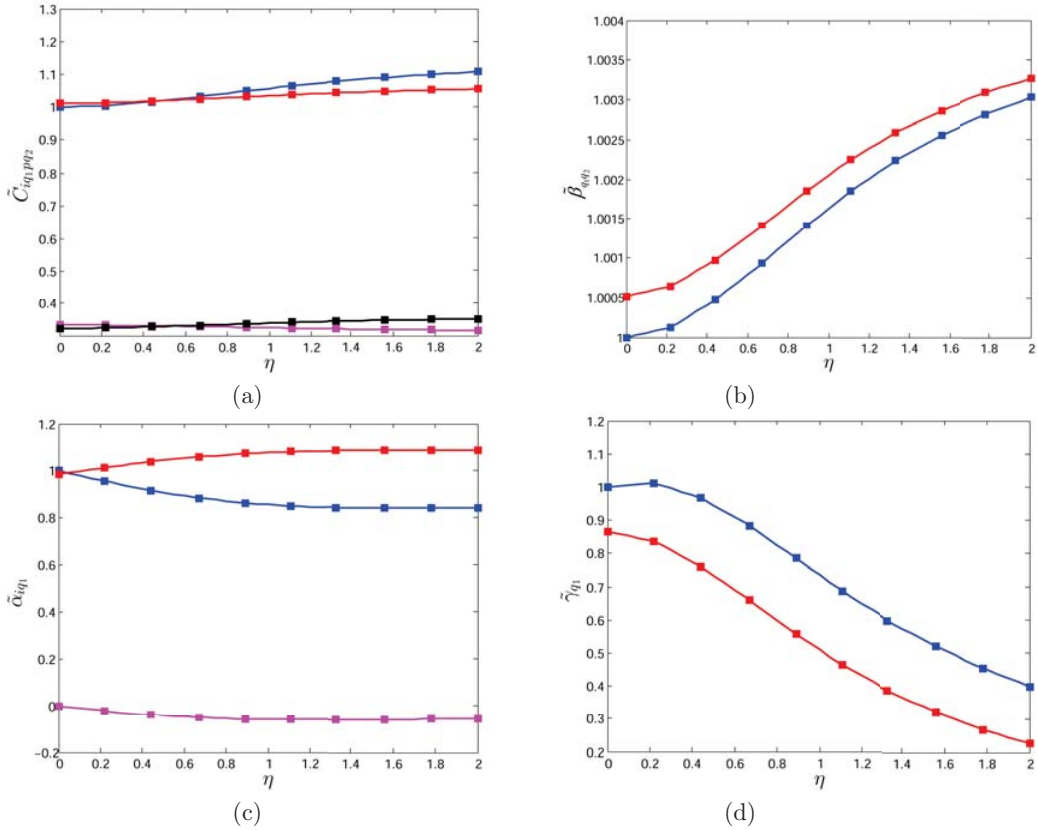

Figure 4: *Dimensionless components of the overall constitutive tensors vs piezo-electric multiplicative factor $\eta$, with $0 \leq \eta \leq 2$. (a) $\tilde{C}_{1111}$ (blue curve), $\tilde{C}_{2222}$ (red curve), $\tilde{C}_{1212}$ (magenta curve), and $\tilde{C}_{1112} = \tilde{C}_{2212}$ (black curve); (b) $\tilde{\beta}_{11}$ (blue curve), and $\tilde{\beta}_{22}$ (red curve); (c) $\tilde{\alpha}_{11}$ (blue curve), $\tilde{\alpha}_{22}$ (red curve), and $\tilde{\alpha}_{12}$ (magenta curve); (d) $\tilde{\gamma}_1$ (blue curve), and $\tilde{\gamma}_2$ (red curve).*

The overall pyroelectric tensor $\boldsymbol{\gamma}$ results to be the most affected by the variation of the piezo-electricity of the inclusion, since at $\eta = 2$ the component $\tilde{\gamma}_1$ shows a variation of about 58% and $\tilde{\gamma}_2$ of 70% with respect to their values at $\eta = 0$. The variation of the overall elastic tensor $\mathbb{C}$ at $\eta = 2$ ranges from 4% for $\tilde{C}_{2222}$ to 10.7% for $\tilde{C}_{1111}$ with respect to their values at $\eta = 0$, while the variation of the overall tensor $\boldsymbol{\alpha}$ ranges from 10% for $\tilde{\alpha}_{11}$ to 15.7% for $\tilde{\alpha}_{22}$.

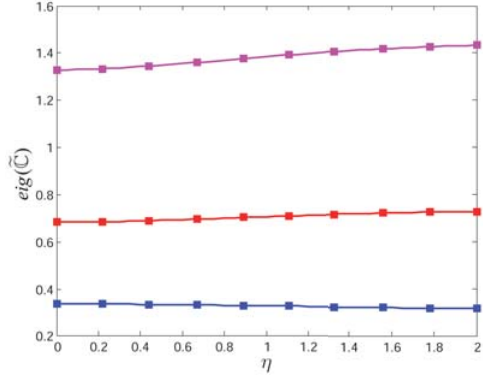

Figure 5: *Eigenvalues of $\tilde{\mathbb{C}}$ vs the piezo-electric multiplicative factor $\eta$.*

The overall dielectric permittivities tensor $\boldsymbol{\beta}$ results to be weakly influenced by the values of $\boldsymbol{e}_{PZT}$, since at $\eta = 2$ $\tilde{\beta}_{11}$ varies of 0.25% and $\tilde{\beta}_{22}$ of 0.3% with respect to their values at $\eta = 0$. In figure 5 the eigenvalues of the dimensionless overall $\tilde{\mathbb{C}}$ tensor are represented with respect to the piezo-electric multiplication factor $\eta$. Specifically, figure 5 shows that the lowest eigenvalue of $\tilde{\mathbb{C}}$ decreases with increasing $\eta$, while the other two eigenvalues increase. The positive definiteness of the overall tensor $\mathbb{C}$ is confirmed by the fact that all the eigenvalues are greater than zero and such a property of $\mathbb{C}$ has been proved in detail in Appendix B.

In the model problem herein considered, the two-phase periodic medium, whose geometry is depicted figure 2(a), is subjected to $\mathcal{L}$-periodic harmonic body forces $\mathbf{b}(\mathbf{x})$, free charge densities $\rho_e(\mathbf{x})$, and heat sources $r(\mathbf{x})$ depending only on $x_1$ and expressed as

$$b_j(x_1) = B_j\, e^{(i\, 2\, \pi\, n_b\, x_1/L)}, \quad \rho_e(x_1) = P_e\, e^{(i\, 2\, \pi\, n_{\rho_e}\, x_1/L)}, \quad r(x_1) = R\, e^{(i\, 2\, \pi\, n_r\, x_1/L)}, \qquad (70)$$

with $j = 1, 2$, $n_b = 1, 2$, $n_{\rho_e} = 1, 2$, and $n_r = 1, 2$. Because of the periodicity of the heterogeneous material and volume forces, only a portion of the heterogeneous model has to be considered and solved. In particular, given the invariance of volume forces (70) with respect to $x_2$, the problem has been modeled using 11 cells along the $\mathbf{e}_1$ direction and 1 cell along the $\mathbf{e}_2$ direction.

In force of expressions (70), the solution in terms of macro displacement, electric potential, and relative temperature will depend only on the $x_1$ variable and, consequently, the homogenized field equations (67a)-(67c) take the form

$$\begin{aligned}
&C_{1i1i}\, U_{i,11}(x_1) + e_{i11}\, \Phi_{,11}(x_1) - \alpha_{i1}\, \Theta_{,1}(x_1) = -b_i(x_1),\\
&e_{111}\, U_{1,11}(x_1) - \beta_{11}\, \Phi_{,11} + \gamma_1\, \Theta_{,1} = \rho_e(x_1),\\
&K_{11}\, \Theta_{,1}(x_1) = -r(x_1),
\end{aligned} \qquad (71)$$

where the index $i$ is not summed. Equations (71) allow to derive the expressions of the analytical solution for the macro fields as

$$\begin{aligned}
U_1(x_1) =&\, B_1\, \frac{\beta_{11}}{C_{1111}\beta_{11} + e_{111}^2}\, \left(\frac{L}{2\pi}\right)^2 \frac{1}{n_b^2}\, e^{(i\, 2\, \pi\, n_b\, x_1/L)} +\\
&- P_e\, \frac{e_{111}}{e_{111}^2 + \beta_{11} C_{1111}}\, \left(\frac{L}{2\pi}\right)^2 \frac{1}{n_{\rho_e}^2}\, e^{(i\, 2\, \pi\, n_{\rho_e}\, x_1/L)} +\\
&- i\, R\, \frac{\alpha_{11}\beta_{11} - \gamma_1 e_{111}}{K_{11}(e_{111}^2 + C_{1111}\beta_{11})}\, \left(\frac{L}{2\pi}\right)^3 \frac{1}{n_r^3}\, e^{(i\, 2\, \pi\, n_r\, x_1/L)},
\end{aligned} \qquad (72\text{a})$$

$$\begin{aligned}
U_2(x_1) =&\, B_2\, \frac{1}{C_{1212}}\, \left(\frac{L}{2\pi}\right)^2 \frac{1}{n_b^2}\, e^{(i\, 2\, \pi\, n_b\, x_1/L)} +\\
&- i\, R\, \frac{\alpha_{21}}{C_{1212} K_{11}}\, \left(\frac{L}{2\pi}\right)^3 \frac{1}{n_r^3}\, e^{(i\, 2\, \pi\, n_r\, x_1/L)},
\end{aligned} \qquad (72\text{b})$$

$$\Phi(x_1) = B_1 \frac{e_{111}}{C_{1111}\beta_{11} + e_{111}^2} \left(\frac{L}{2\pi}\right)^2 \frac{1}{n_b^2} e^{(i\, 2\pi n_b x_1/L)} +$$
$$+ P_e \frac{C_{1111}}{e_{111}^2 + \beta_{11} C_{1111}} \left(\frac{L}{2\pi}\right)^2 \frac{1}{n_{\rho_e}^2} e^{(i\, 2\pi n_{\rho_e} x_1/L)} +$$
$$+ i\, R\, \frac{1}{e_{111} K_{11}} \left[ \frac{C_{1111}(\alpha_{11}\beta_{11} - \gamma_1 e_{111})}{e_{111}^2 + C_{1111}\beta_{11}} - \alpha_{11} \right] \left(\frac{L}{2\pi}\right)^3 \frac{1}{n_r^3} e^{(i\, 2\pi n_r x_1/L)}, \qquad (72c)$$
$$\Theta(x_1) = R\, \frac{1}{K_{11}} \left(\frac{L}{2\pi}\right)^2 \frac{1}{n_r^2} e^{(i\, 2\pi n_r x_1/L)}. \qquad (72d)$$

If only the imaginary part of the macroscopic fields (72a)-(72d) is considered, solutions (72a)-(72d) read

$$U_1(x_1) = B_1 \frac{\beta_{11}}{C_{1111}\beta_{11} + e_{111}^2} \left(\frac{L}{2\pi}\right)^2 \frac{1}{n_b^2} \sin(2\pi n_b x_1/L) +$$
$$- P_e \frac{e_{111}}{e_{111}^2 + \beta_{11} C_{1111}} \left(\frac{L}{2\pi}\right)^2 \frac{1}{n_{\rho_e}^2} \sin(2\pi n_{\rho_e} x_1/L) +$$
$$- R\, \frac{\alpha_{11}\beta_{11} - \gamma_1 e_{111}}{K_{11}(e_{111}^2 + C_{1111}\beta_{11})} \left(\frac{L}{2\pi}\right)^3 \frac{1}{n_r^3} \cos(2\pi n_r x_1/L), \qquad (73a)$$
$$U_2(x_1) = B_2 \frac{1}{C_{1212}} \left(\frac{L}{2\pi}\right)^2 \frac{1}{n_b^2} \sin(2\pi n_b x_1/L) +$$
$$- R\, \frac{\alpha_{21}}{C_{1212} K_{11}} \left(\frac{L}{2\pi}\right)^3 \frac{1}{n_r^3} \cos(2\pi n_r x_1/L), \qquad (73b)$$
$$\Phi(x_1) = B_1 \frac{e_{111}}{C_{1111}\beta_{11} + e_{111}^2} \left(\frac{L}{2\pi}\right)^2 \frac{1}{n_b^2} \sin(2\pi n_b x_1/L) +$$
$$+ P_e \frac{C_{1111}}{e_{111}^2 + \beta_{11} C_{1111}} \left(\frac{L}{2\pi}\right)^2 \frac{1}{n_{\rho_e}^2} \sin(i\, 2\pi n_{\rho_e} x_1/L) +$$
$$+ R\, \frac{1}{e_{111} K_{11}} \left[ \frac{C_{1111}(\alpha_{11}\beta_{11} - \gamma_1 e_{111})}{e_{111}^2 + C_{1111}\beta_{11}} - \alpha_{11} \right] \left(\frac{L}{2\pi}\right)^3 \frac{1}{n_r^3} \cos(2\pi n_r x_1/L), \qquad (73c)$$
$$\Theta(x_1) = R\, \frac{1}{K_{11}} \left(\frac{L}{2\pi}\right)^2 \frac{1}{n_r^2} \sin(i\, 2\pi n_r x_1/L). \qquad (73d)$$

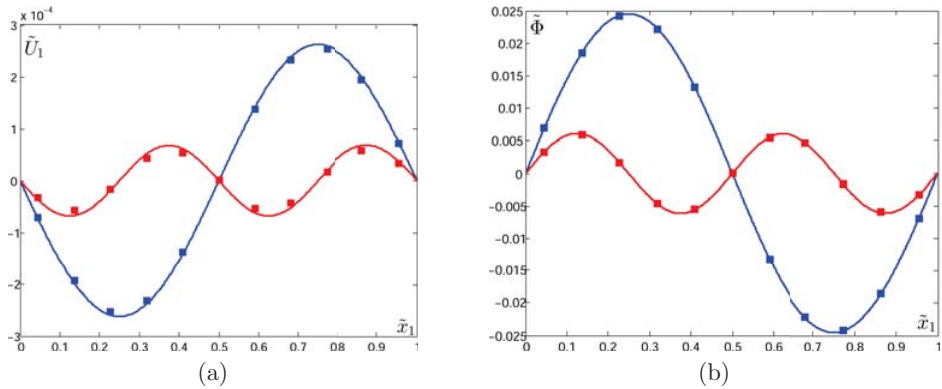

Figure 6: *Dimensionless macro fields $\tilde{U}_1$ and $\tilde{\Phi}$ induced by $\rho_e(x_1)$ vs $\tilde{x}_1 = x_1/L$. The solution of the first-order homogenized model (continuous lines) is compared with the finite element solution of the heterogeneous model via the up-scaling relations (squares). Blue lines and squares correspond to the case $n_{\rho_e} = 1$, red ones to $n_{\rho_e} = 2$.*

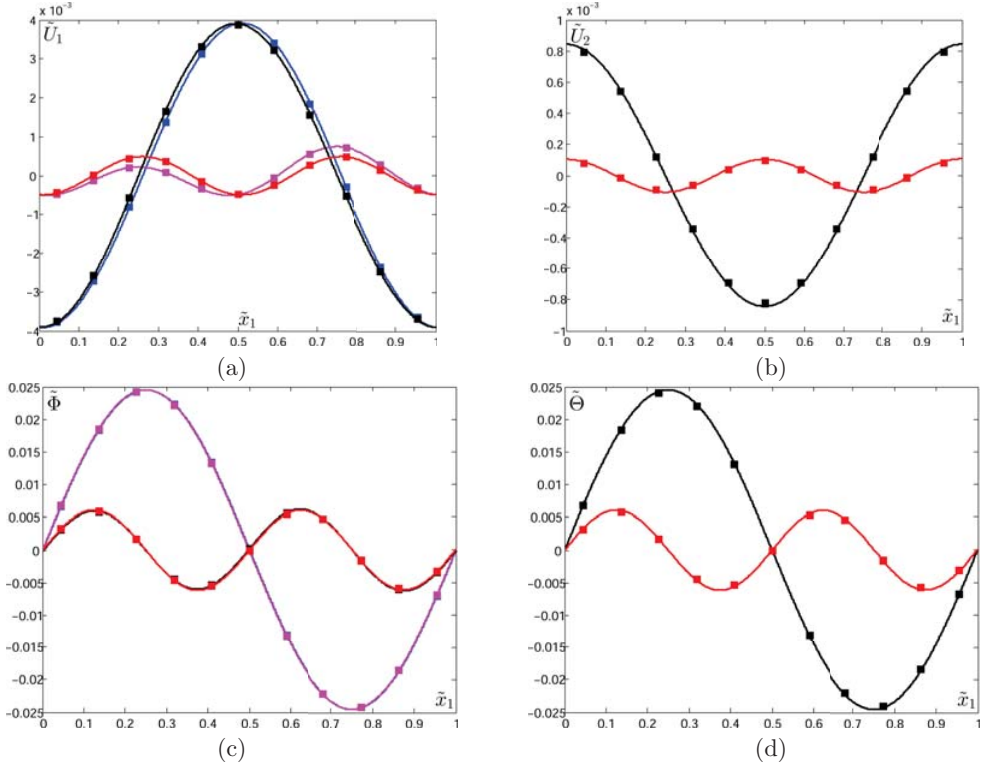

Figure 7: *Dimensionless macro fields $\tilde{U}_1$, $\tilde{U}_2$, $\tilde{\Phi}$ and $\tilde{\Theta}$ induced by $\rho_e(x_1)$ and $r(x_1)$. The solution of the first-order homogenized model (continuous lines) is compared with the finite element solution of the heterogeneous model via the up-scaling relations (squares). Blue lines and squares correspond to $n_{\rho_e} = 1$ and $n_r = 1$, magenta ones to $n_{\rho_e} = 1$ and $n_r = 2$, black ones to $n_{\rho_e} = 2$ and $n_r = 1$, and red ones to $n_{\rho_e} = 2$ and $n_r = 2$.*

The obtained analytical solutions in terms of the macroscopic fields (73a)-(73d) for the homogeneous model with overall constitutive tensors as described in (68), are then compared with those obtained from a finite element analysis of the fully heterogeneous model where periodic boundary conditions have been imposed on the displacement, electric potential, and relative temperature fields. A detailed formulation of the finite element framework that has been developed to solve the cell problems and analyze the heterogeneous model is reported in Appendix C. The macro displacement field $\mathbf{U}(x_1)$, the electric potential $\Phi(x_1)$, and the relative temperature $\Theta(x_1)$, as expressed in (73a)-(73d), are compared to the solutions of the heterogeneous problem, which are obtained from the corresponding microscopic solutions trough the up-scaling relations (52).

One defines the following dimensionless macro-fields

$$\tilde{U}_1(x_1) = \frac{U_1(x_1)}{L}, \quad \tilde{U}_2(x_1) = \frac{U_2(x_1)}{L}, \quad \tilde{\Phi}(x_1) = \frac{\Phi(x_1)\sqrt{\beta_{11}}}{\sqrt{C_{1111}}\,L}, \quad \tilde{\Theta}(x_1) = \frac{\Theta(x_1)\,\alpha_{11}}{C_{1111}}, \tag{74}$$

together with the dimensionless amplitudes of volume forces (70)

$$\tilde{B}_1 = \frac{B_1\,L}{C_{1111}}, \quad \tilde{B}_2 = \frac{B_2\,L}{C_{1111}}, \quad \tilde{P}_e = \frac{P_e\,L}{\sqrt{C_{1111}\,\beta_{11}}}, \quad \tilde{R} = \frac{R\,L^2\,\alpha_{11}}{C_{1111}\,K_{11}}. \tag{75}$$

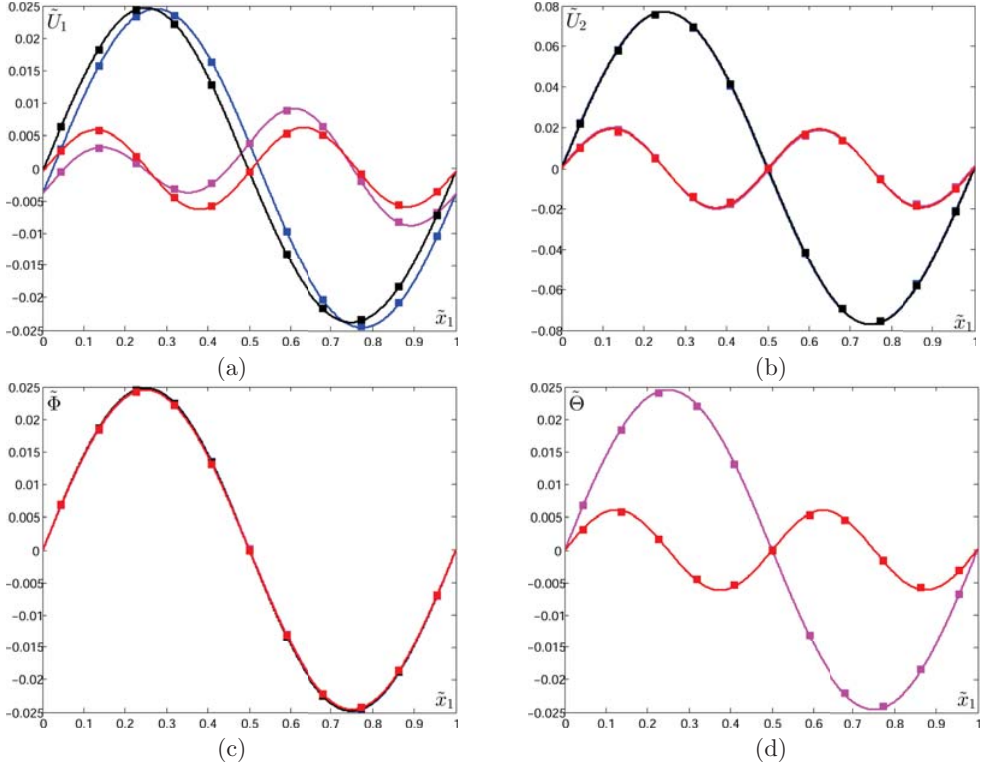

Figure 8: *Macro fields $\tilde{U}_1$, $\tilde{U}_2$, $\tilde{\Phi}$ and $\tilde{\Theta}$ induced by $\rho_e(x_1)$, $r(x_1)$, $b_1(x_1)$ and $b_2(x_1)$. The solution of the first-order homogenized model (continuous lines) is compared with the finite element solution of the heterogeneous model via the up-scaling relations (squares). Results are reported for $n_{\rho_e} = 1$, $n_r = 1$, and $n_b = 1$ (blue curves), $n_{\rho_e} = 1$, $n_r = 1$, and $n_b = 2$ (magenta curves), $n_{\rho_e} = 1$, $n_r = 2$, and $n_b = 1$ (black curves), $n_{\rho_e} = 1$, $n_r = 2$, and $n_b = 2$ (red curves).*

In figure 6, the analytical solutions (73a)-(73d) are compared with the numerical results provided by the heterogeneous model when only free charge densities are applied ($\mathbf{b}(\mathbf{x}) = \mathbf{0}$ and $r(\mathbf{x}) = 0$) with wave number $n_{\rho_e} = 1, 2$ and dimensionless amplitude $\tilde{P}_e = 1$. In this case, the displacement in the $\mathbf{e}_2$ direction and the relative temperature, namely $U_2(x_1)$ and $\Theta(x_1)$ are vanishing, while the dimensionless component of the displacement in the $\mathbf{e}_1$ direction, $\tilde{U}_1(x_1)$, and the dimensionless electric potential $\tilde{\Phi}(x_1)$ show the trend in figure 6 with respect to the dimensionless length $\tilde{x}_1 = \frac{x_1}{L}$.

If heat sources $r(x_1)$ act such that $\tilde{R} = 1$ simultaneously with free charge densities $\rho_e(x_1)$ with $\tilde{P}_e = 1$, the resulting dimensionless components of the displacement field $\tilde{U}_1(x_1)$, $\tilde{U}_2(x_1)$, the dimensionless electric potential $\tilde{\Phi}(x_1)$, and the dimensionless relative temperature $\tilde{\Theta}(x_1)$ are shown in figure 7 vs $\tilde{x}_1$ for wave numbers $n_{\rho_e} = 1, 2$ and $n_r = 1, 2$. The obtained solutions confirm that $\tilde{U}_2(x_1)$ and $\tilde{\theta}(x_1)$ are not influenced by the presence of free charge densities, being the results for $n_{\rho_e} = 1$ coincident with the ones for $n_{\rho_e} = 2$ for these two macro-fields.

Finally, the dimensionless macro-fields in the case of free charge densities $\rho_e(x_1)$, heat sources $r(x_1)$, and body forces $b_1(x_1)$ and $b_2(x_1)$ acting simultaneously with dimensionless amplitudes $\tilde{P}_e = 1$, $\tilde{R} = 1$, $\tilde{B}_1 = 1$, and $\tilde{B}_2 = 1$ and wave numbers $n_{\rho_e} = 1, 2$, $n_r = 1, 2$, and $n_b = 1, 2$ are shown in figures 8 and 9. In all the cases investigated, a good agreement has been obtained between the solution of the first-order homogenized model and the numerical solution of the heterogeneous one, confirming the accuracy of the proposed asymptotic homogenization approach for this class of periodic thermo-piezoelectric materials.

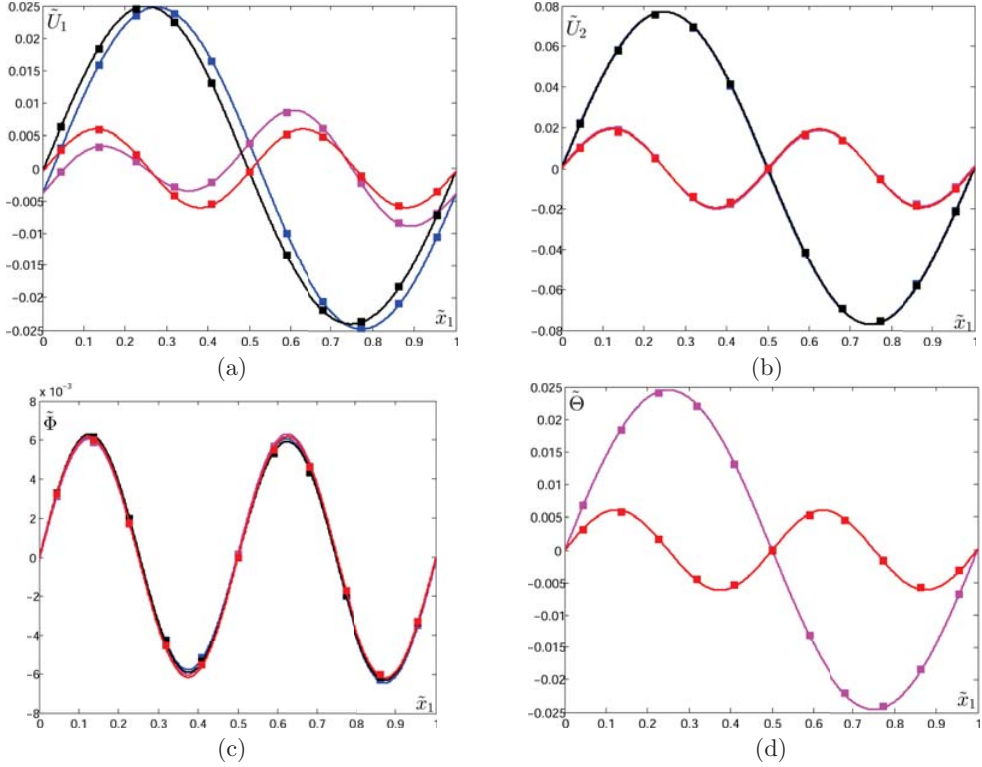

Figure 9: *Macro fields $\tilde{U}_1$, $\tilde{U}_2$, $\tilde{\Phi}$ and $\tilde{\Theta}$ induced by $\rho_e(x_1)$, $r(x_1)$, $b_1(x_1)$ and $b_2(x_1)$. The solution of the first-order homogenized model (continuous lines) is compared with the finite element solution of the heterogeneous model via the up-scaling relations (squares). Results are reported for $n_{\rho_e}=1$, $n_r=1$, and $n_b=1$ (blue curves), $n_{\rho_e}=2$, $n_r=1$, and $n_b=2$ (magenta curves), $n_{\rho_e}=2$, $n_r=2$, and $n_b=1$ (black curves), $n_{\rho_e}=2$, $n_r=2$, and $n_b=2$ (red curves).*

# 7 Conclusions

The present work has been devoted to formulate an asymptotic homogenization model for thermo-piezoelectric materials with periodic microstructure. Down-scaling relations have been derived, which allow to express the micro-displacement, the micro-electric potential, and the micro-relative temperature in terms of the macroscopic fields and their gradients trough perturbation functions. Such functions are obtained from the solution of recursive non homogeneous differential cell problems defined over the unit cell $\mathcal{Q}$.

Perturbation functions are $\mathcal{Q}$-periodic and have vanishing mean values over the unit cell. They reflect the effect of the microstructure on the microscopic fields and on their coupling. The mechanical and the electric problems remain coupled in the asymptotically expanded microscale field equations, resulting in a strong coupling between the micro-displacement field and the micro-electric potential. Substitution of the down-scaling relations into the microscopic field equations of the considered thermo-piezoelectric material led to the average field equations of infinite order, whose formal solution can be obtained by inserting an asymptotic expansion of the macro-fields in powers of the microstructural size. Truncation of the asymptotic expansions of the macro fields to the zeroth order led the field equations of the homogeneous first-order (Cauchy) thermo-piezoelectric medium equivalent to the heterogeneous one in terms of the overall constitutive tensors of the equivalent first-order medium whose exact expressions have been provided.

The accuracy of the first-order homogenization model has been assessed in reference to a model problem in which a two-phase medium with periodic microstructure has been considered. The microstructure of the periodic cell reflected the one of pyroelectric cells employed as energy harvesters, which take advantage of temperature fluctuations to generate electrical outputs with the purpose of powering devices and systems.

The periodic cell was made of a ceramic inclusion on a substrate with negligible piezoelectric and pyroelectric properties. The two-phase composite material was subject to periodic volume forces, free electric charge densities, and heat sources, whose periodicity is much greater than the microstructural size.

The analytical solution for the macro-displacement, electric potential and relative temperature fields obtained from the homogenized field equations of the first-order equivalent continuum subjected to harmonic volume forces has been compared with the numerical solution obtained from a finite element analysis of the entire heterogeneous model. To this purpose, a thermo-piezoelectric finite element has been implemented in the finite element software FEAP and the derived finite element solution in terms of micro fields has been correlated to the macroscopic displacements, electric potential, and relative temperature trough the up-scaling relations.

The very good matching between the homogenized first-order solution and the numerical one of the heterogeneous model in all the load cases examined confirmed the validity and the accuracy of the proposed multi-field homogenization technique for periodic thermo-piezoelectric materials.

Applications of piezoelectric and pyroelectric devices are numerous, especially in the field of thermal detectors, passive infrared sensors, energy harvesters (Graf et al., 2007; Wilson et al., 2007) as well as biomedical devices (De Rossi and Dario, 1983). An accurate characterization of the macroscale constitutive relations can reveal of great importance for the improvement of the efficiency of such materials and the design of new ones. The asymptotic homogenization technique presented in this study constitutes a rigorous tool for the study of thermo-piezoelectric materials whose microstructure has no requirements other than periodicity and can be of any geometry.

The macroscopic thermo-piezoelectric behaviour of a periodic heterogeneous medium can therefore be accurately described by the presented homogenization technique by means of the derivation of the overall constitutive constants of the first-order (Cauchy) equivalent continuum. As such, first-order homogenization techniques show deficiencies in reproducing the behaviour of thermo-piezoelectric materials in the presence of nonlocal phenomena or, equivalently, high gradients of stresses, deformations, electric potential, relative temperature, and volume forces, to cite a few. In fact, nonlocal phenomena related to the microstructural length scale and size effects cannot in general be properly described by homogenization techniques of first-order. In these cases, higher accuracy could be achieved trough the solution of higher order cell problems. Alternatively, nonlocal higher order homogenization techniques, which involve characteristic length scale associated to the microstructural effects, can be conveniently deployed in order to derive constitutive relations of equivalent higher order continua. These topics are left for further research.

## Acknowledgments

This research ha received support from the European Research Council to the ERC Starting Grant "Multifield and multi-scale Computational Approach to Design and Durability of PhotoVoltaic Modules" - CA2PVM, under the European Union's Seventh Framework Programme (FP/2007-2013) / ERC Grant Agreement n. 306622.

Forest, S., Sab, K., 1998. Cosserat overall modeling of heterogeneous materials. Mechanics Research Communications 25(4), 449–454.

Forest, S., Trinh, D., 2011. Generalized continua and nonhomogeneous boundary conditions in homogenisation methods. ZAMMJournal of Applied Mathematics and Mechanics/Zeitschrift fr Angewandte Mathematik und Mechanik 91(2), 90–109.

Gambin, B., Kröner, E., 1989. Higher order terms in the homogenized stressstrain relation of periodic elastic media. physica status solidi (b). International Journal of Engineering Science 151(2), 513–519.

Graf, A., Arndt, M., Sauer, M., Gerlach, G., 2007. Review of micromachined thermopiles for infrared detection. Measurement Science and Technology 18(7), R59.

Guo, Q., Cao, G., Shen, I., 2003. Measurements of piezoelectric coefficient d33 of lead zirconate titanate thin films using a mini force hammer. Journal of Vibration and Acoustics 135(1), 011003.

Hsiao, C., Jia-Wai, J., 2015. Pyroelectric harvesters for generating cyclic energy. Energies 8.5, 3489–3502.

Hsiao, C., Jia-Wai, J., An-Shen, S., 2015. Study on pyroelectric harvesters integrating solar radiation with wind power. Energies 8.7, 7465–7477.

Kanouté, P., Boso, D., Chaboche, J., Schrefler, B., 2009. Multiscale methods for composites: a review. Archives of Computational Methods in Engineering 16(1), 31–75.

Kimata, M., 2013. Trends in small-format infrared array sensors. IEEE SENSORS.

Kommepalli, H., Mateti, K., Rahn, C., Tadigadapa, S., 2010. Displacement and blocking force performance of piezoelectric t-beam actuators. In ASME 2010 International Design Engineering Technical Conferences and Computers and Information in Engineering Conference American Society of Mechanical Engineers, 841–850.

Kouznetsova, V., Geers, M., Brekelmans, W., 2002. Advanced constitutive modeling of heterogeneous materials with a gradient-enhanced computational homogenization scheme. International Journal for Numerical Methods in Engineering 54, 1235–1260.

Kouznetsova, V., Geers, M., Brekelmans, W., 2004. Multi-scale second-order computational homogenization of multi-phase materials: a nested finite element solution strategy. Computer Methods in Applied Mechanics and Engineering 193(48), 5525–5550.

Malmonge, L., Malmonge, J., Sakamoto, W., 2003. Study of pyroelectric activity of pzt/pvdf-hfp composite. Materials Research 6.4, 469–473.

Mehrabadi, M., Cowin, S., 1990. Eigentensors of linear anisotropic elastic materials. The Quarterly Journal of Mechanics and Applied Mathematics 43(1), 15–41.

Miehe, C., Schroder, J., Schotte, J., 1999. Computational homogenization analysis in finite plasticity simulation of texture development in polycrystalline materials. Computer methods in applied mechanics and engineering 171, 387–418.

Mindlin, R., 1974. Equations of high frequency vibrations of thermopiezoelectric crystal plates. International Journal of Solids and Structures 10(6), 625–637.

Moulson, A., Herbert, J., 2003. Electroceramics: materials, properties, applications. John Wiley & Sons.

Nowacki, W., 1986. Thermoelasticity. Second ed. Pergamon Press, Oxford.

Peerlings, R., Fleck, N., 2004. Computational evaluation of strain gradient elasticity constants. International Journal for Multiscale Computational Engineering 2(4).

Salvadori, A., Bosco, E., Grazioli, D., 2014. A computational homogenization approach for Li-ion battery cells. Part 1 - Formulation. Journal of the Mechanics and Physics of Solids 65, 114–137.
28

## A  Solution of higher order recursive differential problems

In this appendix, the general form of higher order differential problems is reported for heat diffusion and piezoelectric problems at the order $\varepsilon^m$ with $m \in \mathbb{Z}$ and $m \geq 1$. Such differential problems have been derived substituting the asymptotic expansions of the micro fields (8a)-(8c) into the local balance equations (6a)-(6c).

**Heat diffusion problem**
For heat diffusion problem one has

$$\left[K_{ij}^m\left(\frac{\partial \theta^{(m+1)}}{\partial x_j} + \theta_{,j}^{(m+2)}\right)\right]_{,i} + \frac{\partial}{\partial x_i}\left[K_{ij}^m\left(\frac{\partial \theta^{(m)}}{\partial x_j} + \theta_{,j}^{(m+1)}\right)\right] = h^{(m+2)}(\mathbf{x}), \tag{76}$$

with interface conditions

$$\left[\left[\theta^{(m+2)}\right]\right]\bigg|_{\boldsymbol{\xi}\in\Sigma_1} = 0, \quad \left[\left[K_{ij}^m\left(\frac{\partial \theta^{(m+1)}}{\partial x_j} + \theta_{,j}^{(m+2)}\right)n_i\right]\right]\bigg|_{\boldsymbol{\xi}\in\Sigma_1} = 0. \tag{77}$$

Solvability condition in the class of $\mathcal{Q}$-periodic functions (Bakhvalov and Panasenko, 1984) implies that

$$h^{m+2}(\mathbf{x}) = \left\langle K_{q_{m+1}q_{m+2}}^m M_{q_1...q_m}^{(m)} + K_{q_{m+2}j}^m M_{q_1...q_{m+1},j}^{(m+1)} \right\rangle \frac{\partial^{m+2}\Theta(\mathbf{x})}{\partial x_{q_1}...\partial x_{q_{m+2}}},$$

and the solution $\theta^{(m+2)}$ takes the form

$$\theta^{(m+2)}(\mathbf{x};\boldsymbol{\xi}) = M_{q_1...q_{m+2}}^{(m+2)}(\boldsymbol{\xi}) \frac{\partial^{m+2}\Theta(\mathbf{x})}{\partial x_{q_1}...\partial x_{q_{m+2}}}. \tag{78}$$

**Piezoelectric problem**

From equations (9), and (10), the piezoelectric differential problem at $\varepsilon^m$ with $m \in \mathbb{Z}$ and $m \geq 1$ is expressed as

$$\left[ C_{ijkl}^m \left( \frac{\partial u_k^{(m+1)}}{\partial x_l} + u_{k,l}^{(m+2)} \right) \right]_{,j} + \frac{\partial}{\partial x_j} \left[ C_{ijkl}^m \left( \frac{\partial u_k^{(m)}}{\partial x_l} + u_{k,l}^{(m+1)} \right) \right] + \left[ e_{ijk}^m \left( \frac{\partial \phi^{(m+1)}}{\partial x_k} + \phi_{,k}^{(m+2)} \right) \right]_{,j}$$

$$+ \frac{\partial}{\partial x_j} \left[ e_{ijk}^m \left( \frac{\partial \phi^{(m)}}{\partial x_k} + \phi_{,k}^{(m+1)} \right) \right] - \left( \alpha_{ij}^m \theta^{(m+1)} \right)_{,j} - \frac{\partial}{\partial x_j} \left( \alpha_{ij}^m \theta^{(m)} \right) = f_i^{(m+2)}(\mathbf{x}),$$

$$\left[ e_{kli}^m \left( \frac{\partial u_k^{(m+1)}}{\partial x_l} + u_{k,l}^{(m+2)} \right) \right]_{,i} + \frac{\partial}{\partial x_i} \left[ e_{kli}^m \left( \frac{\partial u_k^{(m)}}{\partial x_l} + u_{k,l}^{(m+1)} \right) \right] - \left[ \beta_{il}^m \left( \frac{\partial \phi^{(m+1)}}{\partial x_l} + \phi_{,l}^{(m+2)} \right) \right]_{,i}$$

$$- \frac{\partial}{\partial x_i} \left[ \beta_{il}^m \left( \frac{\partial \phi^{(m)}}{\partial x_l} + \phi_{,l}^{(m+1)} \right) \right] + \left( \gamma_i^m \theta^{(m+1)} \right)_{,i} + \frac{\partial}{\partial x_i} \left( \gamma_i^m \theta^{(m)} \right) = g^{(m+2)}(\mathbf{x}),$$
(79)

whose interface conditions have the form

$$\left[\!\left[ u_k^{(m+2)} \right]\!\right]\Big|_{\boldsymbol{\xi} \in \Sigma_1} = 0, \quad \left[\!\left[ \phi^{(m+2)} \right]\!\right]\Big|_{\boldsymbol{\xi} \in \Sigma_1} = 0,$$

$$\left[\!\left[ \left( C_{ijkl}^m \left( \frac{\partial u_k^{(m+1)}}{\partial x_l} + u_{k,l}^{(m+2)} \right) + e_{ijk}^m \left( \frac{\partial \phi^{(m+1)}}{\partial x_k} + \phi_{,k}^{(m+2)} \right) - \alpha_{ij}^m \theta^{(m+1)} \right) n_j \right]\!\right]\Big|_{\boldsymbol{\xi} \in \Sigma_1} = 0,$$

$$\left[\!\left[ \left( e_{kli}^m \left( \frac{\partial u_k^{(m+1)}}{\partial x_l} + u_{k,l}^{(m+2)} \right) - \beta_{il}^m \left( \frac{\partial \phi^{(m+1)}}{\partial x_l} + \phi_{,l}^{(m+2)} \right) + \gamma_i^m \theta^{(m+1)} \right) n_i \right]\!\right]\Big|_{\boldsymbol{\xi} \in \Sigma_1} = 0. \quad (80)$$

Once again, solvability condition imposes that

$$f_i^{(m+2)}(\mathbf{x}) = \left\langle C_{iq_{m+1}kq_{m+2}}^m N_{kpq_1\ldots q_m}^{(m)} + C_{iq_{m+2}kl}^m N_{kpq_1\ldots q_{m+1},l}^{(m+1)} + e_{iq_{m+2}k}^m \tilde{W}_{kpq_1\ldots q_{m+1},k}^{(m+1)} \right\rangle \frac{\partial^{m+2} U_p(\mathbf{x})}{\partial x_{q_m}\ldots \partial x_{q_{m+2}}} +$$

$$+ \left\langle C_{iq_{m+2}kl}^m \tilde{N}_{kpq_1\ldots q_{m+1},l}^{(m)} + e_{iq_{m+1}q_{m+2}}^m W_{q_1\ldots q_m}^{(m)} + e_{iq_{m+2}k}^m W_{q_1\ldots q_{m+1},k}^{(m+1)} \right\rangle \frac{\partial^{m+2} \Phi(\mathbf{x})}{\partial x_{q_m}\ldots \partial x_{q_{m+2}}} +$$

$$+ \left\langle C_{iq_{m+1}kl}^m \hat{N}_{kq_1\ldots q_m,l}^{(m+1)} + e_{iq_{m+1}k}^m \hat{W}_{q_1\ldots q_m,k}^{(m+1)} - \alpha_{iq_{m+1}}^m M_{q_1\ldots q_m}^{(m)} \right\rangle \frac{\partial^{m+1} \Theta(\mathbf{x})}{\partial x_{q_m}\ldots \partial x_{q_{m+1}}},$$

$$g^{(m+2)}(\mathbf{x}) = \left\langle e_{klq_{m+2}}^m N_{kpq_1\ldots q_{m+1},l}^{(m+1)} + e_{iq_{m+2}q_{m+1}}^m N_{kpq_1\ldots q_m}^{(m)} - \beta_{q_{m+2}l}^m \tilde{W}_{kpq_1\ldots q_{m+1},l}^{(m+1)} \right\rangle \frac{\partial^{m+2} U_p(\mathbf{x})}{\partial x_{q_m}\ldots \partial x_{q_{m+2}}} +$$

$$+ \left\langle e_{klq_{m+2}}^m \tilde{N}_{kq_1\ldots q_{m+1},l}^{(m+1)} - \beta_{q_1q_{m+2}}^m W_{q_1\ldots q_m}^{(m)} - \beta_{q_{m+2}l}^m W_{q_1\ldots q_{m+1},l}^{(m+1)} \right\rangle \frac{\partial^{m+2} \Phi(\mathbf{x})}{\partial x_{q_m}\ldots \partial x_{q_{m+2}}} +$$

$$+ \left\langle e_{klq_{m+1}}^m \hat{N}_{kq_1\ldots q_m,l}^{(m+1)} - \beta_{q_{m+1}l}^m \hat{W}_{q_1\ldots q_m,l}^{(m+1)} + \gamma_{q_{m+1}}^m M_{q_1\ldots q_m}^{(m)} \right\rangle \frac{\partial^{m+1} \Theta(\mathbf{x})}{\partial x_{q_m}\ldots \partial x_{q_{m+1}}},$$

and the solutions are expressed as

$$u_k^{(m+2)}(\mathbf{x}; \boldsymbol{\xi}) = N_{kpq_1\ldots q_{m+2}}^{(m+2)}(\boldsymbol{\xi}) \frac{\partial^{m+2} U_p(\mathbf{x})}{\partial x_{q_1}\ldots \partial x_{q_{m+2}}} + \tilde{N}_{kq_1\ldots q_{m+2}}^{(m+2)}(\boldsymbol{\xi}) \frac{\partial^{m+2} \Phi(\mathbf{x})}{\partial x_{q_1}\ldots \partial x_{q_{m+2}}} +$$

$$+ \hat{N}_{kq_1\ldots q_{m+1}}^{(m+2)}(\boldsymbol{\xi}) \frac{\partial^{m+1} \Theta(\mathbf{x})}{\partial x_{q_1}\ldots \partial x_{q_{m+1}}},$$

$$\phi^{(m+2)}(\mathbf{x}; \boldsymbol{\xi}) = W_{q_1\ldots q_{m+2}}^{(m+2)}(\boldsymbol{\xi}) \frac{\partial^{m+2} \Phi(\mathbf{x})}{\partial x_{q_1}\ldots \partial x_{q_{m+2}}} + \tilde{W}_{pq_1\ldots q_{m+2}}^{(m+2)}(\boldsymbol{\xi}) \frac{\partial U_p(\mathbf{x})}{\partial x_{q_1}\ldots \partial x_{q_{m+2}}} +$$

$$+ \hat{W}_{q_1\ldots q_{m+1}}^{(m+2)}(\boldsymbol{\xi}) \frac{\partial^{m+1} \Theta(\mathbf{x})}{\partial x_{q_1}\ldots \partial x_{q_{m+1}}}.$$
(81)

## A.1 Higher order cell problems

From the solution of differential problems (76) and (79) and of those at the previous orders of $\varepsilon$ as detailed in Section 3, it is possible to derive the form of cell problems at the order $\varepsilon^m$ (with $m \in \mathbb{Z}$ and $m \geq 1$) in terms of the perturbation functions from manipulation of differential problems (76) and (79) and of the relative interface conditions (77) and (80).

**Heat diffusion problem**
The heat diffusion cell problem at order $\varepsilon^m$ and the relative interface conditions are expressed has

$$\left(K_{ij}^m M_{q_1\ldots q_{m+2}}^{(m+2)}\right)_{,i} + \frac{1}{2^{m+2}} \sum_{\mathcal{P}(q)} \left[\left(K_{iq_{m+2}}^m M_{q_1\ldots q_{m+1}}^{(m+1)}\right)_{,i} + K_{q_{m+1}q_{m+2}}^m M_{q_1\ldots q_m}^{(m)} + \right.$$
$$\left. + K_{q_{m+2}j}^m M_{q_1\ldots q_{m+1},j}^{(m+1)}\right] = \frac{1}{2} \sum_{\mathcal{P}(q)} \left\langle K_{q_{m+1}q_{m+2}}^m M_{q_1\ldots q_m}^{(m)} + K_{q_{m+2}j}^m M_{q_1q_2q_{m+1},j}^{(m+1)}\right\rangle,$$
$$\left[\left[M_{q_1\ldots q_{m+2}}^{(m+2)}\right]\right]\bigg|_{\boldsymbol{\xi}\in\Sigma_1} = 0,$$
$$\left[\left[\left\{K_{ij}^m \left[M_{q_1\ldots q_{m+2},j}^{(m+2)} + \frac{1}{2^{m+2}} \sum_{\mathcal{P}(q)} \left(\delta_{iq_{m+2}} M_{q_1\ldots q_{m+1}}^{(m+1)}\right)\right]\right\} n_i\right]\right]\bigg|_{\boldsymbol{\xi}\in\Sigma_1} = 0, \quad (82)$$

where $\mathcal{P}(q)$ stands for all the possible permutations of the multi-index $q$. The solution of problem (82) allows to derive the form of the $\mathcal{Q}$-periodic perturbation functions $M_{q_1\ldots q_{m+2}}^{(m+2)}$ which satisfy the normalization condition of type (21) over the unit cell $\mathcal{Q}$.

**Piezoelectric problem**
In the following, the three sets of piezoelectric cell problems at order $m$ (with $m \in \mathbb{Z}$ and $m \geq 1$) resulting from the strong coupling that exist between the displacement and the electric potential fields, are reported with the relative interface conditions. The first cell problem results

$$\begin{cases}
\left(C_{ijkl}^m N_{kpq_1\ldots q_{m+2},l}^{(m+2)}\right)_{,j} + \left(e_{ijk}^m \tilde{W}_{pq_1\ldots q_{m+2},k}^{(m+2)}\right)_{,j} + \frac{1}{2^{m+2}} \sum_{\mathcal{P}(q)} \left[\left(C_{ijkq_{m+2}}^m N_{kpq_1\ldots q_{m+1}}^{(m+1)}\right)_{,j} + \right. \\
+ C_{iq_{m+1}kq_{m+2}}^m N_{kpq_1\ldots q_m}^{(m)} + C_{iq_{m+2}kl}^m N_{kpq_1\ldots q_{m+1},l}^{(m+1)} + \left(e_{ijq_{m+2}}^m \tilde{W}_{pq_1\ldots q_{m+1}}^{(m+1)}\right)_{,j} + \\
+ e_{iq_{m+2}k}^m \tilde{W}_{pq_1\ldots q_{m+1},k}^{(m+1)}\right] = \frac{1}{2^{m+2}} \sum_{\mathcal{P}(q)} \left\langle C_{iq_{m+1}kq_{m+2}}^m N_{kpq_1\ldots q_m}^{(m)} + C_{iq_{m+2}kl}^m N_{kpq_1\ldots q_{m+1},l}^{(m+1)} + \right. \\
+ e_{iq_{m+2}k}^m \tilde{W}_{pq_1\ldots q_{m+1},k}^{(m+1)}\right\rangle \\
\\
\left(e_{kli}^m N_{kpq_1\ldots q_{m+2},l}^{(m+2)}\right)_{,i} - \left(\beta_{il}^m \tilde{W}_{pq_1\ldots q_{m+2},l}^{(m+2)}\right)_{,i} + \frac{1}{2^{m+2}} \sum_{\mathcal{P}(q)} \left[\left(e_{kq_{m+2}i}^m N_{kpq_1\ldots q_{m+1}}^{(m+1)}\right)_{,i} + \right. \\
+ e_{klq_{m+2}}^m N_{kpq_1\ldots q_{m+1},l}^{(m+1)} + e_{kq_{m+2}q_{m+1}}^m N_{kpq_1\ldots q_m}^{(m)} - \left(\beta_{iq_{m+2}}^m \tilde{W}_{pq_1\ldots q_{m+1}}^{(m+1)}\right)_{,i} - \beta_{q_{m+2}l}^m \tilde{W}_{pq_1\ldots q_{m+1},l}^{(m+1)}\right] = \\
= \frac{1}{2^{m+2}} \sum_{\mathcal{P}(q)} \left\langle e_{klq_{m+2}}^m N_{kpq_1\ldots q_{m+1},l}^{(m+1)} + e_{kq_{m+2}q_{m+1}}^m N_{kpq_1\ldots q_m}^{(m)} - \beta_{q_{m+2}l}^m \tilde{W}_{pq_1\ldots q_{m+1},l}^{(m+1)}\right\rangle
\end{cases},$$

(83)

with interface conditions

$$\left[\left[N_{kpq_1\ldots q_{m+2}}^{(m+2)}\right]\right]\bigg|_{\boldsymbol{\xi}\in\Sigma_1} = 0,$$
$$\left[\left[\tilde{W}_{pq_1\ldots q_{m+2}}^{(m+2)}\right]\right]\bigg|_{\boldsymbol{\xi}\in\Sigma_1} = 0,$$
$$\left[\left[\left\{C_{ijkl}^m \left[N_{kpq_1\ldots q_{m+2},l}^{(m+2)} + \frac{1}{2^{m+2}} \sum_{\mathcal{P}(q)} \left(\delta_{q_{m+2}l} N_{kpq_1\ldots q_{m+1}}^{(m+1)}\right)\right] + e_{ijk}^m \left[\tilde{W}_{pq_1\ldots q_{m+2},k}^{(m+2)} + \right.\right.\right.\right.$$

$$+ \; \frac{1}{2^{m+2}} \sum_{\mathcal{P}(q)} \left( \delta_{q_{m+2}k} \tilde{W}^{(m+1)}_{pq_1\ldots q_{m+1}} \right) \Bigg] \Bigg\} n_j \Bigg] \Bigg] \Bigg|_{\boldsymbol{\xi} \in \Sigma_1} = 0,$$

$$\left[\left[\left\{ e^m_{kli} \left[ N^{(m+2)}_{kpq_1\ldots q_{m+2},l} + \frac{1}{2^{m+2}} \sum_{\mathcal{P}(q)} \left( \delta_{q_{m+2}l} N^{(m+1)}_{kpq_1\ldots q_{m+1}} \right) \right] - \beta^m_{il} \left[ \tilde{W}^{(m+2)}_{pq_1\ldots q_{m+2},l} \right. \right.\right.\right.$$

$$\left.\left.\left.\left. + \; \frac{1}{2^{m+2}} \sum_{\mathcal{P}(q)} \left( \delta_{q_{m+2}l} \tilde{W}^{(m+1)}_{pq_1\ldots q_{m+1}} \right) \right] \right\} n_i \right] \right] \Bigg|_{\boldsymbol{\xi} \in \Sigma_1} = 0.$$

It allows the determination of perturbation functions $N^{m+2}_{kpq_1\ldots q_{m+2}}$ and $\tilde{W}^{(m+2)}_{pq_1\ldots q_{m+2}}$.

The second piezoelectric cell problem has the following form

$$\begin{cases} \left( C^m_{ijkl} \tilde{N}^{(m+2)}_{kq_1\ldots q_{m+2},l} \right)_{,j} + \left( e^m_{ijk} W^{(m+2)}_{q_1\ldots q_{m+2},k} \right)_{,j} + \frac{1}{2^{m+2}} \sum_{\mathcal{P}(q)} \left[ \left( C^m_{ijkq_{m+2}} \tilde{N}^{(m+1)}_{kq_1\ldots q_{m+1}} \right)_{,j} + \right. \\ \left. + C^m_{iq_{m+2}kl} \tilde{N}^{(m+1)}_{kq_1\ldots q_{m+1},l} + \left( e^m_{ijq_{m+2}} W^{(m+1)}_{q_1\ldots q_{m+1}} \right)_{,j} + e^m_{iq_{m+1}q_{m+2}} W^{(m)}_{q_1\ldots q_m} + e^m_{iq_{m+2}k} W^{(m+1)}_{q_1\ldots q_{m+1}} \right] = \\ = \frac{1}{2^{m+2}} \sum_{\mathcal{P}(q)} \left\langle C^m_{iq_{m+2}kl} \tilde{N}^{(m+1)}_{kq_1\ldots q_{m+1},l} + e_{iq_{m+1}q_{m+2}} W^{(m)}_{q_1\ldots q_m} + e^m_{iq_{m+2}k} W^{(m+1)}_{q_1\ldots q_{m+1},k} \right\rangle \\ \left( e^m_{kli} \tilde{N}^{(m+2)}_{kq_1\ldots q_{m+2},l} \right)_{,i} - \left( \beta^m_{il} W^{(m+2)}_{q_1\ldots q_{m+2},l} \right)_{,i} + \frac{1}{2^{m+2}} \sum_{\mathcal{P}(q)} \left[ \left( e^m_{kq_{m+2}i} \tilde{N}^{(m+1)}_{kq_1\ldots q_{m+1},l} \right)_{,i} + \right. \\ \left. e^m_{klq_{m+2}} \tilde{N}^{(m+1)}_{kq_1\ldots q_{m+1},l} - \left( \beta^m_{iq_{m+2}} W^{(m+1)}_{q_1\ldots q_{m+1}} \right)_{,i} - \beta^m_{q_{m+1}q_{m+2}} W^{(m)}_{q_1\ldots q_m} - \beta^m_{q_{m+2}l} W^{(m+1)}_{q_1\ldots q_{m+1},l} \right] = \\ = \frac{1}{2^{m+2}} \sum_{\mathcal{P}(q)} \left\langle e^m_{klq_{m+2}} \tilde{N}^{(m+1)}_{kq_1\ldots q_{m+1},l} - \beta^m_{q_{m+1}q_{m+2}} W^{(m)}_{q_1\ldots q_m} - \beta^m_{q_{m+2}l} W^{(m+1)}_{q_1\ldots q_{m+1},l} \right\rangle \end{cases},$$

(84)

with interface conditions

$$\left[\left[ \tilde{N}^{(m+2)}_{kq_1\ldots q_{m+2}} \right]\right]\Big|_{\boldsymbol{\xi} \in \Sigma_1} = 0,$$

$$\left[\left[ W^{(m+2)}_{q_1\ldots q_{m+2}} \right]\right]\Big|_{\boldsymbol{\xi} \in \Sigma_1} = 0,$$

$$\left[\left[\left\{ C^m_{ijkl} \left[ \tilde{N}^{(m+2)}_{kq_1\ldots q_{m+2},l} + \frac{1}{2^{m+2}} \sum_{\mathcal{P}(q)} \left( \delta_{q_{m+2}l} \tilde{N}^{(m+1)}_{kq_1\ldots q_{m+1}} \right) \right] + e^m_{ijk} \left[ W^{(m+2)}_{q_1\ldots q_{m+2},k} \right. \right.\right.\right.$$

$$\left.\left.\left.\left. + \; \frac{1}{2^{m+2}} \sum_{\mathcal{P}(q)} \left( \delta_{q_{m+2}k} W^{(m+1)}_{q_1\ldots q_{m+1}} \right) \right] \right\} n_j \right] \right] \Bigg|_{\boldsymbol{\xi} \in \Sigma_1} = 0,$$

$$\left[\left[\left\{ e^m_{kli} \left[ \tilde{N}^{(m+2)}_{kq_1\ldots q_{m+2},l} + \frac{1}{2^{m+2}} \sum_{\mathcal{P}(q)} \left( \delta_{q_{m+2}l} \tilde{N}^{(m+1)}_{kq_1\ldots q_{m+1}} \right) \right] - \beta^m_{il} \left[ W^{(m+2)}_{q_1\ldots q_{m+2},l} + \right.\right.\right.\right.$$

$$\left.\left.\left.\left. - \; \frac{1}{2^{m+2}} \sum_{\mathcal{P}(q)} \left( \delta_{q_{m+2}l} W^{(m+1)}_{q_1\ldots q_{m+1}} \right) \right] \right\} n_i \right] \right] \Bigg|_{\boldsymbol{\xi} \in \Sigma_1} = 0,$$

which provides perturbation functions $\tilde{N}^{(m+2)}_{kq_1...q_{m+2}}$ and $W^{(m+2)}_{q_1...q_{m+2}}$.

Finally, the third piezoelectric cell problem is expressed with its interface conditions as

$$\begin{cases} \left(C^m_{ijkl}\,\hat{N}^{(m+2)}_{kq_1...q_{m+1},l}\right)_{,j} + \left(e^m_{ijk}\,\hat{W}^{(m+2)}_{q_1...q_{m+1},k}\right)_{,j} + \frac{1}{2^{m+1}}\sum_{\mathcal{P}(q)}\left[\left(C^m_{ijkq_{m+1}}\,\hat{N}^{(m+1)}_{kq_1...q_m}\right)_{,j} + \right. \\ \left. + C^m_{iq_{m+1}kl}\,\hat{N}^{(m+1)}_{kq_1...q_m,l} + \left(e^m_{ijq_{m+1}}\,\hat{W}^{(m+1)}_{q_1...q_m}\right)_{,j} + e^m_{iq_{m+1}k}\,\hat{W}^{(m+1)}_{q_1...q_m,k} + \right. \\ \left. - \left(\alpha^m_{ij}\,M^{(m+1)}_{q_1...q_{m+1}}\right)_{,j} - \alpha^m_{iq_{m+1}}\,M^{(m)}_{q_1...q_m}\right] = \\ = \frac{1}{2^{m+1}}\sum_{\mathcal{P}(q)}\left\langle C^m_{iq_{m+1}kl}\,\hat{N}^{(m+1)}_{kq_1...q_m,l} + e_{iq_{m+1}k}\,\hat{W}^{(m+1)}_{q_1...q_m,k} - \alpha^m_{iq_{m+1}}\,M^{(m)}_{q_1...q_m}\right\rangle \\[4pt] \left(e^m_{kli}\,\hat{N}^{(m+2)}_{kq_1...q_{m+1},l}\right)_{,i} - \left(\beta^m_{il}\,\hat{W}^{(m+2)}_{q_1...q_{m+1},l}\right)_{,i} + \frac{1}{2^{m+1}}\sum_{\mathcal{P}(q)}\left[\left(e^m_{kq_{m+1}i}\,\hat{N}^{(m+1)}_{kq_1...q_m}\right)_{,i} + \right. \\ \left. + e^m_{klq_{m+1}}\,\hat{N}^{(m+1)}_{kq_1...q_m,l} - \left(\beta^m_{iq_{m+1}}\,\hat{W}^{(m+1)}_{q_1...q_m}\right)_{,i} - \beta^m_{q_{m+1}l}\,\hat{W}^{(m+1)}_{q_1...q_m,l} + \left(\gamma^m_i\,M^{(m+1)}_{q_1...q_{m+1}}\right)_{,i} + \right. \\ \left. + \gamma^m_{q_{m+1}}\,M^{(m)}_{q_1...q_m}\right] = \frac{1}{2^{m+1}}\sum_{\mathcal{P}(q)}\left\langle e^m_{klq_{m+1}}\,\hat{N}^{(m+1)}_{kq_1...q_m,l} - \beta^m_{q_{m+1}l}\,\hat{W}^{(m+1)}_{q_1...q_m,l} + \gamma^m_{q_{m+1}}\,M^{(m)}_{q_1...q_m}\right\rangle \\[4pt] \left[\left[\hat{N}^{(m+2)}_{kq_1...q_{m+1}}\right]\right]\Big|_{\boldsymbol{\xi}\in\Sigma_1} = 0, \\[4pt] \left[\left[\hat{W}^{(m+2)}_{q_1...q_{m+1}}\right]\right]\Big|_{\boldsymbol{\xi}\in\Sigma_1} = 0, \\[4pt] \left[\left[\left\{C^m_{ijkl}\left[\hat{N}^{(m+2)}_{kq_1...q_{m+1},l} + \frac{1}{2^{m+1}}\sum_{\mathcal{P}(q)}\left(\delta_{q_{m+1}l}\,\hat{N}^{(m+1)}_{kq_1...q_m}\right)\right] + e^m_{ijk}\left[\hat{W}^{(m+2)}_{q_1...q_{m+1},k} + \right.\right.\right.\right. \\ \left.\left.\left. + \frac{1}{2^{m+1}}\sum_{\mathcal{P}(q)}\left(\delta_{q_{m+1}l}\,\hat{W}^{(m+1)}_{q_1...q_m}\right)\right] - \alpha^m_{ij}\,M^{(m+1)}_{q_1...q_{m+1}}\right\} n_j\right]\Big|_{\boldsymbol{\xi}\in\Sigma_1} = 0, \\[4pt] \left[\left[\left\{e^m_{kli}\left[\hat{N}^{(m+2)}_{kq_1...q_{m+1},l} + \frac{1}{2^{m+1}}\sum_{\mathcal{P}(q)}\left(\delta_{q_{m+1}l}\,\hat{N}^{(m+1)}_{kq_1...q_m}\right)\right] - \beta^m_{il}\left[\hat{W}^{(m+2)}_{q_1...q_{m+1},l} + \right.\right.\right.\right. \\ \left.\left.\left. + \frac{1}{2^{m+1}}\sum_{\mathcal{P}(q)}\left(\delta_{q_{m+1}l}\,\hat{W}^{(m+1)}_{q_1...q_m}\right)\right] + \gamma^m_i\,M^{(m+1)}_{q_1...q_{m+1}}\right\} n_i\right]\Big|_{\boldsymbol{\xi}\in\Sigma_1} = 0, \end{cases} \quad (85)$$

which gives $\hat{N}^{m+2}_{kq_1...q_{m+1}}$ and $\hat{W}^{(m+2)}_{q_1...q_{m+1}}$.

## B  Symmetry and positive definiteness of the overall thermo-piezoelectric tensors

In this appendix, the simmetries of tensors of components $n^{(2)}_{ipq_1q_2}$, $w^{(2)}_{q_1q_2}$, and $m^{(2)}_{q_1q_2}$ that appear in the average field equations of infinite order (54a)-(54c), together with the equality $\tilde{n}^{(2)}_{pq_1q_1} = \tilde{w}^{(2)}_{pq_1q_2}$ and the ellipticity of field equations (59), (63a) and (63b) are demonstrated in order to relate the coefficients $n^{(2)}_{ipq_1q_2}$, $w^{(2)}_{q_1q_2}$, $m^{(2)}_{q_1q_2}$, and $\tilde{n}^{(2)}_{pq_1q_2}$ to the overall thermo-piezoelectric constants of the the equivalent first-order continuum $C_{iq_1pq_2}, \beta_{q_1q_2}, K_{q_1q_2}$ and $e_{pq_2q_2}$.

**Tensor $\boldsymbol{n}^{(2)}$**

One considers the expression of tensor $\boldsymbol{n}^{(2)}$ whose components are derived from the known term of the first equation of cell problem (44) at order $\varepsilon^0$, namely

$$n^{(2)}_{ipq_1q_2} = \frac{1}{2}\left\langle C^m_{iq_1pq_2} + C^m_{iq_2kl}\,N^{(1)}_{kpq_1,l} + e^m_{iq_2k}\,\tilde{W}^{(1)}_{pq_1,k} + C^m_{iq_2pq_1} + C^m_{iq_1kl}\,N^{(1)}_{kpq_2,l} + e^m_{iq_1k}\,\tilde{W}^{(1)}_{pq_2,k}\right\rangle. \quad (86)$$

The weak form of the first equation of cell problem (38) at the order $\varepsilon^{-1}$

$$\left(C^m_{rjkl}\,N^{(1)}_{kpq_1,l}\right)_{,j} + \left(e^m_{rjk}\,\tilde{W}^{(1)}_{pq_1,k}\right)_{,j} + C^m_{rjpq_1,j} = 0, \quad (87)$$

can be written as
$$\left\langle \left( C^m_{rjkl} N^{(1)}_{kpq_1,l} + e^m_{rjk} \tilde{W}^{(1)}_{pq_1,k} + C^m_{rjpq_1} \right) N^{(1)}_{riq_2,j} \right\rangle = 0, \tag{88}$$

where equation (87) has been multiplied by a test function $N^{(1)}_{riq_2}$, and the divergence theorem, together with the $\mathcal{Q}$-periodicity of perturbation functions and of micro constitutive tensors, have been exploited. A permutation of indices $q_1$ and $q_2$ in equation (88) gives

$$\left\langle \left( C^m_{rjkl} N^{(1)}_{kpq_2,l} + e^m_{rjk} \tilde{W}^{(1)}_{pq_2,k} + C^m_{rjpq_2} \right) N^{(1)}_{riq_1,j} \right\rangle = 0. \tag{89}$$

By adding the vanishing terms (88) and (89) to expression (86), after some manipulations $n^{(2)}_{ipq_1q_2}$ takes the form

$$n^{(2)}_{ipq_1q_2} = \frac{1}{2} \left\langle C^m_{rjkl} \left( N^{(1)}_{riq_2,j} + \delta_{ir}\delta_{jq_2} \right) \left( N^{(1)}_{kpq_1,l} + \delta_{pk}\delta_{lq_1} \right) + e^m_{iq_2k} \tilde{W}^{(1)}_{pq_1,k} + e^m_{rjk} \tilde{W}^{(1)}_{pq_1,k} N^{(1)}_{riq_2,j} + \right.$$
$$\left. + C^m_{rjkl} \left( N^{(1)}_{riq_1,j} + \delta_{ir}\delta_{jq_1} \right) \left( N^{(1)}_{kpq_2,l} + \delta_{pk}\delta_{lq_2} \right) + e^m_{iq_1k} \tilde{W}^{(1)}_{pq_2,k} + e^m_{rjk} \tilde{W}^{(1)}_{pq_2,k} N^{(1)}_{riq_1,j} \right\rangle. \tag{90}$$

Analogously, the weak form of the second equation of cell problem (38) at the order $\varepsilon^{-1}$

$$\left( e^m_{klj} N^{(1)}_{kpq_1,l} \right)_{,j} - \left( \beta^m_{jl} \tilde{W}^{(1)}_{pq_1,l} \right)_{,j} + e^m_{pq_1j,j} = 0, \tag{91}$$

is expressed as

$$\left\langle \left( e^m_{klj} N^{(1)}_{kpq_1,l} - \beta^m_{jl} \tilde{W}^{(1)}_{pq_1,l} + e^m_{pq_1j} \right) \tilde{W}^{(1)}_{iq_1,j} \right\rangle = 0, \tag{92}$$

with test function $\tilde{W}^{(1)}_{iq_1}$. The summation of equation (92) and its counterpart, obtained by exchanging $q_1$ and $q_2$, to equation (90) leads to the final expression for $n^{(2)}_{ipq_1q_2}$, which takes the form

$$n^{(2)}_{ipq_1q_2} = \frac{1}{2} \left\langle C^m_{rjkl} \left( N^{(1)}_{riq_2,j} + \delta_{ir}\delta_{jq_2} \right) \left( N^{(1)}_{kpq_1,l} + \delta_{pk}\delta_{lq_1} \right) + \beta^m_{jl} \tilde{W}^{(1)}_{pq_1,l} \tilde{W}^{(1)}_{iq_2,j} + \right.$$
$$\left. + C^m_{rjkl} \left( N^{(1)}_{riq_1,j} + \delta_{ir}\delta_{jq_1} \right) \left( N^{(1)}_{kpq_2,l} + \delta_{pk}\delta_{lq_2} \right) + \beta^m_{jl} \tilde{W}^{(1)}_{pq_2,l} \tilde{W}^{(1)}_{iq_1,j} \right\rangle =$$
$$= \frac{1}{2} \left( C^m_{pq_1iq_2} + C^m_{pq_2iq_1} \right). \tag{93}$$

Equation (93) proves the positive definiteness of tensor $\boldsymbol{n}^{(2)}$. Symmetry properties of $\boldsymbol{n}^{(2)}$ come from the major and minor symmetries of the elastic microscopic tensor $\mathbb{C}^m$, i.e.

$$C^m_{pq_1iq_2} = C^m_{iq_2pq_1}, \qquad C^m_{pq_1iq_2} = C^m_{q_1piq_2} = C^m_{q_1pq_2i} = C^m_{pq_1q_2i}, \tag{94}$$

and from the equality $N^{(1)}_{kpq_1} = N^{(1)}_{kq_1p}$, whose validity is guaranteed by the structure of cell problem (38). In particular, because of the repetition of indices $q_1$ and $q_2$, the following relation holds

$$n^{(2)}_{ipq_1q_2} \frac{\partial^2 U_p(\mathbf{x})}{\partial x_{q_1} \partial x_{q_2}} = \frac{1}{2} \left( C_{pq_1iq_2} + C_{pq_2iq_1} \right) \frac{\partial^2 U_p(\mathbf{x})}{\partial x_{q_1} \partial x_{q_2}} =$$
$$\frac{1}{2} \left( C_{pq_1iq_2} \frac{\partial^2 U_p(\mathbf{x})}{\partial x_{q_1} \partial x_{q_2}} + C_{pq_1iq_2} \frac{\partial^2 U_p(\mathbf{x})}{\partial x_{q_2} \partial x_{q_1}} \right) = C_{pq_1iq_2} \frac{\partial^2 U_p(\mathbf{x})}{\partial x_{q_1} \partial x_{q_2}},$$

with the components of the overall elastic tensor $\mathbb{C}$ equal to

$$C_{pq_1iq_2} = \left\langle C^m_{rjkl} \left( N^{(1)}_{riq_2,j} + \delta_{ir}\delta_{jq_2} \right) \left( N^{(1)}_{kpq_1,l} + \delta_{pk}\delta_{lq_1} \right) + \beta^m_{jl} \tilde{W}^{(1)}_{pq_1,l} \tilde{W}^{(1)}_{iq_2,j} \right\rangle.$$

**Tensor $\boldsymbol{w}^{(2)}$**

For what regards tensor $\boldsymbol{w}^{(2)}$, its components are given by the known term of the second equation of cell problem (44) at order $\varepsilon^0$, and have the form

$$w^{(2)}_{q_1q_2} = \frac{1}{2} \left\langle \beta^m_{q_1q_2} + \beta^m_{q_2l} W^{(1)}_{q_1,l} - e^m_{klq_2} \tilde{N}^{(1)}_{kq_1,l} + \beta^m_{q_2q_1} + \beta^m_{q_1l} W^{(1)}_{q_2,l} - e^m_{klq_1} \tilde{N}^{(1)}_{kq_2,l} \right\rangle. \tag{95}$$

Following the same path previously adopted for $\boldsymbol{n}^{(2)}$, the weak form of the first equation of cell problem (40), namely

$$\left(C^m_{ijkl}\,\tilde{N}^{(1)}_{kq_1,l}\right)_{,j} + \left(e^m_{ijk}\,W^{(1)}_{q_1,k}\right)_{,j} + e^m_{ijq_1,j} = 0, \tag{96}$$

takes the form

$$\left\langle \left(C^m_{ijkl}\,\tilde{N}^{(1)}_{kq_1,l} + e^m_{ijk}\,W^{(1)}_{q_1,k} + e^m_{ijq_1}\right)\tilde{N}^{(1)}_{iq_2,j}\right\rangle = 0, \tag{97}$$

thanks to the divergence theorem and $\mathcal{Q}$-periodicity of perturbation functions and micro constitutive tensors. By summing equation (97) and its counterpart, obtained exchanging $q_1$ and $q_2$, to equation (95) one obtains

$$w^{(2)}_{q_1 q_2} = \frac{1}{2}\Big\langle \beta^m_{q_1 q_2} + \beta^m_{q_2 l}\,W^{(1)}_{q_1,l} + C^m_{ijkl}\,\tilde{N}^{(1)}_{kq_1,l}\,\tilde{N}^{(1)}_{iq_2,j} + e^m_{ijk}\,W^{(1)}_{q_1,k}\,\tilde{N}^{(1)}_{iq_2,j}$$
$$+ \beta^m_{q_2 q_1} + \beta^m_{q_1 l}\,W^{(1)}_{q_2,l} + C^m_{ijkl}\,\tilde{N}^{(1)}_{kq_2,l}\,\tilde{N}^{(1)}_{iq_1,j} + e^m_{ijk}\,W^{(1)}_{q_2,k}\,\tilde{N}^{(1)}_{iq_1,j}\Big\rangle. \tag{98}$$

From the second equation of cell problem (40), namely

$$\left(e^m_{kli}\,\tilde{N}^{(1)}_{kq_1,l}\right)_{,i} - \left(\beta^m_{il}\,\tilde{W}^{(1)}_{q_1,l}\right)_{,i} - \beta^m_{iq_1,i} = 0, \tag{99}$$

one can write the weak form

$$\left\langle \left(e^m_{kli}\,\tilde{N}^{(1)}_{kq_2,l} - \beta^m_{il}\,W^{(1)}_{q_2,l} - \beta^m_{iq_2}\right)W^{(1)}_{q_1,i}\right\rangle = 0, \tag{100}$$

assuming $W^{(1)}_{q_1}$ as a test function. By adding the vanishing term (100) and its counterpart to equation (98), the final expression of $w^{(2)}_{q_1 q_2}$ is derived

$$w^{(2)}_{q_1 q_2} = \frac{1}{2}\Big\langle \beta^m_{q_1 q_2} + \beta^m_{q_2 l}\,W^{(1)}_{q_1,l} + C^m_{ijkl}\,\tilde{N}^{(1)}_{kq_1,l}\,\tilde{N}^{(1)}_{iq_2,j} + \beta^m_{il}\,W^{(1)}_{q_1,l}\,W^{(1)}_{q_2,i} + \beta^m_{iq_1}\,W^{(1)}_{q_2,i} +$$
$$+ \beta^m_{q_2 q_1} + \beta^m_{q_1 l}\,W^{(1)}_{q_2,l} + C^m_{ijkl}\,\tilde{N}^{(1)}_{kq_2,l}\,\tilde{N}^{(1)}_{iq_1,j} + \beta^m_{il}\,W^{(1)}_{q_2,l}\,W^{(1)}_{q_1,i} + \beta^m_{iq_2}\,W^{(1)}_{q_1,i}\Big\rangle =$$
$$= \frac{1}{2}\Big\langle \beta^m_{il}\left(W^{(1)}_{q_1,l} + \delta_{q_1 l}\right)\left(W^{(1)}_{q_2,i} + \delta_{q_2 i}\right) + C^m_{ijkl}\,\tilde{N}^{(1)}_{kq_1,l}\,\tilde{N}^{(1)}_{iq_2,j} +$$
$$+ \beta^m_{il}\left(W^{(1)}_{q_2,l} + \delta_{q_2 l}\right)\left(W^{(1)}_{q_1,i} + \delta^{(1)}_{q_1 i}\right) + C^m_{ijkl}\,\tilde{N}^{(1)}_{kq_1,l}\,\tilde{N}^{(1)}_{iq_2,j}\Big\rangle =$$
$$= \Big\langle \beta^m_{il}\left(W^{(1)}_{q_1,l} + \delta_{q_1 l}\right)\left(W^{(1)}_{q_2,i} + \delta_{q_2 i}\right) + C^m_{ijkl}\,\tilde{N}^{(1)}_{kq_1,l}\,\tilde{N}^{(1)}_{iq_2,j}\Big\rangle, \tag{101}$$

which demonstrates the positive definiteness and the symmetry of tensor $\boldsymbol{w}^{(2)}$. Components $w^{(2)}_{q_1 q_2}$ of tensor $\boldsymbol{w}^{(2)}$ correspond to those of the overall symmetric dielectric permittivities tensor $\boldsymbol{\beta}$, thus having $w^{(2)}_{q_1 q_2} = \beta_{q_1 q_2}$.

**Tensor $\boldsymbol{m}^{(2)}$**

Components of tensor $\boldsymbol{m}^{(2)}$ are the known term of the cell problem (36) and are expressed as

$$m^{(2)}_{q_1 q_2} = \frac{1}{2}\left\langle K^m_{q_1 q_2} + K^m_{q_2 j}\,M^{(1)}_{q_1,j} + K^m_{q_2 q_1} + K^m_{q_1 j}\,M^{(1)}_{q_2,j}\right\rangle. \tag{102}$$

If one considers the thermal cell problem (34) of order $\varepsilon^{-1}$

$$\left(K^m_{ij}\,M^{(1)}_{q_1,j}\right)_{,i} + K^m_{iq_1,i} = 0, \tag{103}$$

and multiplies equation (103) by the test function $M^{(1)}_{q_2}$, the following weak form holds

$$\left\langle \left(K^m_{ij}\,M^{(1)}_{q_1,j} + K^m_{iq_1}\right)M^{(1)}_{q_2,i}\right\rangle = 0. \tag{104}$$

Analogously, exploiting $M^{(1)}_{q_1}$ as test function, one has

$$\left\langle \left(K^m_{ij}\,M^{(1)}_{q_2,j} + K^m_{iq_2}\right)M^{(1)}_{q_1,i}\right\rangle = 0. \tag{105}$$

By adding equations (104) and (105) to (102), one obtains

$$m^{(2)}_{q_1q_2} = \frac{1}{2} \left\langle K^m_{q_1q_2} + K^m_{q_1j} M^{(1)}_{q_2,j} + \left( K^m_{ij} M^{(1)}_{q_1,j} + K^m_{iq_1} \right) M^{(1)}_{q_2,i} + K^m_{q_1q_2} + K^m_{q_2j} M^{(1)}_{q_1,j} + \right.$$
$$\left. + \left( K^m_{ij} M^{(1)}_{q_2,j} + K^m_{iq_2} \right) M^{(1)}_{q_1,j} \right\rangle =$$
$$= \frac{1}{2} \left\langle K^m_{ij} \left( M^{(1)}_{q_1,j} + \delta_{jq_1} \right) \left( M^{(1)}_{q_2,i} + \delta_{q_2i} \right) + K^m_{ij} \left( M^{(1)}_{q_1,i} + \delta_{iq_1} \right) \left( M^{(1)}_{q_2,j} + \delta_{q_2j} \right) \right\rangle =$$
$$= \left\langle K^m_{ij} \left( M^{(1)}_{q_1,j} + \delta_{jq_1} \right) \left( M^{(1)}_{q_2,i} + \delta_{q_2i} \right) \right\rangle, \tag{106}$$

which, once again, is the proof of symmetry and positive definiteness of $\boldsymbol{m}^{(2)}$ whose components correspond to those of the symmetric overall thermal conduction tensor $\boldsymbol{K}$, thus having $m^{(2)}_{q_1q_2} = K_{q_1q_2}$.

**Tensors $\tilde{\boldsymbol{n}}^{(2)}$ and $\tilde{\boldsymbol{w}}^{(2)}$**
In the following it will be shown that $\tilde{n}^{(2)}_{pq_1q_2} = \tilde{w}^{(2)}_{pq_1q_2}$. To prove this equality, one considers the expression of the components of $\tilde{\boldsymbol{n}}^{(2)}$, namely the known term of the first equation of cell problem (46)

$$\tilde{n}^{(2)}_{pq_1q_2} = \frac{1}{2} \left\langle C^m_{pq_2kl} \tilde{N}^{(1)}_{kq_1,l} + e^m_{pq_1q_2} + e^m_{pq_2k} W^{(1)}_{q_1,k} + C^m_{pq_1kl} \tilde{N}^{(1)}_{kq_2,l} + e^m_{pq_2q_1} + e^m_{pq_1k} W^{(1)}_{q_2,k} \right\rangle. \tag{107}$$

Taking the equations (87) and (91) of cell problem (38) at the order $\varepsilon^{-1}$, their weak forms can be written as:
$$\left\langle \left( C^m_{pjkl} N^{(1)}_{kiq_1,l} + e^m_{pjk} \tilde{W}^{(1)}_{iq_1,k} + C^m_{pjiq_1} \right) \tilde{N}^{(1)}_{pq_2,j} \right\rangle = 0, \tag{108}$$

and
$$\left\langle \left( e^m_{klj} N^{(1)}_{kiq_1,l} - \beta^m_{jl} \tilde{W}^{(1)}_{iq_1,l} + e^m_{iq_1j} \right) W^{(1)}_{q_2,j} \right\rangle = 0, \tag{109}$$

with $\tilde{N}^{(1)}_{pq_2}$ and $W^{(1)}_{q_2}$ as test functions. Summation of equations (108), (109) and their counterpart, obtained exchanging indices $q_1$ and $q_2$, to equation (107), leads to

$$\tilde{n}^{(2)}_{pq_1q_2} = \frac{1}{2} \left( \langle e^m_{ijk} \left( \delta_{pi} \delta_{jq_1} \delta_{kq_2} - N^{(1)}_{ipq_2,j} W^{(1)}_{q_1,k} - \tilde{W}^{(1)}_{pq_2,k} \tilde{N}^{(1)}_{iq_1,j} \right) - C^m_{ijkl} N^{(1)}_{ipq_2,j} \tilde{N}^{(1)}_{kq_1,l} + \right.$$
$$+ \beta^m_{il} \tilde{W}^{(1)}_{pq_2,i} W^{(1)}_{q_1,l} + e^m_{ijk} \left( \delta_{pi} \delta_{jq_2} \delta_{kq_1} - Nipq_1, j^{(1)} W^{(1)}_{q_2,k} - \tilde{W}^{(1)}_{pq_1,k} \tilde{N}^{(1)}_{iq_2,j} \right) +$$
$$\left. - C^m_{ijkl} N^{(1)}_{ipq_1,j} \tilde{N}^{(1)}_{kq_2,l} + \beta^m_{il} \tilde{W}^{(1)}_{pq_1,i} W^{(1)}_{q_2,l} \right\rangle. \tag{110}$$

In the same way, $\tilde{w}^{(2)}_{pq_1q_2}$ is the known term of the second equation of cell problem (44) at the order $\varepsilon^0$

$$\tilde{w}^{(2)}_{pq_1q_2} = \frac{1}{2} \left\langle e^m_{klq_2} N^{(1)}_{kpq_1,l} + e^m_{pq_2q_1} - \beta^m_{q_2l} \tilde{W}^{(1)}_{pq_1,l} + e^m_{klq_1} N^{(1)}_{kpq_2,l} + e^m_{pq_1q_2} - \beta^m_{q_1l} \tilde{W}^{(1)}_{pq_1,l} \right\rangle. \tag{111}$$

The weak forms of the first and the second equation of cell problem (40), namely equations (96) and (99), take the form
$$\left\langle C^m_{ijkl} \tilde{N}^{(1)}_{kq_1,l} + e^m_{ijk} W^{(1)}_{q_1,k} + e^m_{ijq_1} \right\rangle N^{(1)}_{ipq_2,j} = 0, \tag{112}$$

and
$$\left\langle e^m_{kli} \tilde{N}^{(1)}_{kq_1,l} - \beta^m_{il} W^{(1)}_{q_1,l} - \beta^m_{iq_1} \right\rangle \tilde{W}^{(1)}_{pq_2,i} = 0, \tag{113}$$

with $N^{(1)}_{ipq_2}$ and $\tilde{W}^{(1)}_{pq_2}$ as test functions. Once again, by adding equations (112) and (113) and their counterparts to equation (111), one finally has

$$\tilde{w}^{(2)}_{pq_1q_2} = \frac{1}{2} \left( \langle e^m_{ijk} \left( \delta_{pi} \delta_{jq_1} \delta_{kq_2} - N^{(1)}_{ipq_2,j} W^{(1)}_{q_1,k} - \tilde{W}^{(1)}_{pq_2,k} \tilde{N}^{(1)}_{iq_1,j} \right) - C^m_{ijkl} N^{(1)}_{ipq_2,j} \tilde{N}^{(1)}_{kq_1,l} + \right.$$
$$+ \beta^m_{il} \tilde{W}^{(1)}_{pq_2,i} W^{(1)}_{q_1,l} + e^m_{ijk} \left( \delta_{pi} \delta_{jq_2} \delta_{kq_1} - N^{(1)}_{ipq_1,j} W^{(1)}_{q_2,k} - \tilde{W}^{(1)}_{pq_1,k} \tilde{N}^{(1)}_{iq_2,j} \right) +$$
$$\left. - C^m_{ijkl} N^{(1)}_{ipq_1,j} \tilde{N}^{(1)}_{kq_2,l} + \beta^m_{il} \tilde{W}^{(1)}_{pq_1,i} W^{(1)}_{q_2,l} \right\rangle, \tag{114}$$

from which is clear that $\tilde{n}^{(2)}_{pq_1q_2} = \tilde{w}^{(2)}_{pq_1q_2}$. Furthermore one has

$$\tilde{n}^{(2)}_{pq_1q_2} \frac{\partial^2 \Phi}{\partial x_{q_1} \partial x_{q_2}} = \frac{1}{2}\left(e_{pq_1q_2} + e_{pq_2q_1}\right) \frac{\partial^2 \Phi}{\partial x_{q_1} \partial x_{q_2}} = e_{pq_1q_2} \frac{\partial^2 \Phi}{\partial x_{q_1} \partial x_{q_2}},$$

where the components of the overall piezoelectric tensor **e** are expressed as

$$e_{pq_1q_2} = \frac{1}{2}\left\langle e^m_{ijk}\left(\delta_{pi}\,\delta_{jq_1}\,\delta_{kq_2} - N^{(1)}_{ipq_2,j}\,W^{(1)}_{q_1,k} - \tilde{W}^{(1)}_{pq_2,k}\,\tilde{N}^{(1)}_{iq_1,j}\right) - C^m_{ijkl}\,N^{(1)}_{ipq_2,j}\,\tilde{N}^{(1)}_{kq_1,l} + \beta^m_{il}\,\tilde{W}^{(1)}_{pq_2,i}\,W^{(1)}_{q_1,l}\right\rangle.$$

## C  Finite Element Formulation

In this appendix, a detailed formulation of the thermo-piezoelectric finite element framework developed to solve the whole heterogeneous problem and the cell problems is given. The expression of the linear constitutive equations (4a)-(4c) for the thermo-piezoelectric material is

$$\begin{aligned}
\sigma_{ij} &= C^m_{ijkl}\,u_{k,l} + e^m_{ijk}\,\phi_{,k} - \alpha^m_{ij}\,\theta,\\
D_i &= e^m_{kli}\,u_{k,l} - \beta^m_{il}\,\phi_{,l} + \gamma^m_i\,\theta,\\
q_i &= -K^m_{ij}\,\theta_{,j},
\end{aligned} \qquad (115)$$

where, in a 2-D space, the rigorous form of constitutive equations (115) in a tensorial fashion is (Mehrabadi and Cowin, 1990)

$$\begin{pmatrix} \sigma_{11} \\ \sigma_{22} \\ \sqrt{2}\,\sigma_{12} \end{pmatrix} = \begin{pmatrix} C_{1111} & C_{1122} & \sqrt{2}\,C_{1112} \\ C_{2211} & C_{2222} & \sqrt{2}\,C_{2212} \\ \sqrt{2}\,C_{1211} & \sqrt{2}\,C_{1222} & 2\,C_{1212} \end{pmatrix} \begin{pmatrix} u_{1,1} \\ u_{2,2} \\ \frac{\sqrt{2}}{2}(u_{1,2}+u_{2,1}) \end{pmatrix} +$$

$$+ \begin{pmatrix} e_{111} & e_{112} \\ e_{221} & e_{222} \\ \sqrt{2}\,e_{121} & \sqrt{2}\,e_{122} \end{pmatrix} \begin{pmatrix} \phi_{,1} \\ \phi_{,2} \end{pmatrix} - \begin{pmatrix} \alpha_{11} \\ \alpha_{22} \\ \sqrt{2}\,\alpha_{12} \end{pmatrix} \theta,$$

$$\begin{pmatrix} D_1 \\ D_2 \end{pmatrix} = \begin{pmatrix} e_{111} & e_{221} & \sqrt{2}\,e_{121} \\ e_{112} & e_{222} & \sqrt{2}\,e_{122} \end{pmatrix} \begin{pmatrix} u_{1,1} \\ u_{2,2} \\ \frac{\sqrt{2}}{2}(u_{1,2}+u_{2,1}) \end{pmatrix} +$$

$$- \begin{pmatrix} \beta_{11} & \beta_{12} \\ \beta_{21} & \beta_{22} \end{pmatrix} \begin{pmatrix} \phi_{,1} \\ \phi_{,2} \end{pmatrix} + \begin{pmatrix} \gamma_1 \\ \gamma_2 \end{pmatrix} \theta,$$

$$\begin{pmatrix} q_1 \\ q_2 \end{pmatrix} = -\begin{pmatrix} K_{11} & K_{12} \\ K_{21} & K_{22} \end{pmatrix} \begin{pmatrix} \theta_{,1} \\ \theta_{,2} \end{pmatrix}. \qquad (116)$$

The stress tensor $\boldsymbol{\sigma}$, the electric displacement vector **D**, and the heat flux vector **q** satisfy the local balance equations that take the following form in the hypothesis of quasi-static processes

$$\begin{aligned}
\left(C^m_{ijkl}\,u_{k,l}\right)_{,j} + \left(e^m_{ijk}\,\phi_{,k}\right)_{,j} - \left(\alpha^m_{ij}\,\theta\right)_{,j} + b_i &= 0,\\
\left(e^m_{kli}\,u_{k,l}\right)_{,i} - \left(\beta^m_{il}\,\phi_{,l}\right)_{,i} + \left(\gamma^m_i\,\theta\right)_{,i} &= \rho_e,\\
\left(K^m_{ij}\,\theta_{,j}\right)_{,i} + r &= 0,
\end{aligned} \qquad (117)$$

in the presence of body forces **b**, free charge densities $\rho_e$, and heat sources $r$. The boundary $\partial\Omega$ of domain $\Omega$ is the union of a Dirichlet ($\partial\Omega_u, \partial\Omega_\phi, \partial\Omega_\theta$) and a Neumann ($\partial\Omega_\sigma, \partial\Omega_D, \partial\Omega_q$) parts with $\partial\Omega = \partial\Omega_u \cup \partial\Omega_\sigma = \partial\Omega_\phi \cup \partial\Omega_D = \partial\Omega_\theta \cup \partial\Omega_q$ and $\partial\Omega_u \cap \partial\Omega_\sigma = \partial\Omega_\phi \cap \partial\Omega_D = \partial\Omega_\theta \cap \partial\Omega_q = \emptyset$. Boundary conditions can therefore be written in the form

$$\begin{cases} u_i = \bar{u}_i \text{ on } \partial\Omega_u \\ \sigma_{ij}\,n_j = \bar{t}_i \text{ on } \partial\Omega_\sigma \end{cases}, \quad \begin{cases} \phi = \bar{\phi}_i \text{ on } \partial\Omega_\phi \\ D_i\,n_i = -\bar{\tau}_e \text{ on } \partial\Omega_D \end{cases}, \quad \begin{cases} \theta = \bar{\theta}_i \text{ on } \partial\Omega_\theta \\ q_i\,n_i = \bar{q}_i \text{ on } \partial\Omega_q \end{cases}, \qquad (118)$$

where $\bar{t}_i, \bar{\tau}_e$ and $\bar{q}$ are the prescribed values of tractions, free surface charge and heat flux, respectively, and **n** is the outward normal to $\partial\Omega$. After introducing the three test functions $\psi_{u_i}, \psi_\phi$ and $\psi_\theta$, local balance equations (117) and boundary conditions (118) allow to write the weak forms

$$-\int_\Omega \left(C^m_{ijkl}\,u_{k,l} + e^m_{ijk}\,\phi_{,k} - \alpha_{ij}\,\theta\right)\psi_{u,j}\,\mathrm{d}V + \int_\Omega b_i\,\psi_{u_i}\,\mathrm{d}V + \int_{\partial\Omega_\sigma} \bar{t}_i\,\psi_{u_i}\,\mathrm{d}S = 0, \quad \forall \psi_{u_i}\text{ s.t. } \psi_{u_i} = 0 \text{ on } \partial\Omega_u,$$

$$-\int_\Omega (e^m_{kli}\, u_{k,l} - \beta^m_{il}\, \phi_{,l} + \gamma_i\, \theta)\, \psi_{\phi,i}\, dV - \int_\Omega \rho_e\, \psi_\phi\, dV - \int_{\partial \Omega_D} \bar{\tau}_e\, \psi_\phi\, dS = 0, \quad \forall \psi_\phi \text{ s.t. } \psi_\phi = 0 \text{ on } \partial \Omega_\phi,$$

$$+\int_\Omega K^m_{ij}\, \theta_{,j}\, \psi_{\theta,i}\, dV + \int_\Omega r\, \psi_\theta\, dV - \int_{\partial \Omega_q} \bar{q}\, \psi_\theta\, dS = 0, \quad \forall \psi_\theta \text{ s.t. } \psi_\theta = 0 \text{ on } \partial \Omega_\theta. \tag{119}$$

In the finite element discretization, the displacement field $\mathbf{u}(\mathbf{x})$, the electric potential field $\phi(\mathbf{x})$, and the relative temperature field $\theta(\mathbf{x})$ are approximated by a linear combination of shape functions $N_j(\mathbf{x})$ and nodal unknowns $u_{ij}, \phi_j$ and $\theta_j$

$$u_i(\mathbf{x}) = \sum_{j=1}^{N_h} N_j(\mathbf{x})\, u_{ij}, \quad \phi(\mathbf{x}) = \sum_{j=1}^{N_h} N_j(\mathbf{x})\, \phi_j, \quad \theta(\mathbf{x}) = \sum_{j=1}^{N_h} N_j(\mathbf{x})\, \theta_j, \tag{120}$$

where $N_h$ is the finite dimension of a space $V_h$ for which $\{N_j | j = 1, 2, ..., N_h\}$ is a basis. Analogously, for test functions one has

$$\psi_{u_i}(\mathbf{x}) = \sum_{j=1}^{N_h} N_j(\mathbf{x})\, \delta u_{ij}, \quad \psi_\phi(\mathbf{x}) = \sum_{j=1}^{N_h} N_j(\mathbf{x})\, \delta\phi_j, \quad \psi_\theta(\mathbf{x}) = \sum_{j=1}^{N_h} N_j(\mathbf{x})\, \delta\theta_{ij}. \tag{121}$$

Being $\mathbf{x} = x_1\, \mathbf{e}_1 + x_2\, \mathbf{e}_2$ in a two-dimensional space, one can define the following matrices on the single finite element $e$

$$\mathbf{B}_u = \mathbf{D}_u\, \mathbf{N}_u, \quad \mathbf{B}_\phi = \mathbf{D}_\phi\, \mathbf{N}_\phi, \quad \mathbf{B}_\theta = \mathbf{D}_\theta\, \mathbf{N}_\theta, \tag{122}$$

where differential matrices $\mathbf{D}_u, \mathbf{D}_\phi$ and $\mathbf{D}_\theta$ are

$$\mathbf{D}_u = \begin{bmatrix} \frac{\partial}{\partial x_1} & 0 \\ 0 & \frac{\partial}{\partial x_2} \\ \frac{\partial}{\partial x_2} & \frac{\partial}{\partial x_1} \end{bmatrix}, \quad \mathbf{D}_\phi = \mathbf{D}_\theta = \begin{bmatrix} \frac{\partial}{\partial x_1} \\ \frac{\partial}{\partial x_2} \end{bmatrix}, \tag{123}$$

and matrices $\mathbf{N}_u, \mathbf{N}_\phi$ and $\mathbf{N}_\theta$ collect the shapes functions

$$\mathbf{N}_u = \begin{bmatrix} N_1 & 0 & N_2 & 0 & ... & N_{Nnod} & 0 \\ 0 & N_1 & 0 & N_2 & ... & 0 & N_{Nnod} \end{bmatrix}, \quad \mathbf{N}_\phi = \mathbf{N}_\theta = \begin{bmatrix} N_1 & N_2 & ... & N_{Nnod} \end{bmatrix}, \tag{124}$$

with $N_{Nnod}$ being the number of element nodes.

Over each element domain $\delta\Omega_e$, weak forms (119) can therefore be written in matrix form as

$$-\boldsymbol{\delta u}^T \int_{\Omega_e} \mathbf{B}_u^T\, \mathbf{C}^m\, \mathbf{B}_u\, dV\, \mathbf{u} - \boldsymbol{\delta u}^T \int_{\Omega_e} \mathbf{B}_u^T\, \mathbf{e}^m\, \mathbf{B}_\phi\, dV\, \boldsymbol{\phi} + \boldsymbol{\delta u}^T \int_{\Omega_e} \mathbf{B}_u^T\, \boldsymbol{\alpha}^m\, \mathbf{N}_\theta\, dV\, \boldsymbol{\theta} +$$

$$+ \boldsymbol{\delta u}^T \int_{\Omega_e} \mathbf{N}_u^T\, \mathbf{b}\, dV + \boldsymbol{\delta u}^T \int_{\partial \Omega_{e_\sigma}} \mathbf{N}_u^T\, \bar{\mathbf{t}}\, dS = 0,$$

$$-\boldsymbol{\delta\phi}^T \int_{\Omega_e} \mathbf{B}_\phi^T\, \tilde{\mathbf{e}}^m\, \mathbf{B}_u\, dV\, \mathbf{u} + \boldsymbol{\delta\phi}^T \int_{\Omega_e} \mathbf{B}_\phi^T\, \boldsymbol{\beta}^m\, \mathbf{B}_\phi\, dV\, \boldsymbol{\phi} - \boldsymbol{\delta\phi}^T \int_{\Omega_e} \mathbf{B}_\phi^T\, \boldsymbol{\gamma}^m\, \mathbf{N}_\theta\, dV\, \boldsymbol{\theta} = 0 +$$

$$-\boldsymbol{\delta\phi}^T \int_{\Omega_e} \mathbf{N}_\phi^T\, \rho_e\, dV - \boldsymbol{\delta\phi}^T \int_{\partial \Omega_{e_D}} \mathbf{N}_\phi^T\, \bar{\tau}_e\, dS = 0,$$

$$\boldsymbol{\delta\theta}^T \int_{\Omega_e} \mathbf{B}_\theta^T\, \mathbf{K}^m\, \mathbf{B}_\theta\, dV\, \boldsymbol{\theta} + \boldsymbol{\delta\theta}^T \int_{\Omega_e} \mathbf{N}_\theta^T\, r\, dV - \boldsymbol{\delta\theta}^T \int_{\partial \Omega_{e_q}} \mathbf{N}_\theta^T\, \bar{q}\, dS = 0, \tag{125}$$

which must be satisfied for all $\boldsymbol{\delta u}, \boldsymbol{\delta\phi}$ and $\boldsymbol{\delta\theta}$. In equations (125), constitutive tensors $\mathbb{C}^m, \boldsymbol{\beta}^m, \boldsymbol{K}^m, \boldsymbol{e}^m, \boldsymbol{\alpha}^m$, and $\boldsymbol{\gamma}^m$ have been turned to the corresponding matrix representation $\mathbf{C}^m$, $\beta^m$, $\mathbf{K}^m$, $\mathbf{e}^m$, $\alpha^m$, and $\gamma^m$. After defining the following elemental stiffness matrices

$$\mathbf{K}^e_{uu} = \int_{\Omega_e} \mathbf{B}_u^T\, \mathbf{C}^m\, \mathbf{B}_u\, dV, \quad \mathbf{K}^e_{u\phi} = \int_{\Omega_e} \mathbf{B}_u^T\, \mathbf{e}^m\, \mathbf{B}_\phi\, dV, \quad \mathbf{K}^e_{u\theta} = -\int_{\Omega_e} \mathbf{B}_u^T\, \alpha^m\, \mathbf{B}_\theta\, dV,$$

$$\mathbf{K}^e_{\phi u} = \int_{\Omega_e} \mathbf{B}_\phi^T\, \tilde{\mathbf{e}}^m\, \mathbf{B}_u\, dV, \quad \mathbf{K}^e_{\phi\phi} = -\int_{\Omega_e} \mathbf{B}_\phi^T\, \beta^m\, \mathbf{B}_\phi\, dV, \quad \mathbf{K}^e_{\phi\theta} = \int_{\Omega_e} \mathbf{B}_\phi^T\, \gamma^m\, \mathbf{N}_\theta\, dV,$$

$$\mathbf{K}_{\theta\theta}^e = \int_{\Omega_e} \mathbf{B}_\theta^T \, \mathbf{K}^m \, \mathbf{B}_\theta \, \mathrm{d}V, \tag{126}$$

and the external force vectors as

$$\mathbf{f}_{u_{ext}}^e = \int_{\Omega_e} \mathbf{N}_u^T \, \mathbf{b} \, \mathrm{d}V + \int_{\partial\Omega_{e_\sigma}} \mathbf{N}_u^T \, \bar{\mathbf{t}} \, \mathrm{d}S,$$

$$\mathbf{f}_{\phi_{ext}}^e = -\int_{\Omega_e} \mathbf{N}_\phi^T \, \rho_e \, \mathrm{d}V - \int_{\partial\Omega_{e_D}} \mathbf{N}_\phi^T \, \bar{\tau}_e \, \mathrm{d}S,$$

$$\mathbf{f}_{\theta_{ext}}^e = \int_{\Omega_e} \mathbf{N}_\theta^T \, r \, \mathrm{d}V - \int_{\partial\Omega_{e_q}} \mathbf{N}_\theta^T \, \bar{q} \, \mathrm{d}S, \tag{127}$$

the following linear system provides the finite element solution $\mathbf{z} = \{\mathbf{u}, \boldsymbol{\phi}, \boldsymbol{\theta}\}^T$

$$\begin{bmatrix} \mathbf{K}_{uu} & \mathbf{K}_{u\phi} & \mathbf{K}_{u\theta} \\ \mathbf{K}_{\phi u} & \mathbf{K}_{\phi\phi} & \mathbf{K}_{\phi\theta} \\ \mathbf{0} & \mathbf{0} & \mathbf{K}_{\theta\theta} \end{bmatrix} \begin{Bmatrix} \boldsymbol{u} \\ \boldsymbol{\phi} \\ \boldsymbol{\theta} \end{Bmatrix} = \begin{Bmatrix} \mathbf{f}_{u_{ext}} \\ \mathbf{f}_{\phi_{ext}} \\ \mathbf{f}_{\theta_{ext}} \end{Bmatrix}, \tag{128}$$

where the elemental stiffness matrices (126) and the external forces vectors (127) have been assembled in the corresponding global ones.

The thermo-piezoelectric element has been implemented in the finite element software FEAP for numerically solving the coupled thermo-electromechanical problem, exploiting the isoparametric concept to approximate the element geometry and using triangular finite elements.

## C.1 Periodic boundary conditions

In order to impose periodic boundary conditions on displacements, electric potential, and relative temperature fields, and on tractions $\bar{\mathbf{t}}$, free surface charge densities $\bar{\tau}_e$, and heat fluxes $\bar{q}$, a static condensation of the global stiffness matrix, as defined in equation (128), has been performed.

All the nodes of the considered discretization can be classified in three different sets: the internal nodes, indicated with $i$; the nodes belonging to the left and upper boundaries of the system, indicated with $b^+$; and the nodes belonging to the right and lower boundaries of the system, indicated with $b^-$. It is therefore possible to reorder the degrees of freedom $\mathbf{u}, \boldsymbol{\phi}$, and $\boldsymbol{\theta}$ of system (128), collected in the generalized vector $\mathbf{z}$, in three different vectors referring, respectively, to the internal nodes $\mathbf{z}_i$, to the left and upper boundaries nodes $\mathbf{z}_{b^+}$, and to the right and lower boundaries nodes $\mathbf{z}_{b^-}$, in order to rewrite the system (128) in the form

$$\begin{bmatrix} \mathbf{K}_{ii} & \mathbf{K}_{ib^+} & \mathbf{K}_{ib^-} \\ \mathbf{K}_{b^+i} & \mathbf{K}_{b^+b^+} & \mathbf{K}_{b^+b^-} \\ \mathbf{K}_{b^-i} & \mathbf{K}_{b^-b^+} & \mathbf{K}_{b^-b^-} \end{bmatrix} \begin{Bmatrix} \mathbf{z}_i \\ \mathbf{z}_{b^+} \\ \mathbf{z}_{b^-} \end{Bmatrix} = \begin{Bmatrix} \mathbf{f}_{i_{ext}} \\ \mathbf{f}_{b_{ext}^+} \\ \mathbf{f}_{b_{ext}^-} \end{Bmatrix}. \tag{129}$$

The following conditions have to be imposed in order to assure the periodicity of the unknown fields and external forces

$$\begin{cases} \mathbf{b}^+ = \mathbf{b}^- \\ \mathbf{f}_{b_{ext}^+} = -\mathbf{f}_{b_{ext}^-} \end{cases}. \tag{130}$$

From conditions (130), one finally has

$$\mathbf{z}_i = \left[ \mathbf{K}_{ii} - (\mathbf{K}_{ib^+} + \mathbf{K}_{ib^-}) \left( \mathbf{K}_{b^-b^+} + \mathbf{K}_{b^-b^-} + \mathbf{K}_{b^+b^+} + \mathbf{K}_{b^+b^-} \right)^{-1} (\mathbf{K}_{b^+i}\mathbf{K}_{b^-i}) \right]^{-1} \mathbf{f}_{i_{ext}}$$

$$\mathbf{z}_{b^+} = \mathbf{z}_{b^-} = -\left( \mathbf{K}_{b^-b^+} + \mathbf{K}_{b^-b^-} + \mathbf{K}_{b^+b^+} + \mathbf{K}_{b^+b^-} \right)^{-1} (\mathbf{K}_{b^+i} + \mathbf{K}_{b^-i}) \mathbf{z}_i \tag{131}$$